\documentclass[a4paper,fleqn,usenatbib]{mnras}

\usepackage{mathptmx}
\usepackage{ae,aecompl}

\usepackage{graphicx}	% Including figure files
\usepackage{amsmath}	% Advanced maths commands
\usepackage{amssymb}	% Extra maths symbols

\usepackage{float}

\usepackage{color,epsfig} % to use epsfig 
\usepackage[utf8]{inputenc} % to get the accent in my name

\usepackage{multirow}

\usepackage[flushleft]{threeparttable}
\makeatletter
\newlength{\abovecaptionskip}%
\makeatother

\newcommand{\be}{\begin{equation}}
\newcommand{\ee}{\end{equation}}

\definecolor{orange}{RGB}{236,133,40}
\providecommand{\comment}[1]{{\textbf{\textcolor{red}{[#1]}}}}
\providecommand{\comment}[1]{} % comment to other line to hide comments
\providecommand{\caveat}[1]{}

\newcommand*\diff{\mathop{}\!\mathrm{d}}
\newcommand{\vect}[1]{\mathbfit{#1}}

\newcommand{\mstar}{M_{\star}}
\newcommand{\rstar}{R_{\star}}

\newcommand{\rp}{R_{\rm p}}
\newcommand{\rt}{R_{\rm t}}

\newcommand{\rint}{R_{\rm int}}
\newcommand{\racc}{R_{\rm acc}}

\newcommand{\mdot}{\dot{M}}

\newcommand{\mh}{M_{\rm h}}
\newcommand{\tmin}{t_{\rm min}}

\newcommand{\tp}{t_{\rm p}}

\newcommand{\tstar}{t_{\star}}

\newcommand{\rg}{R_{\rm g}}

\def\msun{\, \mathrm{M}_{\hbox{$\odot$}}}
\def\rsun{\, \mathrm{R}_{\hbox{$\odot$}}}
\def\gpercm3{\, \rm g \, cm^{-3}}
\def\ergcm3{\, \rm erg \, cm^{-3}}
\def\ergcm3s{\, \rm erg \, cm^{-3} \, s^{-1}}
\def\kmpers{\, \rm km \, s^{-1}}
\def\days{\, \rm d}
\def\hours{\, \rm h}

\def\kelvin{\, \rm K}
\def\gauss{\, \rm G}

\def\ergpers{\, \rm erg\, s^{-1}}
\def\msunperyr{\, \rm \mathrm{M}_{\hbox{$\odot$}} \, {yr}^{-1}}
\def\cm2g{\, \rm cm^{2} \, g^{-1}}
\usepackage{pdfrender}

\newcommand*{\boldgreek}[1]{%
  \textpdfrender{%
    TextRenderingMode=FillStroke,%
    LineWidth=.35pt,%
  }{#1}%
}

\title[Nozzle shock in TDEs]{The nozzle shock in tidal disruption events}

\author[C. Bonnerot and W. Lu]{Clément Bonnerot$^{1,2}$%
\thanks{E-mail: clement.bonnerot@nbi.ku.dk} and Wenbin Lu$^{1,3}$%
\thanks{E-mail: wenbinlu@astro.princeton.edu}
\\
$^{1}$TAPIR, Mailcode 350-17, California Institute of Technology, Pasadena, CA 91125, USA\\
$^{2}$Niels Bohr International Academy, Niels Bohr Institute, Blegdamsvej 17, DK-2100 Copenhagen Ø, Denmark\\
$^{3}$Department of Astrophysical Sciences, Princeton University, NJ 08544, USA\\
}
\date{Accepted XXX. Received YYY; in original form ZZZ}

\pubyear{2021}
\begin{document}
\label{firstpage}
\pagerange{\pageref{firstpage}--\pageref{lastpage}}
\maketitle

\begin{abstract}

Tidal disruption events (TDEs) occur when a star gets torn apart by the strong tidal forces of a supermassive black hole, which results in the formation of a debris stream that partly falls back towards the compact object. This gas moves along inclined orbital planes that intersect near pericenter, resulting in a so-called ``nozzle shock''. We perform the first dedicated study of this interaction, making use of a two-dimensional simulation that follows the transverse gas evolution inside a given section of stream. This numerical approach circumvents the lack of resolution encountered near pericenter passage in global three-dimensional simulations using particle-based methods. As it moves inward, we find that the gas motion is purely ballistic, which near pericenter causes strong vertical compression that squeezes the stream into a thin sheet. Dissipation takes place at the resulting nozzle shock, inducing a rise in pressure that causes the collapsing gas to bounce back, although without imparting significant net expansion. As it recedes to larger distances, this matter continues to expand while remaining thin despite the influence of pressure forces. This gas evolution specifies the strength of the subsequent self-crossing shock, which we find to be more affected by black hole spin than previously estimated. We also evaluate the impact of general-relativistic effects, viscous dissipation, magnetic fields and radiative processes on the nozzle shock. This study represents an important step forward in the theoretical understanding of TDEs, bridging the gap between our robust knowledge of the fallback rate and the more complex following stages, during which most of the emission occurs.

\end{abstract}


%%%%%%%%%%%%%%%%%%%%%%%%%%%%%%%%%%%%%%%%%%%%%%%%%%

%%%%%%%%%%%%%%%%% BODY OF PAPER %%%%%%%%%%%%%%%%%%

\vspace{-10cm}

\section{Introduction}

Stars near the center of galaxies occasionally get launched on a plunging near-parabolic trajectory that leads to their disruption by the strong tidal forces of the central supermassive black hole. This phenomenon called tidal disruption event (TDE) results in the formation of a debris stream that progressively fuels the compact object, leading to strong electromagnetic emission outshining the entire host galaxy \citep{rees1988}. This signal represents a powerful probe of the majority of supermassive black holes in the local universe that are otherwise starved of accreting material, and therefore undetectable.

Many TDE candidates have already been discovered as powerful flares of radiation lasting from months to years and originating from the innermost region of otherwise quiescent galaxies. This emission can cover a wide range of the electromagnetic spectrum, which usually includes components in the optical/UV or the soft X-ray bands \citep{komossa2008-j095,gezari2012,chornock2014,saxton2017,leloudas2016,gezari2017-15oi,leloudas2019,holoien2019,van_velzen2019,hung2020,short2020} that likely have different physical origins. Several of these observations additionally present strong evidence of matter present on large scales with outflowing motion, which is expected given the violent interactions that the stellar matter can undergo as it reaches the vicinity of the black hole to eventually get accreted.

The stellar disruption itself is overall well-understood that makes quantitative predictions possible for the hydrodynamics of this early phase. For deep encounters, this process may be accompanied by strong compression potentially triggering nuclear reactions, as first proposed by \citet{carter1982}. The stellar debris then evolves into an elongated stream, whose transverse profile can be confined by the gas self-gravity \citep{kochanek1994,coughlin2015-variability,coughlin2016-structure,steinberg2019}. About half of this stream is bound to the black hole, thus returning to its vicinity according to a predictable fallback rate \citep{evans1989,lodato2009,guillochon2013} that chiefly depends on the stellar structure. However, the gas is not expected to emit observable radiation\footnote{During the stellar disruption, it has been proposed that an X-ray shock breakout signal gets emitted due to strong compression, although with a very short duration \citep{guillochon2009,kobayashi2004}. It is also possible that the debris stream radiates due to hydrogen recombination \citep{kasen2010} but at a luminosity much lower than that expected when the gas returns near the black hole.} until it passes at pericenter for the second time.

Starting from the return of this matter to the black hole, our understanding of the hydrodynamics becomes much less secure. The main difficulty is a computational one that prevents from accurately following the pericenter passage of the stream within a global three-dimensional simulation. An early simulation by \cite{lee1996_tvd} displays a large expansion of the gas at this location that they suspect could be caused by a lack of resolution. Later on, \cite{ayal2000} similarly finds that the gas gets strongly heated during pericenter passage, which has the more extreme consequence of unbinding most of the mass. It now appears likely that the gas evolution found in these works is significantly affected by numerical artefacts caused by a too low resolution. This issue at least partly stems from the fact that the stream of debris displays a very large aspect ratio with a longitudinal extent larger than its transverse width by several orders of magnitude, which makes it difficult to discretize with enough resolution elements in both directions. Although numerical errors seem to become less catastrophic when resolution is improved, there is so far no convincing evidence that the problem completely disappears, even when approaching the limit of currently available computational resources.

In order to alleviate this computational burden, most numerical investigations have so far relied on simplifications of the problem compared to the physically-motivated situation involving a parabolic stellar trajectory and supermassive black hole (for a recent review, see \citet{bonnerot2021-review}). This is usually achieved by either decreasing the stellar eccentricity \citep{hayasaki2013,bonnerot2016-circ,sadowski2016} or the black hole mass \citep{rosswog2009,guillochon2014-10jh,shiokawa2015}, which both tend to reduce the longitudinal size of the stream, making it numerically easier to resolve. The advantage of these simulations is that they can follow the global evolution of the gas in three dimensions. In particular, they capture the self-crossing shock induced by relativistic precession that results from an intersection between the part of the stream going away from the black hole after pericenter passage and that still moving inward. These works find that this collision can initiate the formation of an accretion flow, which is likely associated with most of the radiation observed from TDEs. However, it is not generally possible to extrapolate these results obtained for simplified initial conditions to the more physically realistic situation. More recently, \cite{andalman2020} performed a simulation that does not rely on these simplifications, but still considers a less likely deeply-penetrating disruption. Due to the high computational needs involved, the gas could however only be followed for a short duration even making use of their efficient GPU-accelerated code.

In another class of works, the authors have chosen to treat the passage of the stream near the black hole in an analytical way, which is most commonly achieved by assuming that the center of the stream evolves ballistically with either a full or approximate calculation of general-relativistic geodesics \citep{dai2013,guillochon2015,dai2015,bonnerot2017-stream,lu2020}. Such calculations were mainly used to determine the properties of the two stream components involved in the self-crossing shock. This first interaction was then studied by means of local simulations \citep{lee1996,kim1999,jiang2016,lu2020} to determine the properties of the resulting outflow. Making use of this information, \citet{bonnerot2020-realistic} were able to perform the first simulation of the subsequent accretion disc formation for astrophysically realistic parameters, by treating the outflow through an injection of matter. However, a significant uncertainty in the above analytical treatments concerns the level of stream expansion occurring during pericenter passage. This effect could lead to different properties for the colliding streams, which may reduce the efficiency of the self-crossing shock and cause the outflow to become less spherical than for identical streams (see figure 3 of \cite{bonnerot2021-review} for an illustration of this difference). This mechanism is also fundamental if the black hole has a non-zero spin, which can cause the streams to be offset at the intersection point due to relativistic nodal precession. While the collision may be entirely missed if the streams are both very thin, some of the gas could nevertheless collide if the vertical offset is compensated by an increase in stream thickness \citep{dai2013,guillochon2015,hayasaki2016-spin,liptai2019-spin}.

The main interaction taking place when the stream passes near the black hole is usually referred to as a `nozzle shock'. It results from a strong vertical compression induced by an intersection of the orbital planes of the returning gas near pericenter. This phenomenon is similar to that involved in the process of deep stellar disruption, during which the entire star gets squeezed \citep{carter1982,luminet1985,brassart2008,stone2013}. However, the nozzle shock has not received as much attention so far. One of the earliest investigations of this process has been done in the pioneering work by \citet{kochanek1994} that studies the stream evolution at pericenter in a semi-analytical way. In analogy with the case of deep disruptions, analytical estimates by \cite{guillochon2014-10jh} suggest that the specific kinetic energy dissipated by the nozzle shock is of order the squared stellar sound speed. This conclusion has been used to formulate an analytical prescription for the resulting stream expansion in the semi-analytical study by \citet{guillochon2015} that focuses on the impact of black hole spin. The nozzle shock has also been captured by three-dimensional simulations of stellar disruption by low-mass black holes \citep{ramirez-ruiz2009,shiokawa2015} or involving deep penetrations \citep{sadowski2016} but without a detailed analysis of the underlying mechanism. More recently, it has also been studied in the context of eccentric accretion discs through a semi-analytical treatment of this type of flows \citep{zanazzi2020,lynch2021,lynch2021_magnetic}.

In this paper, we perform the first systematic study dedicated to the nozzle shock in TDEs. Our approach consists in first simulating the stellar disruption in three dimensions to obtain the property of a given section of the fallback stream. To circumvent numerical issues encountered in past works, the subsequent evolution of this stream element is studied with a two-dimensional simulation. This is achieved thanks to a method we developed that follows the dynamics of this gas along the two transverse directions in the frame co-moving with its center of mass and co-rotating with the local direction of stream elongation. With this technique, we are able to study the hydrodynamics of this stream element throughout its evolution around the black hole at sufficient resolution, including the passage at pericenter where the nozzle shock occurs.

This paper is arranged as follows. In Section \ref{sec:simulations}, we describe our initial conditions and the numerical method used to follow the transverse evolution of a stream element in two dimensions. The results of this simulation are presented in Section \ref{sec:results}, from the gas infall towards pericenter where the nozzle shock takes place to the subsequent recession of this matter back to larger distances. In Section \ref{sec:discussion}, we carry out a convergence study, demonstrate the validity of our two-dimensional treatment, determine the consequences of our results on the later stages of evolution, and evaluate the impact of several additional physical processes on the nozzle shock. We include a summary of our main findings in Section \ref{sec:summary}

\section{Meshless simulations}
\label{sec:simulations}

Our numerical strategy consists first in performing a three-dimensional simulation of stellar disruption, from which we obtain the properties of a given section of stream. The subsequent transverse evolution of this stream element is then followed in two dimensions as it continues to approach the black hole.

\begin{figure}
\centering
\includegraphics[width=\columnwidth]{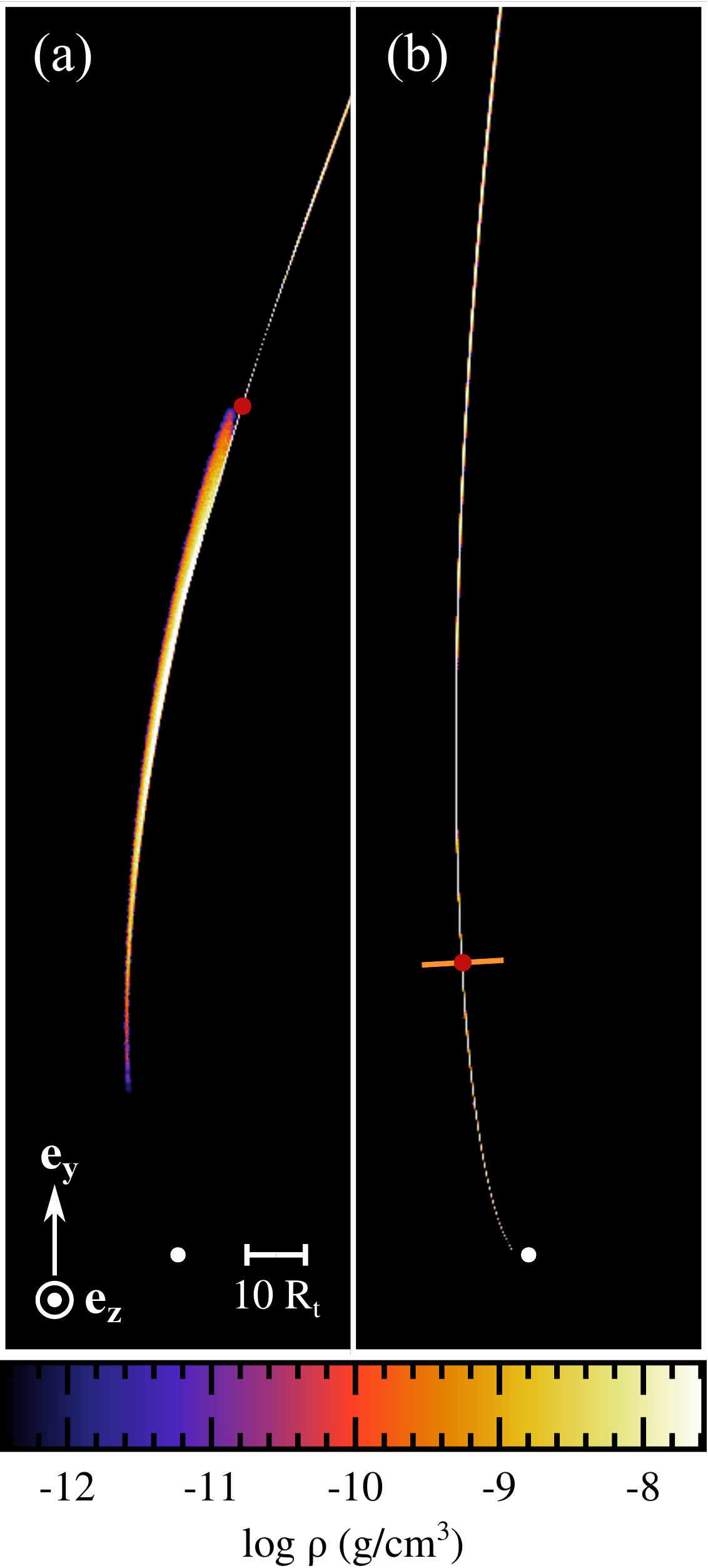}
\caption{Snapshots taken from the preliminary three-dimensional simulation of stellar disruption, showing the gas density inside the tip of the stream of debris as it falls back towards the black hole at times $t/\tmin =$ 0.59 (left panel) and 2.2 (right panel). The value of the density increases from black to white, as shown on the colour bar. The white segment on the first snapshot indicates the scale used and the orientation is given by the white arrows. The red dot indicates the location of the stream element selected to follow its transverse evolution at later times as it continues to move towards the black hole. Its properties are recorded at the time of the second snapshot when it reaches the surface depicted with the orange segment that is orthogonal to the longitudinal stream direction.}
\label{fig:stream}
\end{figure}

\subsection{Initial conditions}
\label{sec:initial}

We consider the disruption of a star with solar mass and radius $\mstar = 1 \msun$ and $\rstar = 1 \rsun$ by a black hole of mass $\mh = 10^6 \msun$. For this encounter, the tidal radius is located at 
\be
\rt = \rstar \left(\frac{\mh}{\mstar}\right)^{1/3} \approx 7 \times 10^{12} \, \rm cm,
\label{eq:tidal_radius}
\ee
which corresponds to the distance within which the tidal force from the black hole exceeds the stellar self-gravity. We assume that the stellar pericenter is equal to the tidal radius that corresponds to a penetration factor $\beta = 1$, defined by the ratio of tidal radius to pericenter. This choice of parameters is that of a typical tidal disruption, for which we are able to study the passage of the stream near pericenter for the first time. Since we already consider this work as a significant step forward, we defer a more complete exploration of the parameter space to the future while providing a discussion of possible effects in Section \ref{sec:extrapolation}.

The stellar disruption is followed with the hydrodynamics code \textsc{gizmo} \citep{hopkins2015}, making use of the meshless-finite-mass technique that consists in dividing the gas into particles of fixed mass. The star has a solar mass and radius with a density profile assumed to be polytropic with an exponent $\gamma_{\star} = 5/3$. It is constructed by first placing the particles in a closed-pack lattice and then radially stretching their position to obtain the desired profile. This distribution is then evolved for a few dynamical times including a damping term until the amplitude of its oscillations become negligible. We use $N_{\rm D} = 10^7$ particles with mass $M_{\rm D} = 10^{-7} \msun$ that is among the largest resolution used for this type of study. The star is initially placed at a distance $R_0 = 3 \rt$ from the black hole on a parabolic orbit with pericenter equal to the tidal radius of equation \eqref{eq:tidal_radius}. Self-gravity is included to simulate the encounter and the gas is assumed to follow an adiabatic equation of state with $\gamma = 5/3$ as expected for this phase of evolution. A Newtonian potential is used to describe the black hole's gravity, which amounts to neglecting the impact of general relativity. This approximation is justified by the fact that the pericenter of the orbit is much larger than the gravitational radius, with a ratio of $\rp/\rg \approx 47 \gg 1$. Gas particles that get within an accretion radius $\racc = 3 \rt$ from the black hole are removed from the simulation.

\begin{figure}
\centering
\includegraphics[width=\columnwidth]{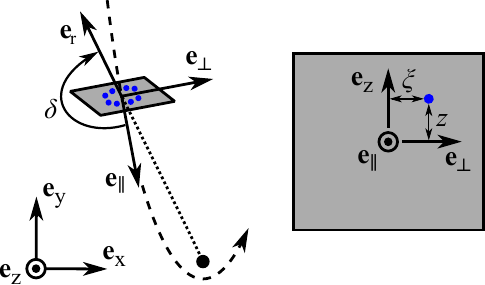}
\caption{Sketch illustrating the two-dimensional treatment used to follow the transverse evolution of a stream element around the black hole. The particles (blue points) used to describe the gas are confined at all times to the grey surface, which follows the trajectory (dashed line) of its center of mass while remaining orthogonal to the longitudinal direction of the stream defined by the unit vector $\vect{e}_{\parallel}$. This direction makes an angle $\delta<0$ with the outward radial direction, which evolves in time as the stream element moves. The transverse position of a given particle is specified by the coordinates $\xi$ and $z$ that correspond to offsets with respect to the center of mass along the in-plane and vertical directions specified by the unit vectors $\vect{e}_{\perp}$ and $\vect{e}_{\rm z}$, respectively.}
\label{fig:sketch}
\end{figure}

\begin{figure}
\centering
\includegraphics[width=\columnwidth]{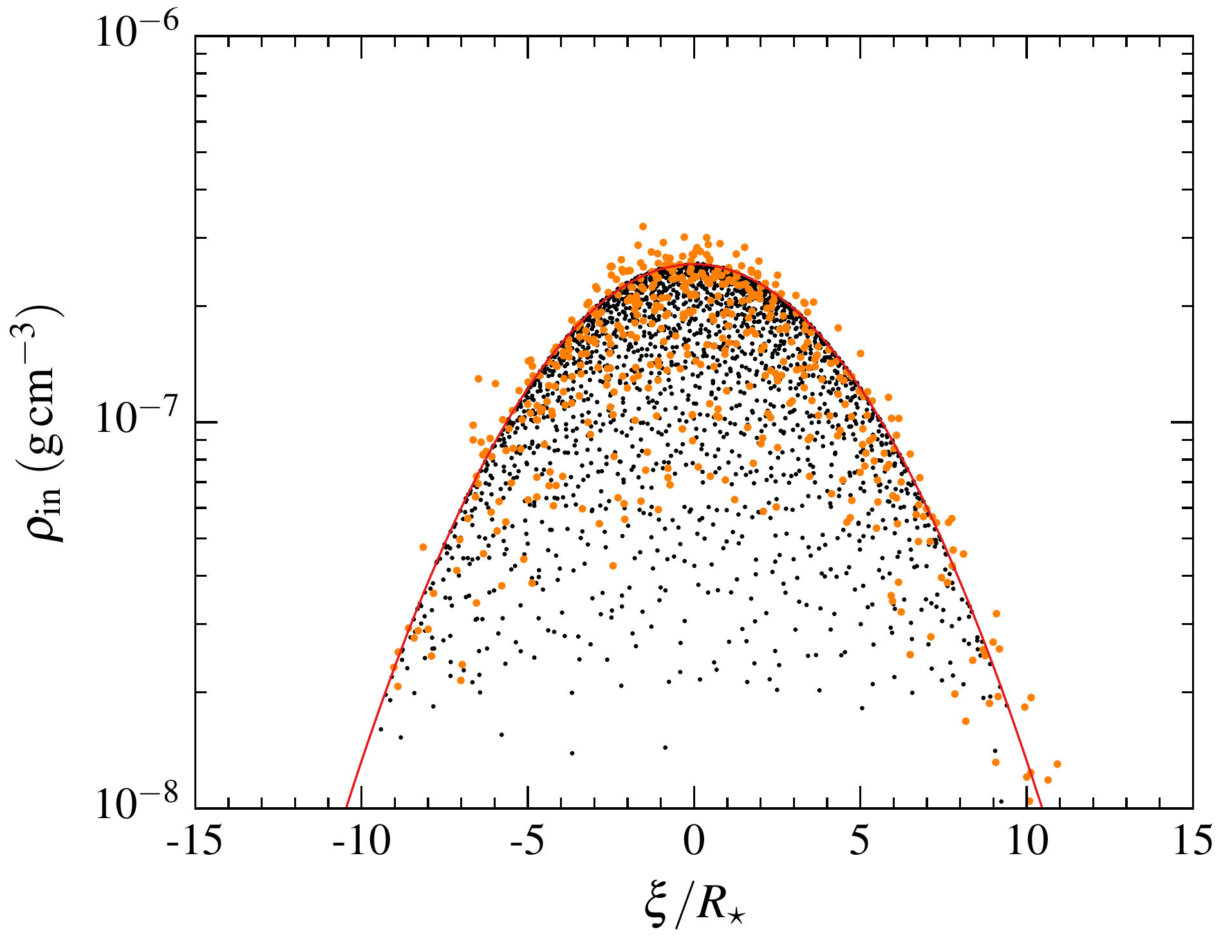}
\includegraphics[width=\columnwidth]{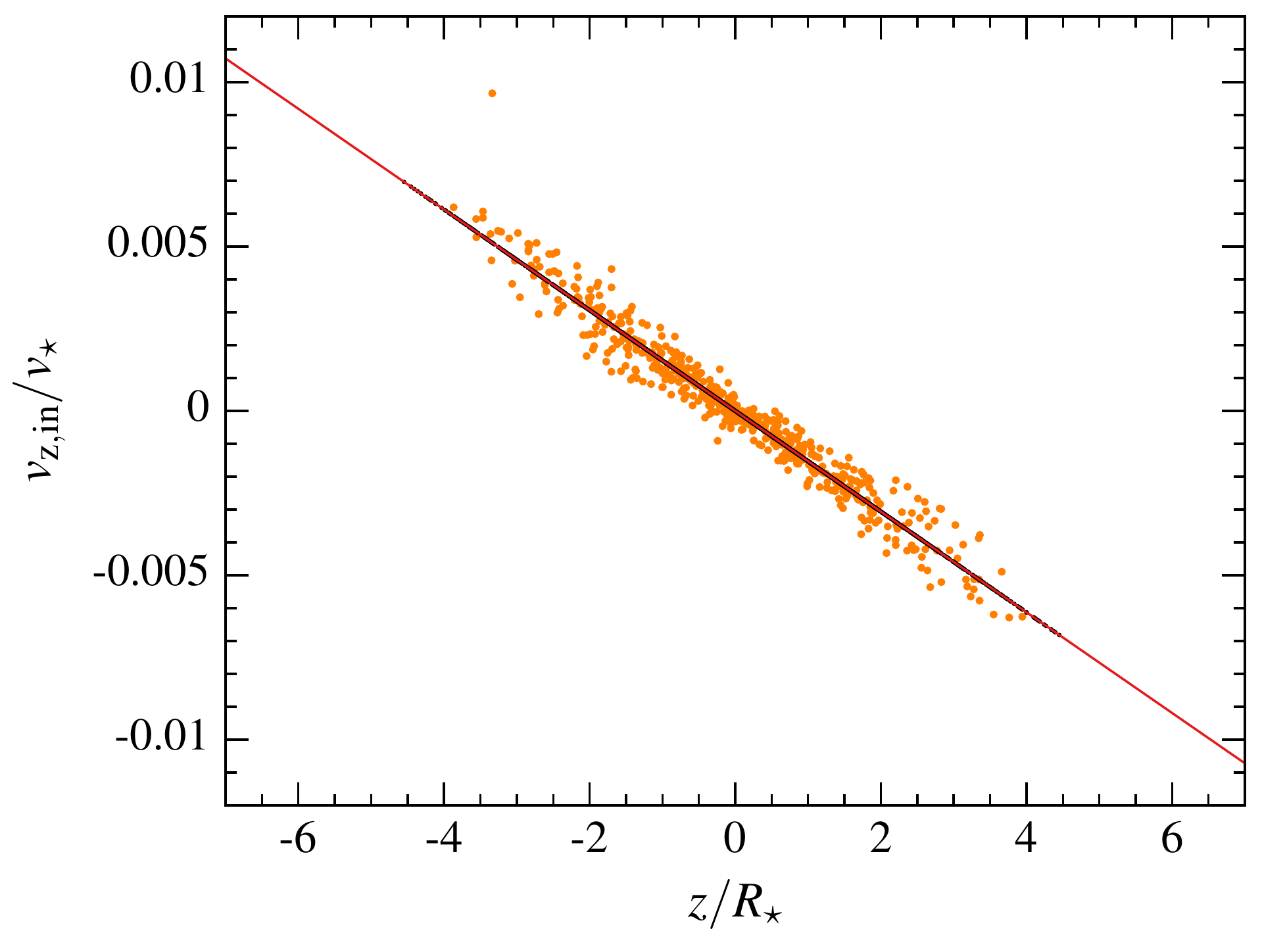}
\caption{Initial density (upper panel) and vertical velocity (lower panel) profiles inside the stream element along the in-plane and vertical directions, respectively. The orange dots correspond to individual particles selected from the three-dimensional simulation of stellar disruption that are used to obtain the fitted profiles shown with the red lines. The smaller black dots correspond to the particles used as initial conditions of the two-dimensional simulation, whose properties are determined by the fitted profiles. We only display $1\%$ of the $N_{\rm p} \approx 2 \times 10^5$ particles used in the simulation so that individual points can be distinguished.}
\label{fig:profiles}
\end{figure}

In this initial phase, the gas evolution is similar to that found in previous studies \citep{lodato2009,guillochon2013,guillochon2014-10jh,coughlin2015-variability}, and we therefore only provide a brief description. Following the encounter, the stellar debris gets imparted a spread in orbital energy a few times larger than that $\Delta \epsilon = G \mh \rstar / \rt^2$ analytically expected from the frozen-in approximation \citep{stone2013}. Half of this gas gets bound to the black hole with its extremity coming back the earliest to pericenter, after a time slightly shorter than $\tmin = 2 \pi G \mh (2 \Delta \epsilon)^{-3/2} \approx 41 \days$ according to Kepler's third law. The tip of the returning stream can be seen from the gas density shown at a time $t/\tmin = 0.59$ since disruption in the left panel (a) of Fig. \ref{fig:stream}. At large radii, the stream displays a thin profile due to its confinement by self-gravity following the disruption \citep{kochanek1994,coughlin2016-structure}. Because it originates from the stellar outermost layers, the gas closer to the black hole is more tenuous, which reduces the impact of self-gravity. As a result, this more bound matter presents, in addition to the central part, a wing several orders of magnitude less dense that extends towards the left. This asymmetry and wider range of densities makes the transverse gas properties of this part of the stream more difficult to describe and to accurately follow numerically. For this reason, we choose to study instead the near-cylindrical and more compact gas that arrives immediately after. The specific element we select to follow its later evolution around the black hole is indicated with the red point in Fig. \ref{fig:stream}.

This element has an orbital energy $\epsilon = -0.575 \Delta \epsilon$ that corresponds to a fallback rate of $\mdot = 0.685 \msunperyr$, only a factor of $\sim 2$ lower than the peak value. Note that this choice is mostly made for computational convenience but we discuss its possible influence on our results in Section \ref{sec:extrapolation}. The two-dimensional transverse properties of the element are recorded at $t / \tmin = 0.22$ when it reaches a distance $R_{\rm in} = 51.2 \rt$ from the black hole. This is done by  defining a surface orthogonal to the stream longitudinal direction at this location, which is displayed with the orange segment on the right panel (b) of Fig. \ref{fig:stream}. The properties of a few hundred particles are recorded as they cross this surface, which are used to determine by interpolation the transverse profiles of the hydrodynamical variables. The location of a given particle on this surface is parametrized by the two coordinates $\xi$ and $z$ shown in the sketch of Fig. \ref{fig:sketch} that correspond to offsets with respect to the center of mass along the stellar orbital plane and the vertical direction, respectively.

The initial density profile can be approximated by a binormal distribution $\rho_{\rm in} (\xi,z) = \Lambda (2 \pi \sigma_{\perp} \sigma_{\rm z})^{-1} \exp(-((\xi/\sigma_{\perp})^2 + (z/\sigma_{\rm z})^2)/2 )$, with fitted values of the standard deviations of $\sigma_{\perp} = 4.1 \rstar$ and $\sigma_{\rm z} =2.0 \rstar$. Note that this element is narrower along the vertical direction, which is caused by earlier oscillations around hydrostatic equilibrium seeded by the passage of the star at pericenter \citep{coughlin2016-pancakes}. To ensure that the density distribution is consistent with the fallback rate, the normalization is fixed by $\Lambda = \int_{-\infty}^{\infty} \int_{-\infty}^{\infty} \rho_{\rm in}(\xi,z) \diff \xi \diff z = \dot{M} / v_{\rm c, in}$. The center of mass speed has a value of $v_{\rm c, in} = 16.6 v_{\star}$, where $v_{\star} = \sqrt{G \mstar / \rstar} = 440 \kmpers$ denotes the approximate stellar sound speed, such that the central density is given by $ \rho_{\rm in}(0,0) = \dot{M}/ (2 \pi v_{\rm c, in} \sigma_{\perp} \sigma_{\rm z}) = 2.4 \times 10^{-7} \gpercm3$. Due to the proximity to the black hole, the center of mass has its velocity completely aligned with the stream longitudinal direction,\footnote{Note that the center of mass velocity is not aligned with the direction of stream elongation before the gas comes back near the black hole. When a part of the stream is for example at the apocenter of its orbit, its velocity is tangential to the black hole while the longitudinal direction is almost radial.} which itself is almost along the inward radial direction. Specifically, these two directions are initially inclined by an angle $\delta_{\rm in} \approx -0.95 \pi < 0$ (see Fig. \ref{fig:sketch}). The initial velocity profiles in the co-moving frame can be approximated by homologous transverse compression along both directions given by $v_{\rm z,in}/v_{\rm c,in} = A_{\rm z,in} z / R_{\rm in}$ and $v_{\rm \perp,in}/v_{\rm c,in} = A_{\rm\perp,in} \xi / R_{\rm in}$, where $A_{\rm z,in} = -0.47$ and $A_{\rm\perp,in} = -0.39$ are obtained by fitting. Similarly, the longitudinal velocity changes across the stream as $v_{\rm\parallel,in}/v_{\rm c,in} = A_{\rm \parallel,in} \xi / R_{\rm in}$ with $A_{\rm \parallel,in} = 0.12>0$ since the material with $\xi >0$ is closer to the black hole. As representative examples, Fig. \ref{fig:profiles} shows the initial density (upper panel) and vertical velocity (lower panel) profiles along the orbital plane and orthogonal to it, respectively. The orange dots correspond to the particles picked from the three-dimensional simulation that are used for the interpolation while the red lines are the fitted profiles. For simplicity, we adopt a uniform specific thermal energy distribution given by its average value of $u = 4.8 \times 10^{-6} v^2_{\star}$ for the selected particles. Note that the corresponding gas temperature $T = 2 m_{\rm p} u /(3 k_{\rm B}) \approx 70 \, \rm K$ is very small, which in reality may be prevented by hydrogen recombination that injects energy into the gas at earlier times \citep{coughlin2016-structure}. This choice is nevertheless appropriate because pressure forces are found to be irrelevant until the gas reaches close to pericenter where strong heating occurs that makes the matter forget about its initially much lower internal energies.

The initial conditions of the two-dimensional simulation are then created by generating a higher resolution version of this stream element along the orthogonal surface. The initial mass of gas is $m_{\rm in} = \Lambda l \approx 2 \times 10^{-6} \msun$, which is specified by a longitudinal length  arbitrarily fixed to $l=\rstar$. Even though the gas only has a surface density $\Sigma$ defined on the orthogonal surface, it can be converted into a three-dimensional density given by $\rho = \Sigma / l$, which is used instead throughout the paper. Importantly, this density is independent of our choice of length $l$ since this parameter affects the mass and enclosing volume in the same proportional way. The particle mass $M_{\rm p}$ specifies the number of particles used in the simulation given by $N_{\rm p} = \left[m_{\rm in}/M_{\rm p}\right]$, where the brackets denote the nearest integer function. We choose a value of $M_{\rm p} = 10^{-11} \msun$ corresponding to $N_{\rm p} \approx 2 \times 10^5$ particles. In Section \ref{sec:convergence}, we determine the influence of resolution on our results to demonstrate that the above particle number is enough to attain numerical convergence.

The particles are first positioned randomly on the orthogonal surface according to the probability distribution associated with the binormal function fitted for the density. To reduce density interpolation errors resulting from random noise, we find it necessary to relax the particle distribution. This is realized by evolving the gas for a short duration inside a periodic box whose vaccuum is filled with ghost particles to prevent artificial expansion. To impose the particles to move towards the desired density profile, they are also subject to a pressure force that vanishes when this configuration is attained. After the particles have reached their final position, we assign their velocities according to the fitted profiles as well as its uniform specific internal energy. The result of this procedure is displayed with the black points in Fig. \ref{fig:profiles}, which by comparison with the red lines demonstrates that the gas properties are correctly imposed. The stream element thus obtained is then used to initialize a two-dimensional simulation that studies its transverse gas evolution at later times, according to the numerical method we now explain.

\subsection{Two-dimensional treatment of hydrodynamics} 

\label{sec:treatment}

We aim at studying the two-dimensional transverse evolution of the selected stream element during its passage near the black hole in the frame co-moving with its ballistic center of mass and co-rotating with the local longitudinal direction of elongation. We rely on the orientation indicated in Fig. \ref{fig:sketch} that is specified by three orthogonal unit vectors $\vect{e}_{\parallel}$, $\vect{e}_{\perp}$ and $\vect{e}_{\rm z}$, aligned with the longitudinal direction and along the two transverse ones, respectively. The stream element has its longitudinal direction aligned with the velocity $\vect{v}_{\rm c}$ of its center of mass such that $\vect{e}_{\parallel} = \vect{v}_{\rm c} / v_{\rm c}$, where $v_{\rm c} = |\vect{v}_{\rm c}|$. The orientation of the above unit vectors is then specified at all times by this velocity, which we evaluate along with the position $\vect{R} = R \vect{e}_{\rm r}$ of the center of mass, assuming that it follows a Keplerian orbit under the gravitational acceleration $\vect{a}_{\rm c} = -G \mh \vect{R} / R^3$. Here, $R = |\vect{R}|$ while $\vect{e}_{\rm r}$ is a unit vector pointing outward in the radial direction.

We examine the evolution of the gas along the two transverse directions by evolving its position relative to the center of mass given by  the vector $\vect{r} = \xi \vect{e}_{\perp} + z \vect{e}_{\rm z}$. By construction, this motion remains confined at all times to the surface orthogonal to the direction of stream elongation, shown in grey in the sketch of Fig. \ref{fig:sketch}. Note that the matter located on this surface is made of different fluid elements at each time because of shearing but this effect does not impact the hydrodynamical properties of the stream element due to their invariance along the longitudinal direction. In the inertial frame, the relative acceleration experienced by this gas is $\vect{a}_{\rm i} = \vect{a}_{\rm t} + \vect{a}_{\rm p}$ that contains the contribution from both pressure gradients and the tidal force, given by
\be
\vect{a}_{\rm p} = - \frac{\boldgreek{\nabla} P}{ \rho},
\ee
\be
\vect{a}_{\rm t} = - \frac{G \mh}{R^3} \left( \vect{r} - 3 \frac{\vect{R} \cdot \vect{r} }{R^2} \vect{R}  \right),
\label{eq:tidal}
\ee
respectively. Equation \eqref{eq:tidal} relies on a first order expansion of the tidal potential, which is valid since $|\vect{r}| \ll R$ at all times. As we derive in Appendix \ref{ap:equations}, the equations of motion that determine the transverse gas evolution in the frame co-rotating with the longitudinal direction are 
\be
\ddot{\xi} = -\frac{G \mh}{R^3} \xi (1-3 \sin^2 \delta) + \xi \Omega^2 - 2 V \Omega - \boldgreek{\nabla} P \cdot \vect{e}_{\perp} / \rho,
\label{eq:xiacc}
\ee
\be
\ddot{z} = -\frac{G \mh}{R^3} z - \boldgreek{\nabla} P \cdot \vect{e}_{\rm z} / \rho,
\label{eq:zacc}
\ee
\be
\dot{v}_{\parallel} = 3 \frac{G \mh}{R^3} \xi \cos \delta \sin \delta + \dot{\xi} \Omega - V \lambda,
\label{eq:vparadot}
\ee
where $v_{\parallel}$ represents the longitudinal component of the gas velocity that arises due to shearing and we introduced the quantity $V = \xi \Omega + v_{\parallel}$ that has the dimension of a speed. The two remaining coefficients involved are defined by $\lambda = -G \mh \cos \delta / (R^2 v_{\rm c})$ and $\Omega = -G \mh \sin \delta / (R^2 v_{\rm c})$ that represent the rate of longitudinal stream stretching and the angular frequency of rotation of the orthogonal surface, respectively. Here, the angle $\delta < 0$ is measured between the longitudinal and outward radial directions (see Fig. \ref{fig:sketch}). Equation \eqref{eq:zacc} is directly obtained from the inertial acceleration while equations \eqref{eq:xiacc} and \eqref{eq:vparadot} additionally contain non-inertial terms associated with the co-rotation of the frame with the longitudinal direction. In equation \eqref{eq:vparadot}, it is assumed that the parallel pressure gradient is negligible due to the larger extent of the gas in the longitudinal direction compared to the transverse ones. This assumption is checked in Section \ref{sec:validity}, where we demonstrate the validity of the two-dimensional treatment used in our simulation.

We implemented equations \eqref{eq:xiacc} and \eqref{eq:zacc} in the code \textsc{gizmo} \citep{hopkins2015} by adding external acceleration terms to the pressure gradients directly obtained by this code.\footnote{The code \textsc{gizmo} uses a second order Leapfrog integrator with adaptive timestepping. Each particle is evolved according to an individual timestep  $\Delta t$ computed from the local properties. One of the main constraints is specified by the Courant condition $\Delta t \leq \eta_{\rm C} d/c_{\rm s}$, where $d$ is a measure of the local distance between particles, $c_{\rm s}$ is the sound speed, and $\eta_{\rm C}$ is a dimensionless parameter. The simulation presented uses $\eta_{\rm C} = 0.4$ but we checked that the results are unaffected when this parameter is reduced.} Equation \eqref{eq:vparadot} is also solved but the longitudinal velocity of the fluid element is not used to update the position of the gas that remains confined to the orthogonal surface. Because we consider a stream element of longitudinal length constant in time, its mass varies due to stretching and elongation along this direction as
\be
m = m_{\rm in} \frac{v_{\rm c, in}}{v_{\rm c}},
\label{eq:mass_change}
\ee
which we implement through a continuous change of the mass of every particle.\footnote{Note that equation \eqref{eq:mass_change} is equivalent to $\dot{m} = - \lambda m$, which justifies that $\lambda$ represents the rate of stream stretching.} This variation in turn affects the density evaluated by the code as well as the specific thermal energy specified by adiabaticity. Note that this effect acts in a symmetric way with respect to pericenter passage, such that the mass of the stream element decreases due to stretching as it approaches the black hole but increases again due to longitudinal compression as this matter recedes back to larger distances.

We assume that the gas thermodynamical evolution is adiabatic with $\gamma = 5/3$. This is justified by the inability of the gas to significantly cool through radiation due the large optical depths involved, as demonstrated in Section \ref{sec:radiative}. Additionally, the gas pressure dominates that provided by radiation with a ratio of $P_{\rm rad} / P_{\rm gas} = (a T^4 /3) / (\rho k_{\rm B} T / m_{\rm p}) \lesssim 0.01$. Anticipating the results, this numerical value uses the maximal temperature and density of $T \approx 10^6 \kelvin$ and $\rho \approx 10^{-3} \gpercm3$ reached by the gas throughout its evolution. Shocks are treated in \textsc{gizmo} by solving the Riemann problem at faces reconstructed at the boundaries between particles. This is done by using the numerical scheme proposed by \citet{panuelos2020}, which we choose for its ability to maintain good particle order, as required for our problem. The gas contained in the element initially satisfies $\rho_{\rm in} < \mh /(2 \pi R^3_{\rm in}) = 1.2 \times 10^{-6} \gpercm3 $ (see upper panel of Fig. \ref{fig:profiles}), implying that self-gravity is overwhelmed by the tidal force \citep{coughlin2016-structure}. Because this is also the case closer to the black hole, the gas self-gravity is not taken into account in the two-dimensional simulation.

\subsection{Ballistic motion of the stream element}

\label{sec:ballistic}

\begin{figure*}
\centering
\includegraphics[width=0.75\textwidth]{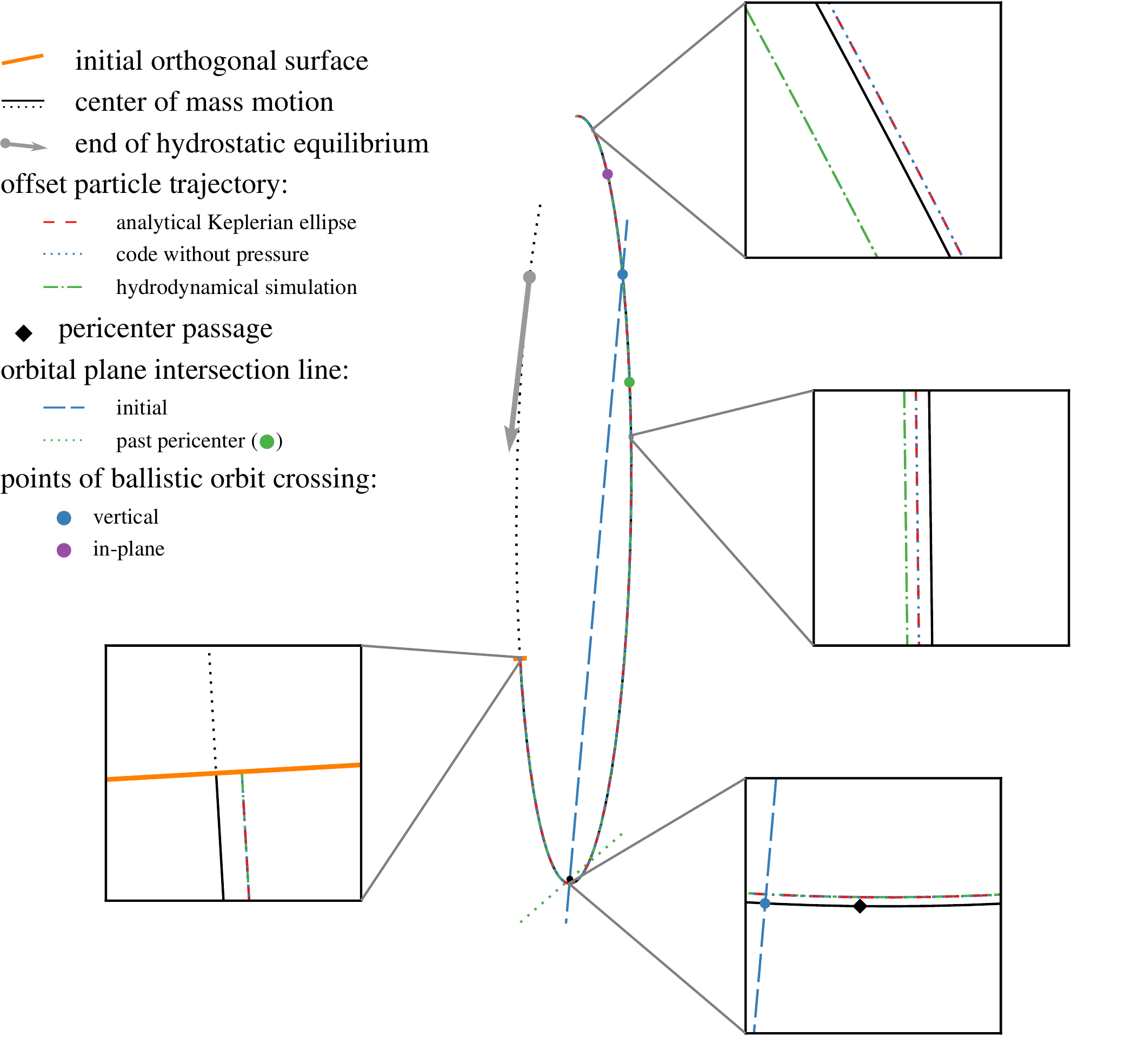}
\caption{Trajectories followed by different parts of the selected stream element during its passage near the black hole (black dot). Differences between them can be distinguished from the four rectangles that are insets zooming-in on specific regions. The black solid line represents the center of mass trajectory, starting from the orthogonal surface (orange segment) where the two-dimensional simulation is initialized. Its pericenter is indicated by the black diamond in the lowermost inset, which is located close to that of the star. The dotted black curve represents the same trajectory at earlier times, as obtained by backward integration. The grey point shows the location where the tidal force overcomes the gas self-gravity, while the arrow of the same color represents the orientation of the corresponding velocity vector. We consider a single particle inside the stream element with initial relative positions $\xi \approx  5 \rstar$ and $z \approx  2 \rstar$, and velocities given by the fitted homologous profiles. Its trajectory is shown with the coloured curves that are computed in different ways: analytically from a Keplerian ellipse (red dashed line) and by integrating equations \eqref{eq:xiacc} and \eqref{eq:zacc} with pressure forces switched off (blue dotted line). These two orbits perfectly overlap as seen from the insets, which proves the correct implementation in \textsc{gizmo}. The particle considered evolves on an orbital plane inclined with respect to that of the center of mass. These two planes intersect along the blue long-dashed line, which is parallel to the grey velocity vector, as expected geometrically. The stream collapses vertically when it crosses this line at the location of the blue points. This happens for example slightly before pericenter as seen from the lowermost inset, that is where the nozzle shock is expected to take place. The green dash-dotted line shows the trajectory of the exact same initial particle but including pressure forces from the rest of the matter within the stream element. It deviates from the two other ballistic orbits (red dashed and blue dotted lines) past pericenter, as seen from the two uppermost insets. The intersection line of its orbital plane with that of the center of mass is shown with the green dotted line, which is evaluated when the stream element has reached the green point.}
\label{fig:trajectory}
\end{figure*}

Before presenting the hydrodynamical results, we start by better characterizing the ballistic motion of the gas. The center of mass trajectory of the stream element is shown with the black solid line in Fig. \ref{fig:trajectory}, which starts at the orthogonal surface represented by the orange segment, where the two-dimensional simulation is initialized. The pericenter of this trajectory is denoted by the black diamond in the lowermost inset, whose location is essentially identical\footnote{Specifically, we find that the pericenter distance of the stream element is slightly lower than that of the star with a numerical value of $\rp = 0.975 \rt$.} to that $\rp = \rt$ of the star due to the small angular momentum spread imparted by the stellar disruption \citep{cheng2014}. The orbital energy of the selected element specifies its semi-major axis $a \approx 87 \rt$, which corresponds to a high eccentricity $e \approx  0.99$. Before reaching the initial orthogonal surface, the stream element was moving on the trajectory indicated with the black dotted line obtained by backwards integration of a Keplerian orbit. We find that the tidal force overcomes the stream self-gravity during this earlier phase at a distance $R_{\rm eq} \approx 1.8 a$ from the black hole. This location is indicated by the grey point while the arrow of the same colour represents the direction of the corresponding velocity vector. The gas is still close to hydrostatic equilibrium at this moment but starts getting tidally compressed as it continues to move inward.

The coloured curves correspond to the trajectory of a single particle belonging to the stream element whose relative positions are initially of $\xi \approx  5 \rstar$ and $z \approx  2 \rstar$ with velocities given by the fitted profiles. The red dashed line is evaluated analytically as a Keplerian ellipse from these initial conditions. The blue dotted line is obtained in a very different way by using the method described in Section \ref{sec:treatment}, switching off pressure forces in equations \eqref{eq:xiacc} and \eqref{eq:zacc}. These two trajectories completely overlap, as can be seen from all the insets. Similarly, the vertical motion is found to be identical using the two methods of calculation. This proves the correct implementation of the numerical solution designed to follow the transverse gas evolution with \textsc{gizmo} as well as the validity of the tidal approximation used in equation \eqref{eq:tidal}. The green dash-dotted line displays the trajectory of the exact same initial particle but taking into account pressure forces. It can already be seen to deviate from the ballistic trajectories, as we describe in detail in Section \ref{sec:results}.

During the infalling phase, we will see that pressure forces are negligible and the gas motion is essentially ballistic with equations \eqref{eq:xiacc} and \eqref{eq:zacc} simplifying to 
\be
\ddot{\xi}|_{P=0} \approx B_{\perp} \frac{G \mh \xi}{R^3},
\label{eq:xiacc_bal}
\ee
\be
\ddot{z}|_{P=0} = -\frac{G \mh z}{R^3},
\label{eq:zacc_bal}
\ee
where $B_{\perp} = 5 \sin^2 \delta /2 -1 + 2 A_{\parallel} \sin \delta$, approximating in the first equation the center of mass speed by its parabolic value of $v_{\rm c} \approx (2G \mh /R)^{1/2}$. Here, the last coefficient uses that the longitudinal velocity evolves close to homologously as $v_{\parallel} = A_{\parallel} v_{\rm c} \xi / R$. Away from pericenter, $\delta \approx -\pi$ such that $B_{\perp} \approx -1$, which implies that the two transverse positions obey the exact same equation. Because the center of mass moves on a near-radial parabolic trajectory, we show in Appendix \ref{ap:scaling} that the two transverse widths then follow an analytic solution given by equation \eqref{eq:zaccbalsol} that in our situation leads to a parabolic scaling $H \propto R^{1/2}$ (see also equations 31 of \citet{sari2010}) shortly after the moment when the stream starts moving ballistically at $R = R_{\rm eq}$. Writing the velocity profiles as $v_{\perp} = A_{\perp} v_{\rm c} \xi / R$ and $v_{\rm z} = A_{\rm z} v_{\rm c} z / R$ for homologous compression, this width evolution also implies that $A_{\perp} = A_{\rm z} = -0.5$. This analytical solution is already almost attained at $R = R_{\rm in}$ since the fits of Section \ref{sec:initial} yield $A_{\rm z,in}=-0.47 \approx -0.5$ and $A_{\rm \perp,in}= -0.39 \approx -0.5$. We will see from our simulation that the width keeps approaching the parabolic scaling as the gas continues to move inward.

The gas inside the stream element with an initial vertical offset evolves on orbital planes inclined with respect to that of the center of mass. These planes intersect along the same line for all the gas owing to the homologous nature of the compression \citep{stone2013}. It is shown with the blue long-dashed line in Fig. \ref{fig:trajectory} that is obtained from the test particle considered above. As we further explain in Appendix \ref{ap:intersections}, its orientation must be parallel to the velocity vector of the stream when the tidal force starts dominating self-gravity, which we indeed find to be the case by comparing with the grey arrow. This geometrical construction was already noticed in the early work of \citet{luminet1985} in the context of deep stellar disruptions. The stream gets vertically compressed\footnote{This vertical collapse can also be anticipated from the fact that the vertical acceleration induced by the tidal force in equation \eqref{eq:zacc_bal} always leads to $\ddot{z}/z<0$.} when it crosses this line at the location of the blue points. This occurs for example slightly before pericenter passage, as seen from the lowermost inset of Fig. \ref{fig:trajectory}. The stream is expected to reach its maximal level of compression near this location that is at the origin of the nozzle shock. Another consequence is that the orbital planes intersect close to apocenter, which would lead to another vertical collapse of the stream element, but only \textit{if} its evolution is entirely ballistic during and after pericenter passage.

The situation is different for the in-plane motion since the fluid elements never intersect with the trajectory of the center of mass during the infalling motion. This is because the gas with $\xi>0$ has a lower pericenter distance than the center of mass (see lowermost inset in Fig. \ref{fig:trajectory}), which is permitted by a change of sign of the in-plane acceleration.\footnote{The pre-factor in equation \eqref{eq:xiacc_bal} takes the value $B_{\perp} \approx -1 < 0$ at large distances, which implies a compression of the stream element. Close to pericenter, $\sin \delta \approx -1$ and the longitudinal homology factor has increased to $A_{\parallel} \approx 0.5$, as explained in Section \ref{sec:pericenter}. This implies that $B_{\perp} \approx 1/2 > 0$, which means that the in-plane acceleration has changed its sign such that the ballistic collapse along this direction is reversed.} However, ballistic motion implies that an intersection does take place along the stellar orbital plane after pericenter passage. Its exact location is displayed with the purple point in Fig. \ref{fig:trajectory} located slightly before the apocenter of the trajectory. Looking at the two insets directly before and after this point, it is clear that the ballistic trajectories (red dashed and blue dotted lines) have changed side compared to that of the center of mass (black solid line). As explained further in Appendix \ref{ap:intersections}, this intersection mainly results from the initial conditions that impose a slightly larger apocenter distance for fluid elements with $\xi > 0$.

\begin{figure}
\centering
\includegraphics[width=\columnwidth]{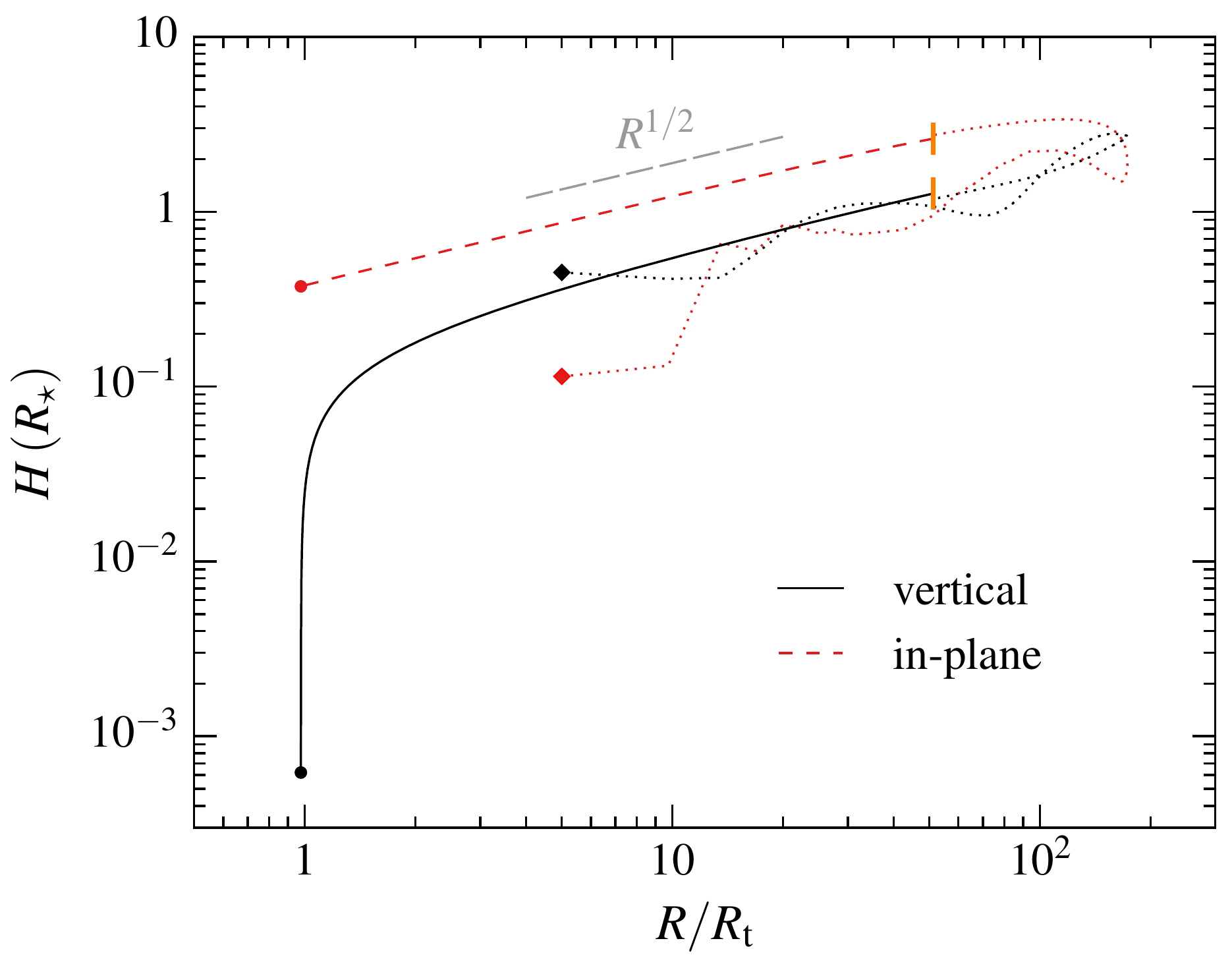}
\includegraphics[width=\columnwidth]{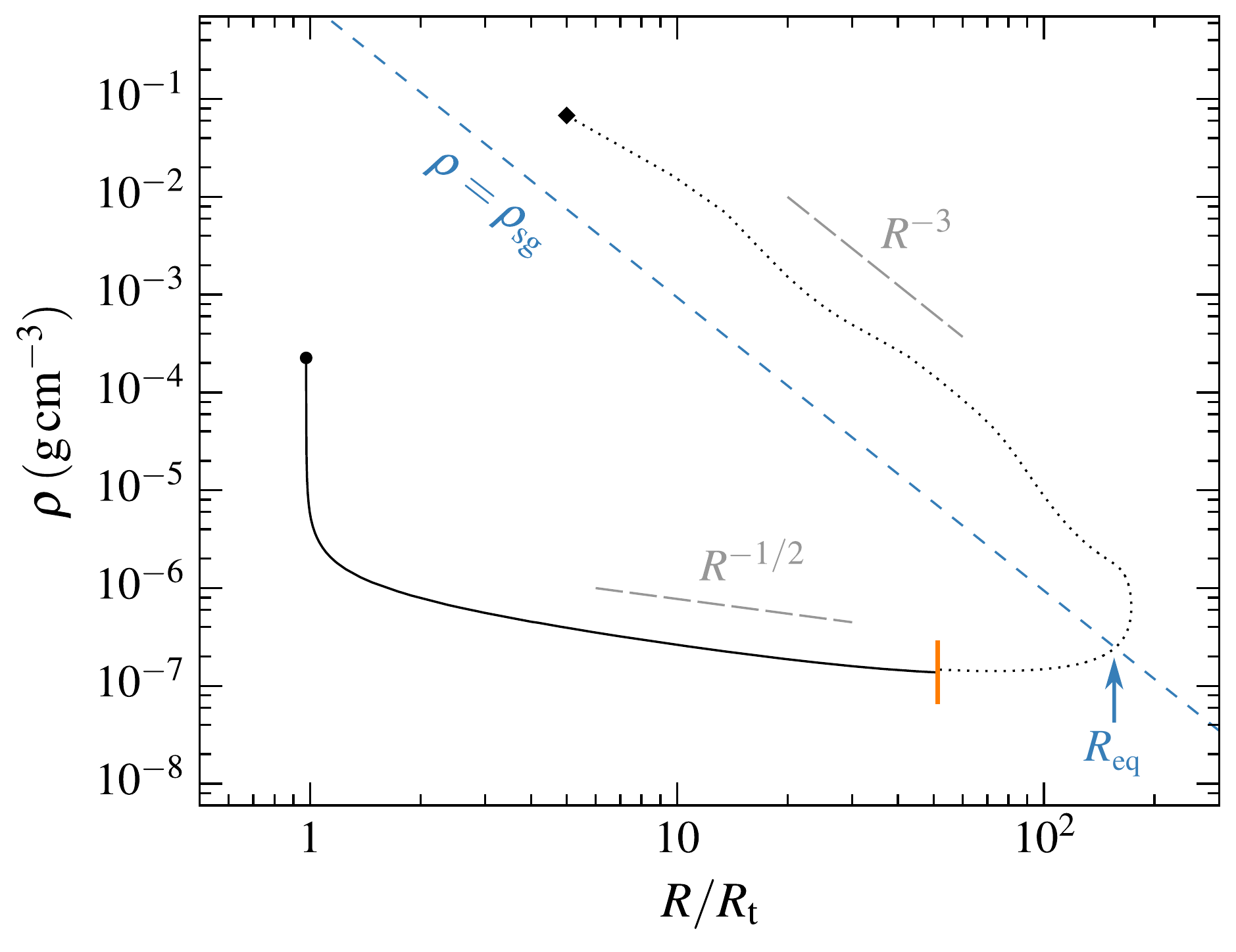}
\caption{Widths (upper panel) and density (lower panel) of the selected stream element as a function of distance from the black hole as it evolves from shortly after the stellar disruption (diamonds) to its return to pericenter (circles). The density is averaged over the element while the widths are defined as the distance from the center of the stream that encloses half of its mass along the vertical (black line) and in-plane (red line) direction. The early phase of outgoing motion and early fallback past apocenter (dotted lines) is obtained from the preliminary simulation of disruption performed in three dimensions. The gas used to compute the above properties has its orbital energies inside a narrow interval centered around that $\epsilon = -0.575 \Delta \epsilon$ of the stream element considered. To compute the widths, we determine the line crossing the center of this element from the positions of a small portion of gas located at its extremities. The subsequent phase of infall (solid black and dashed red lines) is evaluated from the two-dimensional simulation that starts at the location of the orange vertical segments where the transverse properties of the stream element are recorded. The dashed blue line in the lower panel delimits the region inside which the tidal force from the black hole dominates the stream self-gravity, which is given by $\rho_{\rm sg} = \mh/(2\pi R^3)$ \citep{coughlin2016-structure}. It is reached by the stream element at a radius $R = R_{\rm eq} \approx 1.8 a$ indicated by the blue arrow. The grey long-dashed segments indicate the scalings that the widths and density are analytically expected to follow in each phase.}
\label{fig:hrhovsr_infall}
\end{figure}

\begin{figure*}
\centering
\includegraphics[width=\textwidth]{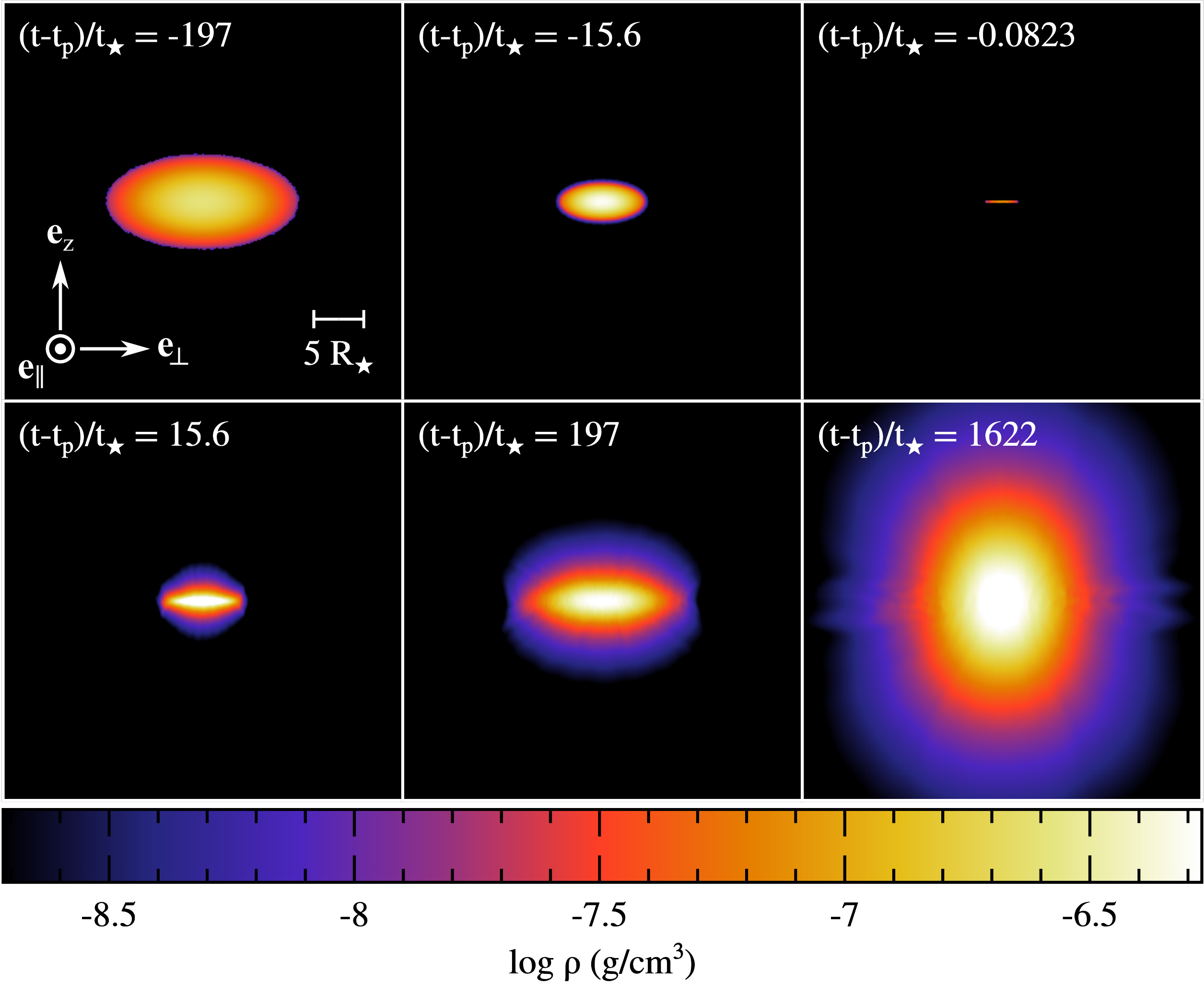}
\caption{Snapshots showing the gas density inside the stream element along the surface orthogonal to its longitudinal direction at different times $(t - \tp)/\tstar = $ -197, -15.6, -0.0823, 15.6, 197 and 1622, which are offset with respect to that $t=t_{\rm p}$ of pericenter passage and normalized by the stellar dynamical timescale $\tstar \approx 0.44 \hours$. The value of the density increases from black to white, as shown on the colour bar. The white segment on the first snapshot indicates the scale used and the orientation is given by the white arrows. The first snapshot corresponds to the initial condition of the two-dimensional simulation whose density and velocity profiles are displayed on Fig. \ref{fig:profiles}. Snapshots with exactly opposite times like the second and fourth ones with $(t - \tp)/\tstar = \pm 15.6$ are taken when the stream element is located at the same distance from the black hole before and after pericenter passage.}
\label{fig:density}
\end{figure*}

\section{Results}

\label{sec:results}

We now present the results of the simulation following the transverse gas evolution in the stream element.\footnote{A movie made from the simulation is available online at \url{http://www.tapir.caltech.edu/~bonnerot/nozzle-shock.html}.} This is done in a chronological manner, particularly focusing on the nozzle shock taking place near pericenter.

\begin{figure*}
\centering
\includegraphics[width=\textwidth]{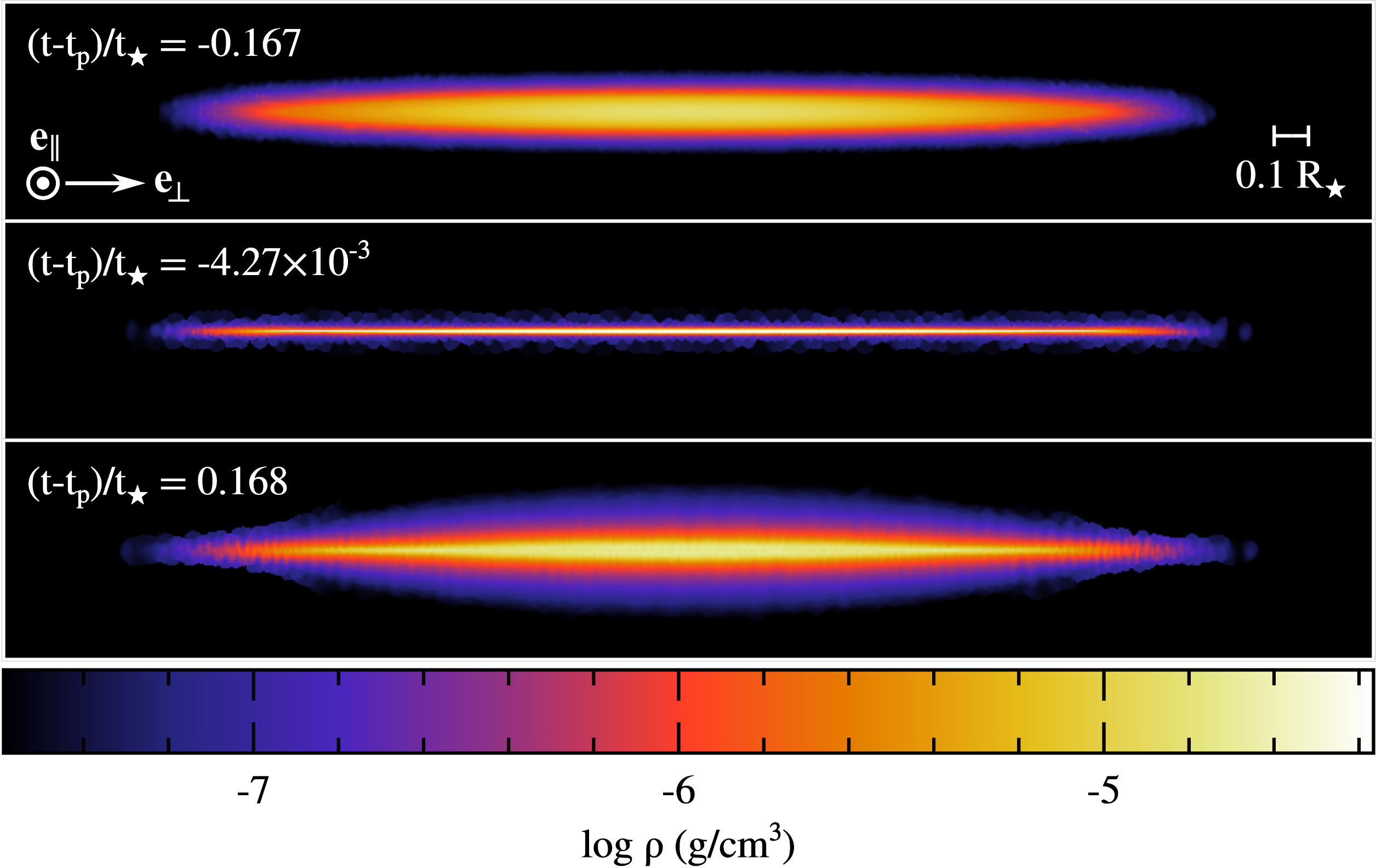}
\caption{Snapshots showing the gas density along the transverse directions inside the stream element at different times $(t - \tp)/\tstar = $ -0.167, $-4.27\times 10^{-3}$ and 0.168, which are offset with respect to that $t=t_{\rm p}$ of pericenter passage and normalized by the stellar dynamical timescale $\tstar \approx 0.44 \hours$. The value of the density increases from black to white, as shown on the colour bar. The white segment on the first snapshot indicates the scale used and the orientation is given by the white arrows. These snapshots are taken around the point where the gas reaches maximal collapse that results in a bounce, which reverses the infalling motion to make the gas expand after pericenter passage.}
\label{fig:density_pericenter}
\end{figure*}

\subsection{Approach to the black hole}
\label{sec:approach}

We start by briefly describing the evolution of the stream element found in the three-dimensional simulation of disruption. Fig. \ref{fig:hrhovsr_infall} shows with dotted lines its transverse widths (upper panel) and density (lower panel) as a function of distance from the black hole during this early phase, from shortly after the stellar disruption (diamonds) to the beginning of the fallback. Both the vertical (black dotted line) and in-plane (red dotted line) widths increase on average according to the scaling $H \propto R^{1/2}$ (grey segment), which is expected from gas in hydrostatic equilibrium with self-gravity compensating gas pressure forces for an adiabatic evolution with $\gamma = 5/3$ \citep{coughlin2016-structure}. In addition, the stream displays oscillations that are almost exactly out-of-phase between the two transverse directions, which implies that the stream gets deformed according to an approximately quadrupolar mode. We attribute these oscillations to in-plane pancake shocks taking place during the stellar disruption, as proposed by \cite{coughlin2016-pancakes}. During this outgoing phase, the density is expected to evolve as $\rho \propto (H^2 \ell)^{-1} \propto R^{-3}$, which is indeed what we find from the simulation (grey segment). This comes from using the width scaling and the fact that the longitudinal extent of a portion of stream with fixed mass increases due to stretching as $\ell\propto R^2$ \citep{coughlin2016-structure}. After reaching its apocenter, the stream element proceeds to move inwards again, which corresponds to a turnaround in Fig. \ref{fig:hrhovsr_infall}. The density then gets lower than the critical value $\rho_{\rm sg} = \mh/(2\pi R^3)$ (blue dashed line), causing the tidal force from the black hole to become larger than the stream self-gravity and pressure forces. The gas therefore starts moving ballistically under external gravity from this radius of $R=R_{\rm eq} \approx 1.8 a$ indicated with the blue arrow in the lower panel, corresponding to the grey point of Fig. \ref{fig:trajectory}.

The remaining of the gas evolution is followed with a two-dimensional simulation that starts at the location of the orange vertical segments in Fig. \ref{fig:hrhovsr_infall}. During this phase, the vertical (black solid line) and in-plane (red dashed line) widths still follow $H \propto R^{1/2}$, but the physical reason for this scaling is different from before. It is due to the fact that the transverse motion is entirely ballistic such that the compression takes place with homology factors that tend to $A_{\rm z } = A_{\perp} = -0.5$, as explained in Section \ref{sec:ballistic}. This compression causes the density (black solid line) to increase towards lower radii as $\rho \propto R^{-1/2}$ because $\rho = \mdot/(\pi H^2 v_{\rm c})$, where the center of mass velocity scales as $v_{\rm c} \propto R^{-1/2}$ for a near-parabolic trajectory. This compression is directly visible in the first two snapshots of Fig. \ref{fig:density}, which show the gas density along the orthogonal surface at the start of the simulation when $(t-\tp)/\tstar = -197$ and later during the infalling phase at $(t-\tp)/\tstar = -15.6$. Here, $\tstar = \rstar/v_{\star} \approx 0.44 \hours$ denote the stellar dynamical timescale while $t = t_{\rm p}$ corresponds to pericenter passage. The above scalings imply that the tidal force evolves as $F_{\rm t} \approx G \mh H /R^3 \propto R^{-5/2}$ and pressure forces as $F_{\rm p} \approx P / (H \rho) \propto R^{-5/6}$, such that their ratio scales as $F_{\rm t} /F_{\rm p} \propto R^{-5/3}$.  This confirms the expectation that the gas transverse motion keeps being specified by external gravity as the stream element continues to approach the black hole. At $R\lesssim 2 \rt$, it can be seen from Fig. \ref{fig:hrhovsr_infall} that the scalings we just derived break down due to a much faster vertical compression at the origin of the nozzle shock.

\subsection{Passage at pericenter}
\label{sec:pericenter}

As explained in Section \ref{sec:ballistic}, the ballistic trajectories of the gas on opposite sides of the stellar orbital plane intersect shortly before pericenter passage. This causes the stream element to get squeezed until its vertical width becomes $H_{\rm z} \approx 10^{-3} \rstar \ll \rstar$, as seen from the upper panel of Fig. \ref{fig:hrhovsr_infall} (black circle) and also visible from the gas density displayed in Fig. \ref{fig:density} shortly before pericenter passage at $(t-\tp)/\tstar = -0.0823$. This strong compression leads to the formation of the nozzle shock that dissipates the kinetic energy associated with this vertical motion. As a result, pressure forces sharply increase until they reverse this collapse, causing the gas to expand again as it moves away from the black hole. This bounce can be seen more closely from Fig. \ref{fig:density_pericenter}, which shows a zoom-in on the gas density distribution during pericenter passage. Because we have seen that ballistic motion does not predict an intersection of trajectories along the in-plane direction, the corresponding width of the stream element undergoes a smaller decrease down to $H_{\perp} \approx 0.1 R_{\star}$, meaning that the stream takes the geometry of a very thin sheet of matter. Pressure gradients are therefore much larger along the vertical direction, which causes the immediate expansion to be primarily orthogonal to the equatorial plane.

\begin{figure}
\centering
\includegraphics[width=\columnwidth]{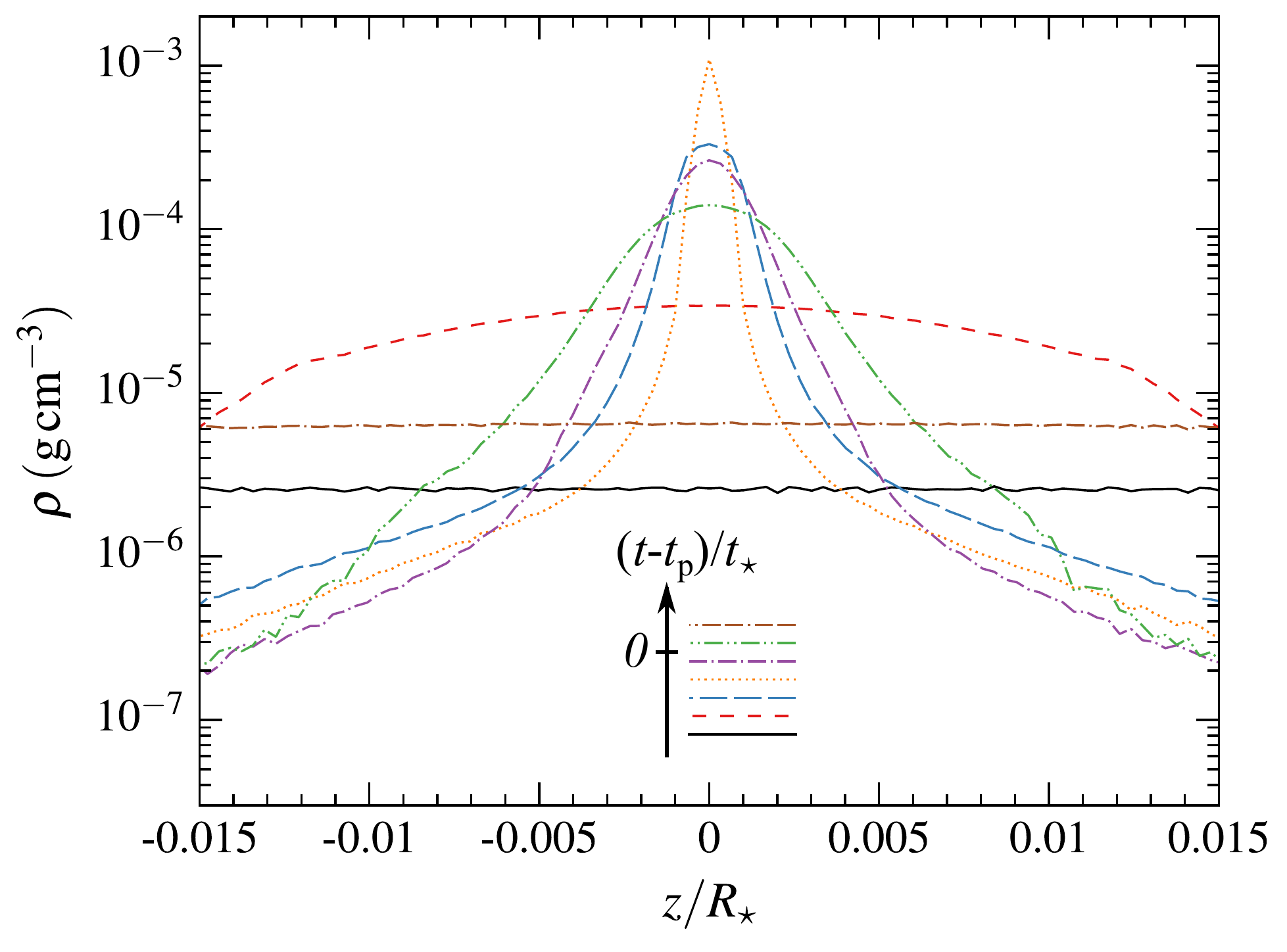}
\includegraphics[width=\columnwidth]{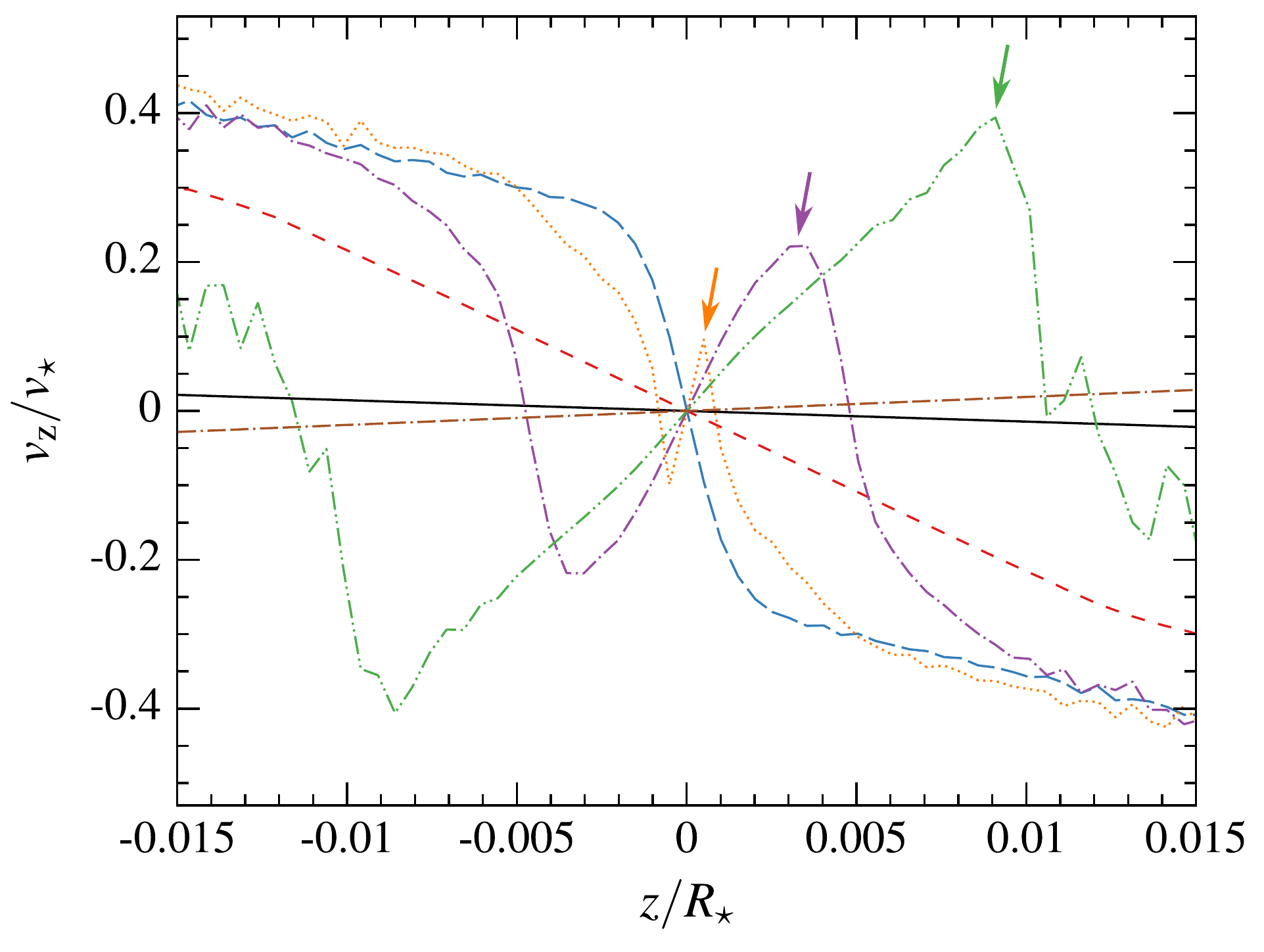}
\caption{Density (upper panel) and vertical velocity (lower panel) profiles along the vertical direction at different times $(t - \tp)/\tstar = -0.65$ (black solid line), -0.066 (red dashed line), -0.024 (blue long-dashed line), -0.016 (orange dotted line), -0.0075 (purple dash-dotted line), 0.00076 (green dash-dot-dotted line), 0.47 (brown dash-dash-dotted line). This chronological ordering of the different curves is indicated schematically at the bottom of the upper panel. The profiles are obtained by dividing the gas distribution into slices parallel to the equatorial plane. The coloured arrows in the lower panel indicates the position of the shockwave that vertically sweeps through the stream, reversing the direction of motion of the collapsing matter.}
\label{fig:rhovzvsz}
\end{figure}

\begin{figure}
\centering
\includegraphics[width=\columnwidth]{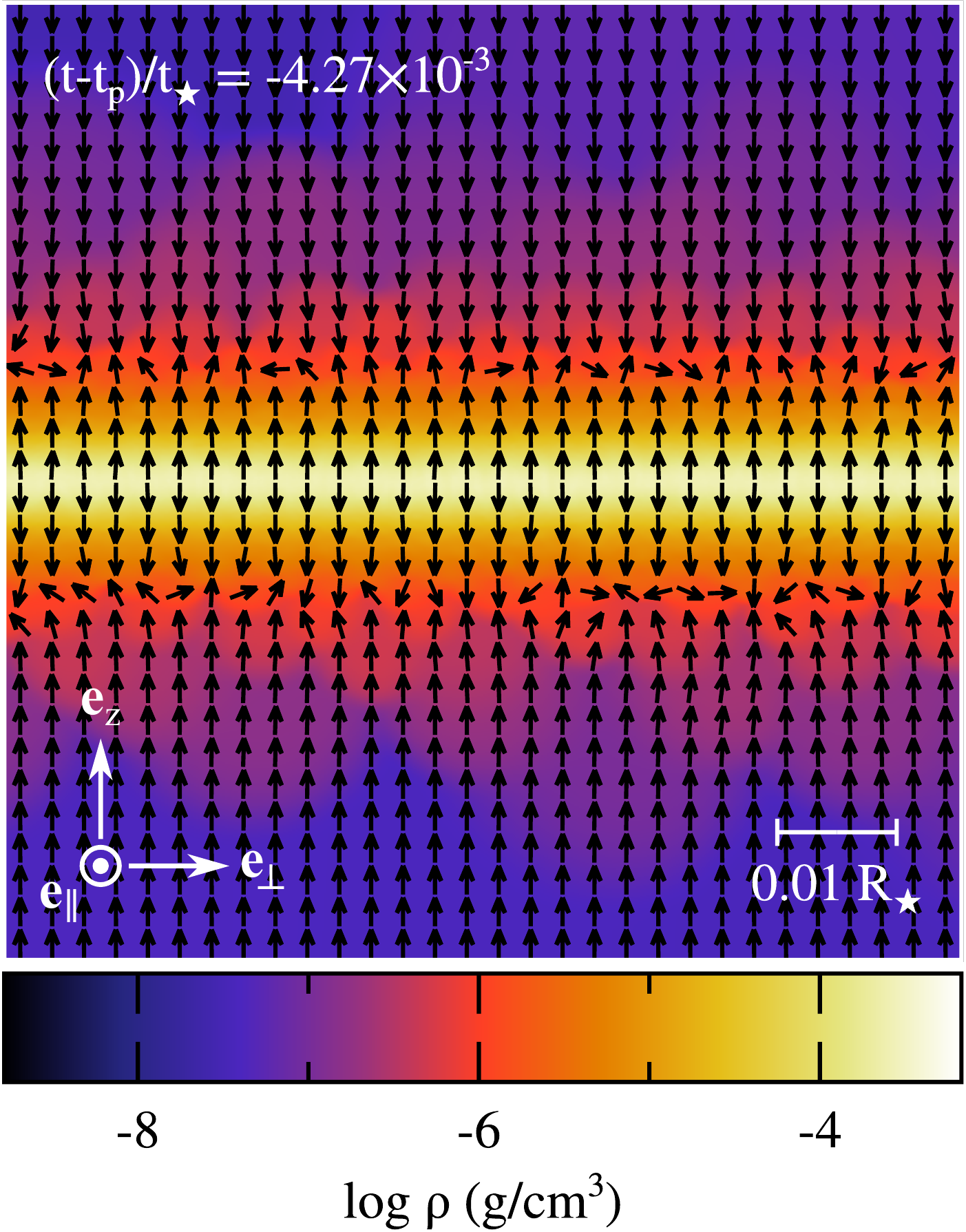}
\caption{Zoom-in on the center of the stream element showing its density at $(t - \tp)/\tstar = -4.27 \times 10^{-3}$, with black arrows of the same length denoting the direction of the velocity field. The value of the density increases from black to white, as shown on the colour bar. The white segment indicates the scale used and the orientation is given by the white arrows. This snapshot is taken shortly after the start of the nozzle shock as the shockwave ahead of the expanding gas passes through matter still approaching the equatorial plane, which has the effect of reverting its direction of motion.}
\label{fig:density_wave}
\end{figure}

\begin{figure}
\centering
\includegraphics[width=\columnwidth]{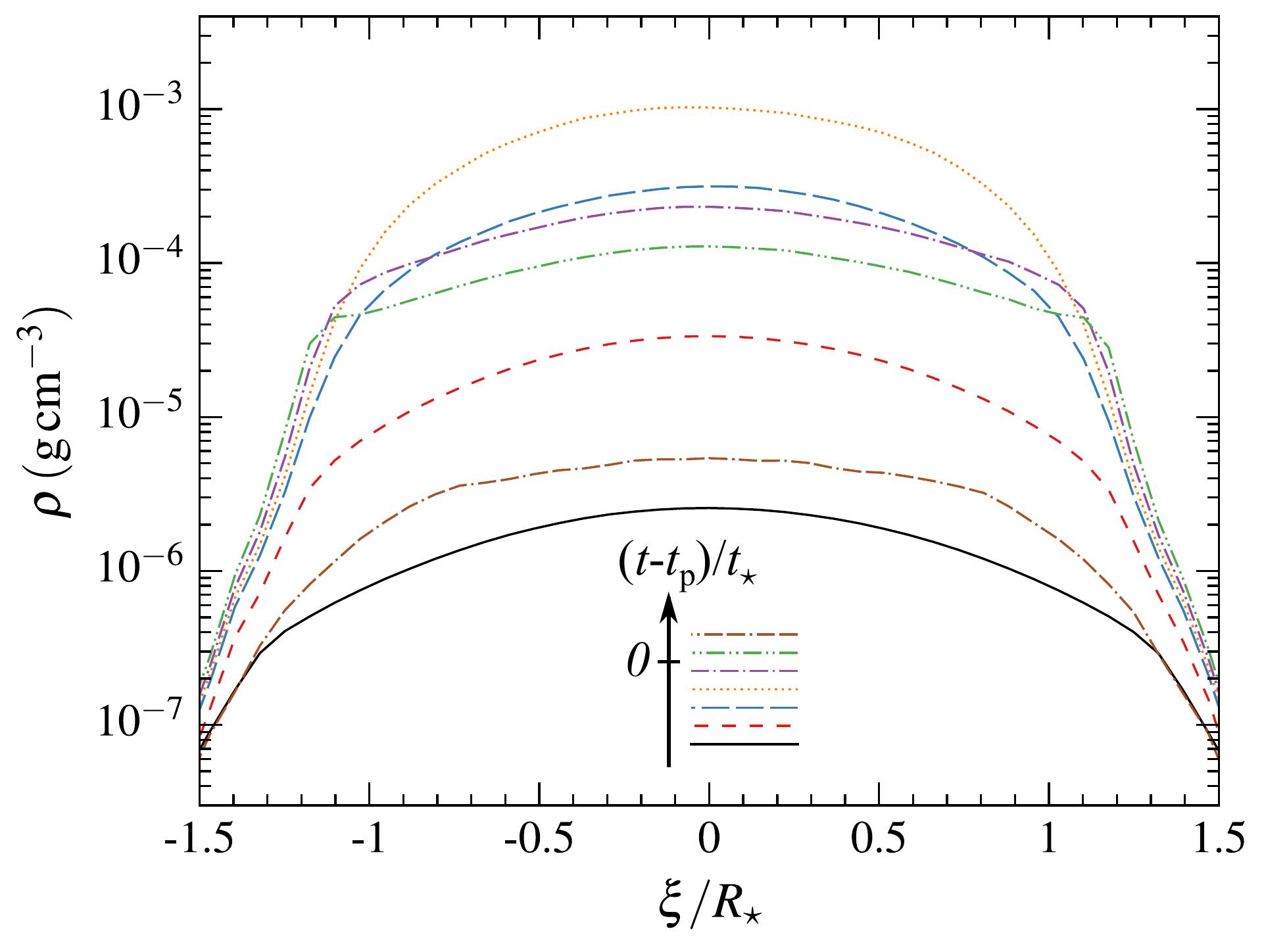}
\includegraphics[width=\columnwidth]{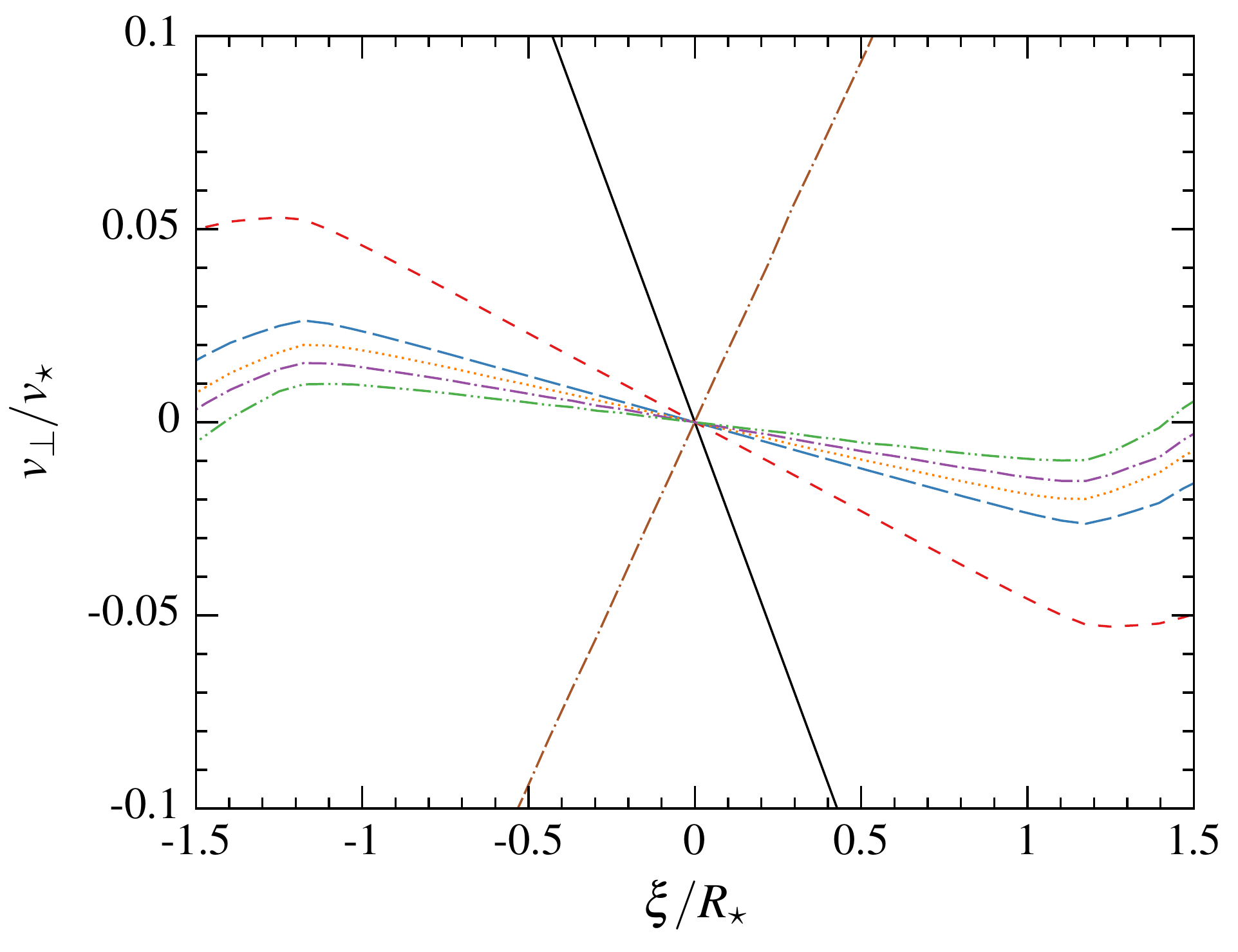}
\caption{Density (upper panel) and perpendicular velocity (lower panel) profiles along the in-plane direction at the same times as used in Fig. \ref{fig:rhovzvsz}. The chronological ordering of the different curves is indicated schematically at the bottom of the upper panel. The profiles are obtained by dividing the gas distribution into slices perpendicular to the equatorial plane.}
\label{fig:rhovperpvsxi}
\end{figure}

\begin{figure}
\centering
\includegraphics[width=\columnwidth]{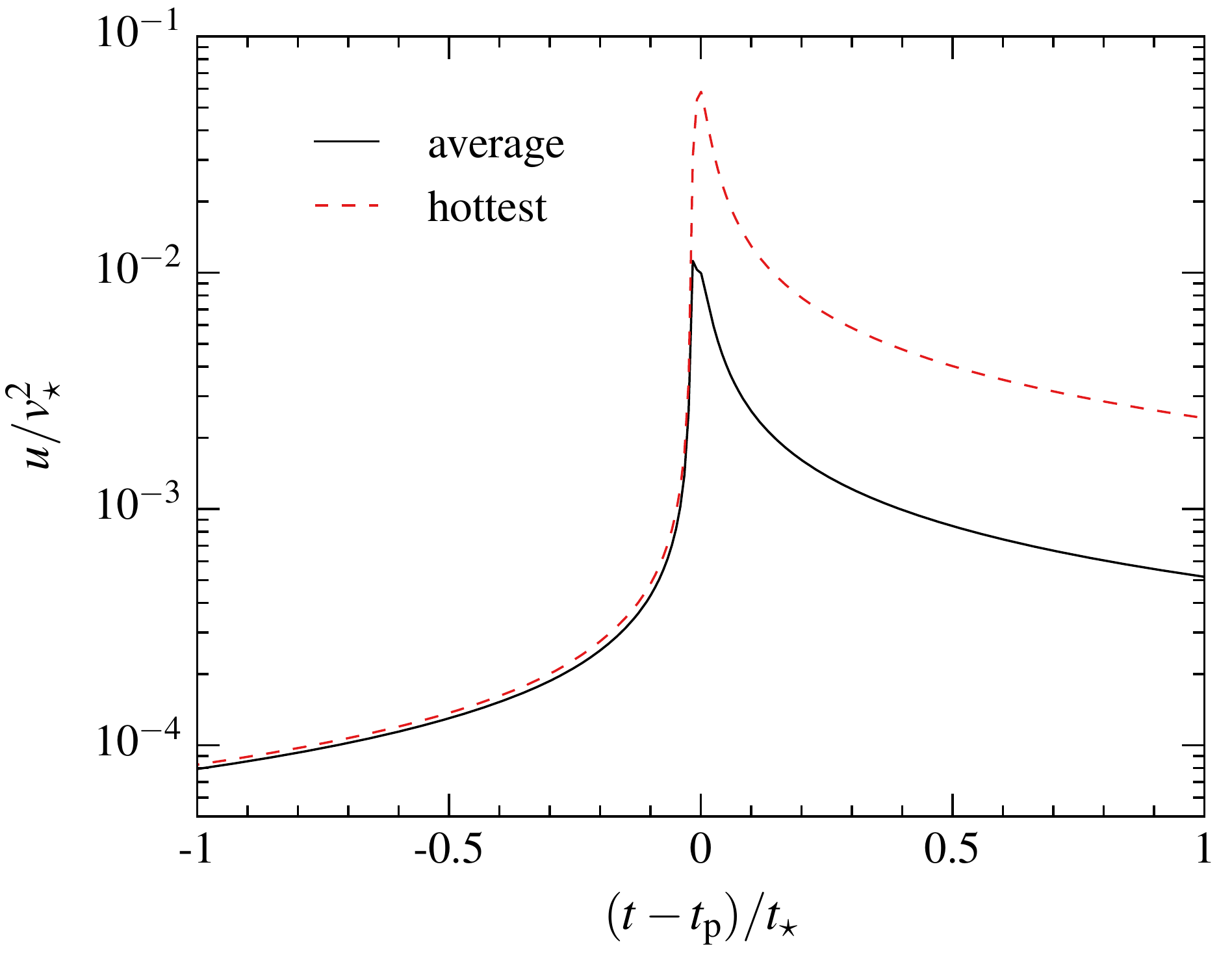}
\flushright
\includegraphics[width=0.99\columnwidth]{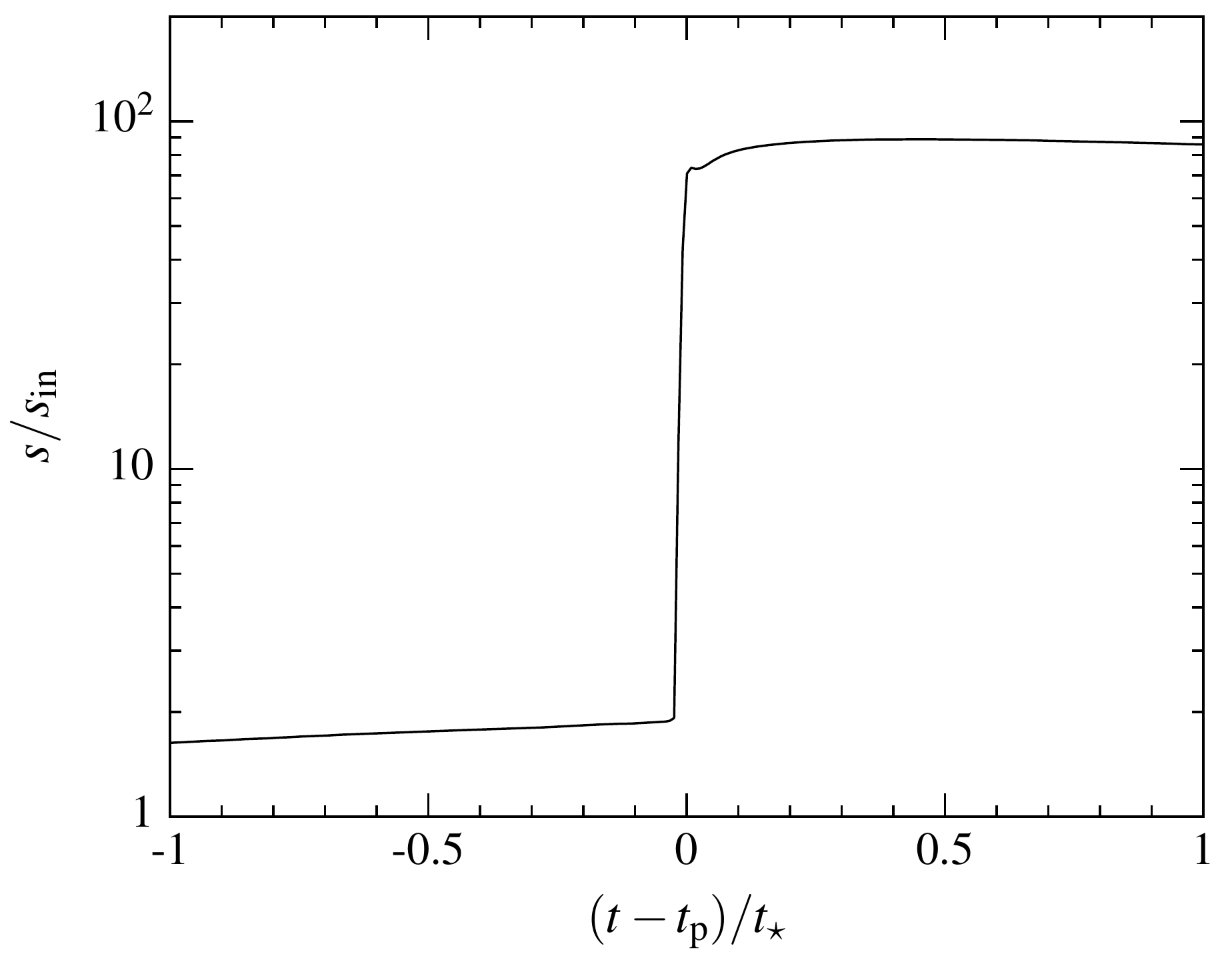}
\caption{Evolution of the specific internal energy (upper panel) and entropy (lower panel) during the nozzle shock. The energy is obtained by averaging over the whole stream element (black solid line) and using only the $10\%$ of particles with the largest energies (red dashed line). The entropy is computed as $s = u \rho^{-2/3}$, which we normalize by its initial value. The time is offset with respect to that $t=t_{\rm p}$ of pericenter passage and normalized by the stellar dynamical timescale $\tstar \approx 0.44 \hours$. }
\label{fig:uvst}
\end{figure}

The hydrodynamics of the nozzle shock can be examined more precisely from Fig. \ref{fig:rhovzvsz}, which displays the density (upper panel) and vertical velocity (lower panel) profiles along the vertical direction that are obtained at different times by dividing the gas within the stream element into slices parallel to the equatorial plane. The chronological ordering of the different curves is indicated schematically at the bottom of the upper panel. At $(t-\tp)/\tstar = -0.65$ (black solid line), the density is still uniform on the scale of $|z| \leq 0.015 \rstar$, with a maximal value larger than initially as a result of compression during the infall phase. The velocity profile is also perfectly homologous due to the ballistic gas motion that is unaffected by small pressure forces. As matter gets closer to pericenter at $(t-\tp)/\tstar = -0.066$ (red dashed line) and -0.024 (blue long-dashed line), the central density starts rising while the gas gets decelerated by growing central pressure gradients, as can be seen from a steepening of the velocity profiles. The gas reaches its maximal density of $\rho \approx 10^{-3} \gpercm3$ for a range of vertical distances $|z| \lesssim 10^{-3} \rstar$ at $(t-\tp)/\tstar = -0.016<0$ (orange dotted line), that is before pericenter passage as predicted in Section \ref{sec:ballistic}. At the same time, the collapse of the matter closest to the equatorial plane gets reverted by pressure forces as indicated by a change in sign of the vertical speeds. Additionally, a shockwave gets launched that propagates outward (coloured arrows) at later times $(t-\tp)/\tstar = -0.0075$ (purple dash-dotted line) and 0.00076 (green dash-dot-dotted line) through matter still collapsing at supersonic speeds. Gas entering this shocked region at increasing vertical heights have their infalling motion reversed, which continues until the shockwave has swept through the entire stream. This process is displayed in Fig. \ref{fig:density_wave}, which zooms in on the gas density near the center of the stream shortly after the start of the nozzle shock at $(t-\tp)/\tstar = -4.27 \times 10^{-3}$. As seen from the black arrows that indicate the direction of the velocity field, matter near the equatorial plane undergoes expansion behind the shockwave, while the gas further away still moves inward. Following this bounce, the density drops back to its value before pericenter passage with a velocity profile that becomes homologous again, as seen in Fig. \ref{fig:rhovzvsz} at $(t-\tp)/\tstar = 0.47$ (brown dash-dash-dotted line). The hydrodynamics of this shock is similar to that taking place during deep tidal disruptions due to the strong compression of the star at pericenter \citep{guillochon2009,brassart2010}.

Fig. \ref{fig:rhovperpvsxi} shows the gas density (upper panel) and perpendicular velocity (lower panel) profiles along the in-plane direction at the same times as in Fig. \ref{fig:rhovzvsz}. The gas density increases for $|\xi| \lesssim \rstar$, which is entirely due to the much stronger vertical compression. Due to low in-plane pressure gradients, the velocity profiles remain close to homologous at all times.\footnote{Note however that small deviations are visible in the outer layers, which we attribute to an early impact of pressure forces on this low-density gas.} In contrast with the vertical direction, these profiles get shallower as the stream element moves inward to reach $v_{\perp} \approx 0$ at pericenter since all the gas velocities are aligned with that of the center of mass along the tangential direction. Later on, the velocities switch sign due to a divergence of the gas trajectories as the stream element moves away from the black hole.

The dissipation taking place during pericenter passage can be analysed from Fig. \ref{fig:uvst}, which shows the evolution of the specific internal energy (upper panel) and the entropy (lower panel) normalized by its initial value. During the infalling phase, the compression experienced by the gas results in close to adiabatic heating at approximately constant entropy. Dissipation at the nozzle shock results in a sharp entropy jump close to pericenter associated with an increase of the internal energy by several orders of magnitude. The maximum value reached is $u \approx 10^{-2} v^2_{\star}$ when averaging over the whole stream element (black solid line) and a factor of a few larger when considering only the hottest parts of the gas (red dashed line). The corresponding maximal temperature is of order $T = m_{\rm p} u / k_{\rm B} \approx 10^5 \kelvin$, which is high enough to ionize hydrogen, as we further explain in Section \ref{sec:radiative}. As expected, the injected internal energy is similar to the kinetic energy of the infalling gas with vertical speeds $v_{\rm z} \approx 0.3 v_{\star}$ (see lower panel of Fig. \ref{fig:rhovzvsz}). Just upstream from the shock, the Mach number is $\mathcal{M} = v_{\rm z}/c_{\rm s} \approx 30 \gg 1$, using the sound speed $c_{\rm s} = (10 u/9)^{1/2} \approx 0.01 v_{\star}$ of the cold infalling gas. The magnitude of the entropy jump is consistent with that obtained from the jump conditions at a strong shock, which predicts an increase by a factor $\sim (2\gamma / \gamma +1)(\gamma-1/\gamma+1)^{\gamma} \mathcal{M}^2 \approx 0.12 \, \mathcal{M}^2 \approx 100$ \citetext{equation (1.85) of \citealt{zeldovich1967}} for $\gamma = 5/3$. After the shock, the gas cools adiabatically through expansion that induces a fast decrease of the internal energy. Importantly, it however remains on average hotter than before pericenter due to the entropy jump.

\begin{figure}
\centering
\includegraphics[width=\columnwidth]{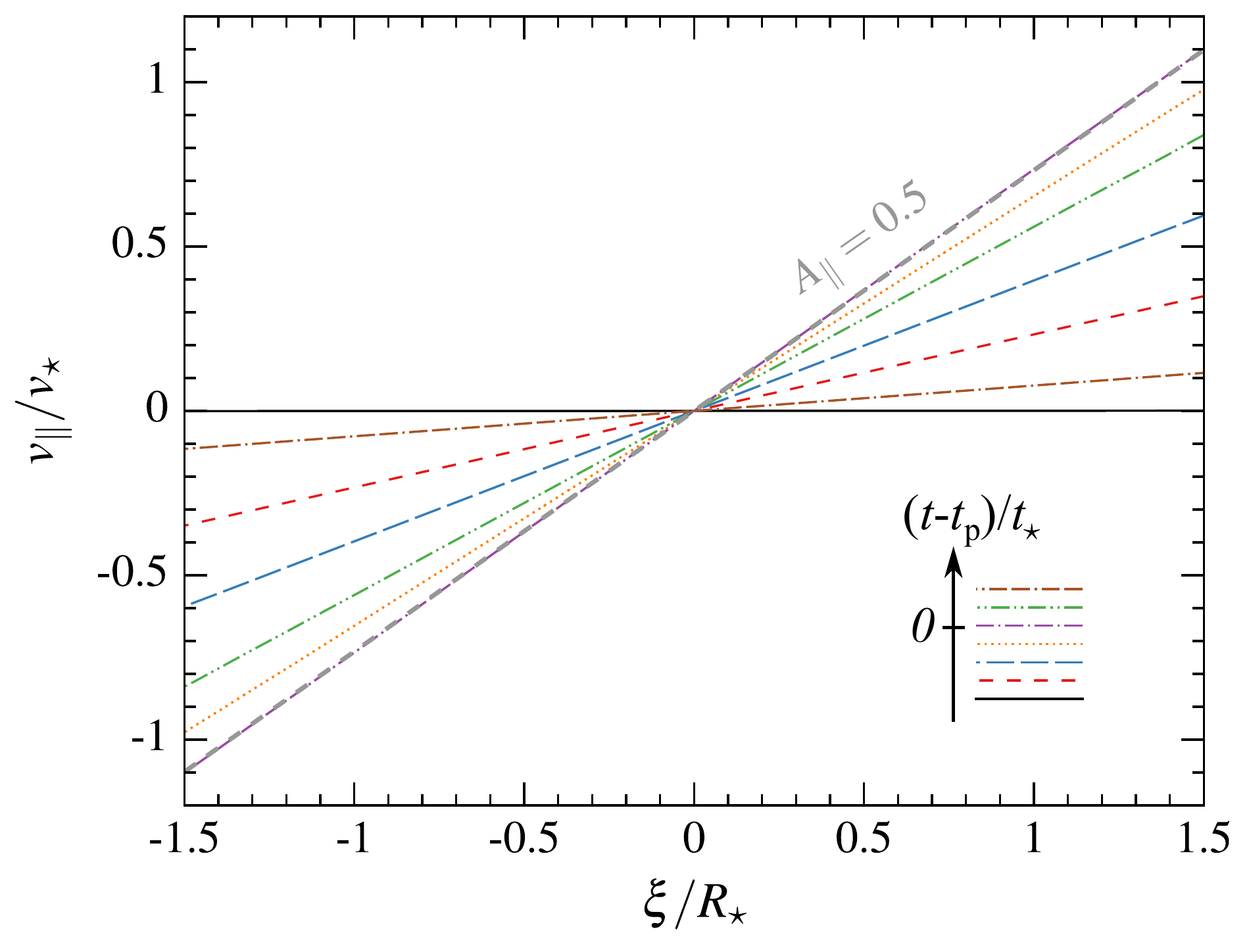}
\caption{Parallel velocity profiles along the in-plane direction at different times $(t - \tp)/\tstar =$ -197 (black solid line), -1.5 (red dashed line), -0.92 (blue long-dashed line), -0.34 (orange dotted line), 0.00076 (purple dash-dotted line), 0.55 (green dash-dot-dotted line) and 3.4 (brown dash-dash-dotted line). This chronological ordering of the different curves is indicated schematically in the lower-right corner. The profiles are obtained by dividing the gas distribution into slices perpendicular to the equatorial plane. The thick grey dashed line represents a homologous velocity profile evaluated from $v_{\parallel} = A_{\parallel} \xi v_{\rm p} / R_{\rm p}$ with $A_{\parallel} = 0.5$, where $v_{\rm p}$ denotes the center of mass velocity at pericenter.}
\label{fig:vparavsxi}
\end{figure}

Due to their distinct locations with respect to the black hole, different parts of the stream element experience relative accelerations along the longitudinal direction that cause shearing motion. This effect can be analysed from Fig. \ref{fig:vparavsxi} that shows the profile of the parallel velocity along the in-plane direction at different times. The chronological ordering of the different curves is indicated schematically in the lower-right corner. During the infalling phase at $(t - \tp)/\tstar =$ -197 (black solid line), -1.5 (red dashed line), -0.92 (blue long-dashed line) and -0.34 (orange dotted line), the profiles progressively steepen as shearing becomes stronger due to the increase in relative acceleration across the stream element at lower radii. This trend continues until pericenter passage where the $e_{\perp}$ direction is radial and all the fluid elements have velocities tangential with respect to the black hole. At this point, the parallel speed can therefore be approximated as $v_{\parallel} = \xi \diff v / \diff R \approx A_{\parallel} \xi v_{\rm p} / R_{\rm p}$ with $A_{\parallel} =0.5$ because all the gas inside the stream element is on close to parabolic orbits with $v \propto R^{-1/2}$. Here, we have set the center of mass velocity to its value $v_{\rm c} = v_{\rm p}$ at pericenter. This analytical homologous profile (thick grey dashed line) coincides as expected with that obtained from the simulation closest to pericenter at $(t - \tp)/\tstar =$ 0.00076 (purple dash-dotted line).\footnote{One can however notice a small deviation due to the fact that the trajectories are not exactly parabolic, which we show in Appendix \ref{ap:intersections} to be related to the occurrence of in-plane intersections of ballistic orbits later in the evolution.} Importantly, we find that the parallel speed is limited to $v_{\parallel} \lesssim v_{\star}$ because of the relation $v_{\rm p} \approx R_{\rm p} v_{\star} / \rstar$ at the tidal radius and the fact that the gas is confined to $\xi \lesssim \rstar$. After pericenter passage at $(t - \tp)/\tstar =$ 0.55 (green dash-dot-dotted line) and 3.4 (brown dash-dash-dotted line), shearing weakens due to a decrease of the tidal force until the velocities becomes similar again across the stream element. However, note that the profile never reverses due to the rotation of the frame of reference that imposes the fluid elements with $\xi>0$ to always remain closer to the black hole. As discussed in Section \ref{sec:viscosity}, shearing taking place near pericenter may lead to additional dissipation if the gas has a viscosity, which could increase the level of heating compared to that induced by the nozzle shock only.

\begin{figure}
\centering
\includegraphics[width=\columnwidth]{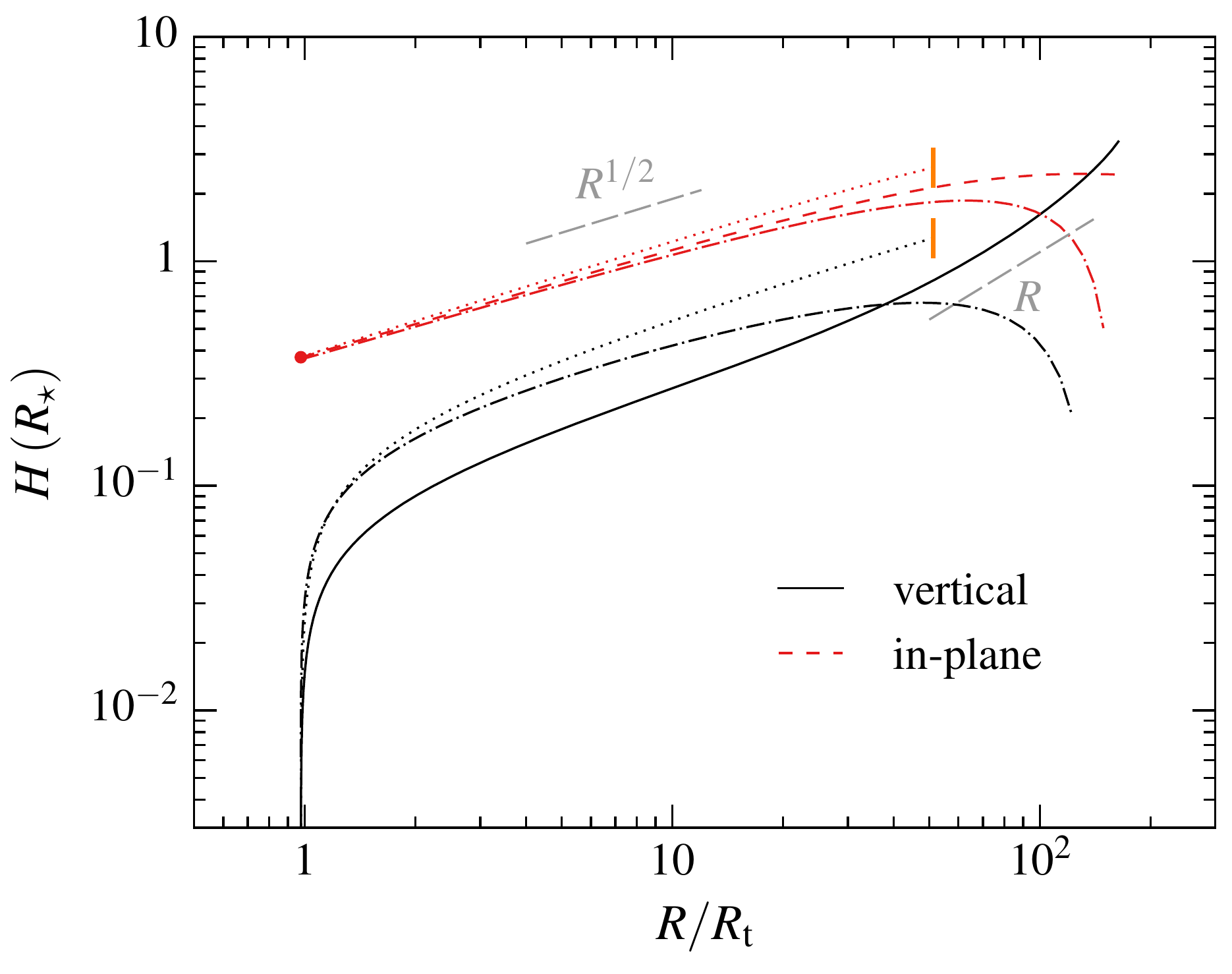}
\caption{Vertical (black solid line) and in-plane (red dashed line) widths as a function of distance from the black hole as the stream element moves outward after pericenter passage (red circle). They are computed from the two-dimensional simulation as the distances from the center of mass that enclose half of the mass in each direction. The dash-dotted lines of the same colours are obtained in the same way but setting pressure forces to zero in the simulation such that the gas moves ballistically. For ease of comparison, we also reproduce with dotted lines these widths for the previous infalling phase, which were already plotted in the upper panel of Fig. \ref{fig:hrhovsr_infall}. The grey long-dashed segments indicate the scalings that the widths are analytically expected to follow in each phase.}
\label{fig:hvsr_outflow}
\end{figure}

\subsection{Recession to larger distances}
\label{sec:recession}

We now turn our attention to the evolution of the stream element past pericenter to determine how its properties are affected by the passage through the nozzle shock. Fig. \ref{fig:hvsr_outflow} displays the vertical (black solid line) and in-plane (red dashed line) widths as a function of distance from the black hole, which are computed from the simulation as the distances that contain half of the mass of the element. The dash-dotted lines of the same colours are obtained in the same way but switching off pressure forces in the simulation such that the gas moves on ballistic trajectories. Owing to the symmetry of the gas properties with respect to the equatorial plane, this situation is remarkably equivalent to an instantaneous sign reversal of the vertical gas velocities at the point where its trajectories intersect near pericenter. We also reproduce with dotted lines the stream widths for the previous infalling phase, which were already plotted in the upper panel of Fig. \ref{fig:hrhovsr_infall}. At low radii $R\lesssim 10 \rt$, the different widths evolve approximately as $H \propto R^{1/2}$ (grey segment) in  both transverse directions that is the same scaling as during the infalling phase. Note however that the vertical component is a factor of a few lower for the hydrodynamical evolution. This means that the outgoing stream element is more concentrated than before pericenter passage, as we explain later. As the stream element moves to larger radii $R\gtrsim 10 \rt$, the widths obtained from the hydrodynamical simulation (black solid and red-dashed lines) start to significantly differ from those assuming ballistic motion (dash-dotted lines). If the gas moved ballistically, we find that it would undergo a vertical collapse followed shortly after by a strong in-plane compression. This is due to the two intersections of trajectories we identified in Section \ref{sec:ballistic} that occur when the center of mass reaches the blue long-dashed line and then the purple point in Fig. \ref{fig:trajectory}. These episodes of compression are not present in the hydrodynamical evolution, which implies that the gas motion departs from purely ballistic trajectories owing to a non-instantaneous influence of pressure. 

To understand more precisely these deviations from ballistic motion, it is useful to determine how pressure forces modify the orbital parameters. Due to vertical pressure gradients, the outgoing gas undergoes a deflection of its velocity away from the equatorial plane. This induces a modification of its orbital plane whose intersection line with that of the center of mass gets rotated along the direction of motion. This can be seen from the green dotted line displayed in Fig. \ref{fig:trajectory} that represents this line computed when the center of mass has reached the green point. Almost a full rotation of the intersection line is completed by this time compared to its initial location (blue long-dashed line). The intersection of trajectories predicted from ballistic motion is therefore prevented, which explains the absence of vertical collapse in our simulation. Another consequence is that the stream element moves perpendicularly to the rotated intersection line. As explained in Appendix \ref{ap:scaling}, this causes the gas to expand faster in the vertical direction as seen from Fig \ref{fig:hvsr_outflow} at large radii where the scaling becomes close to $H_{\rm z} \propto R$ (grey segment).

\begin{figure}
\centering
\includegraphics[width=\columnwidth]{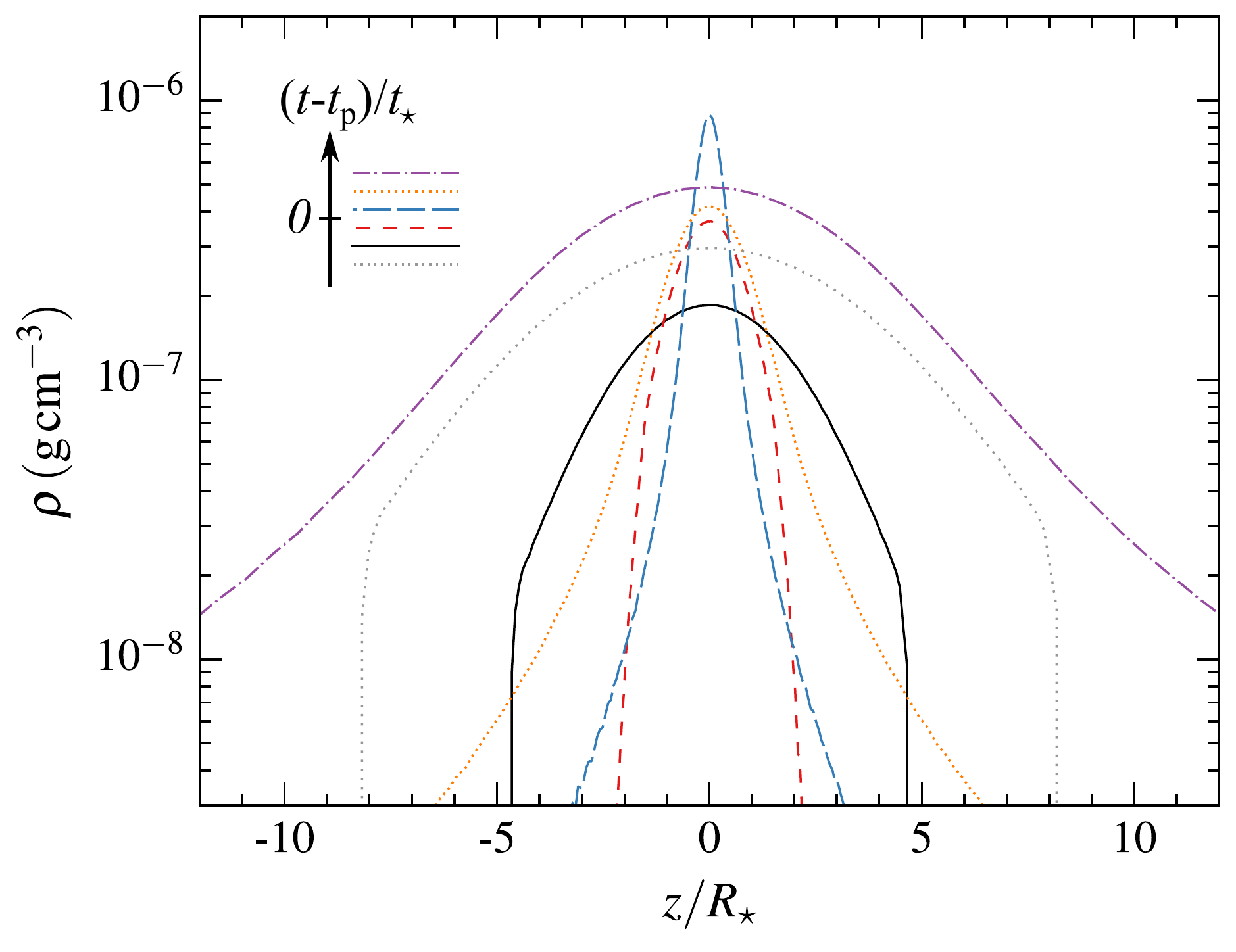}
\includegraphics[width=\columnwidth]{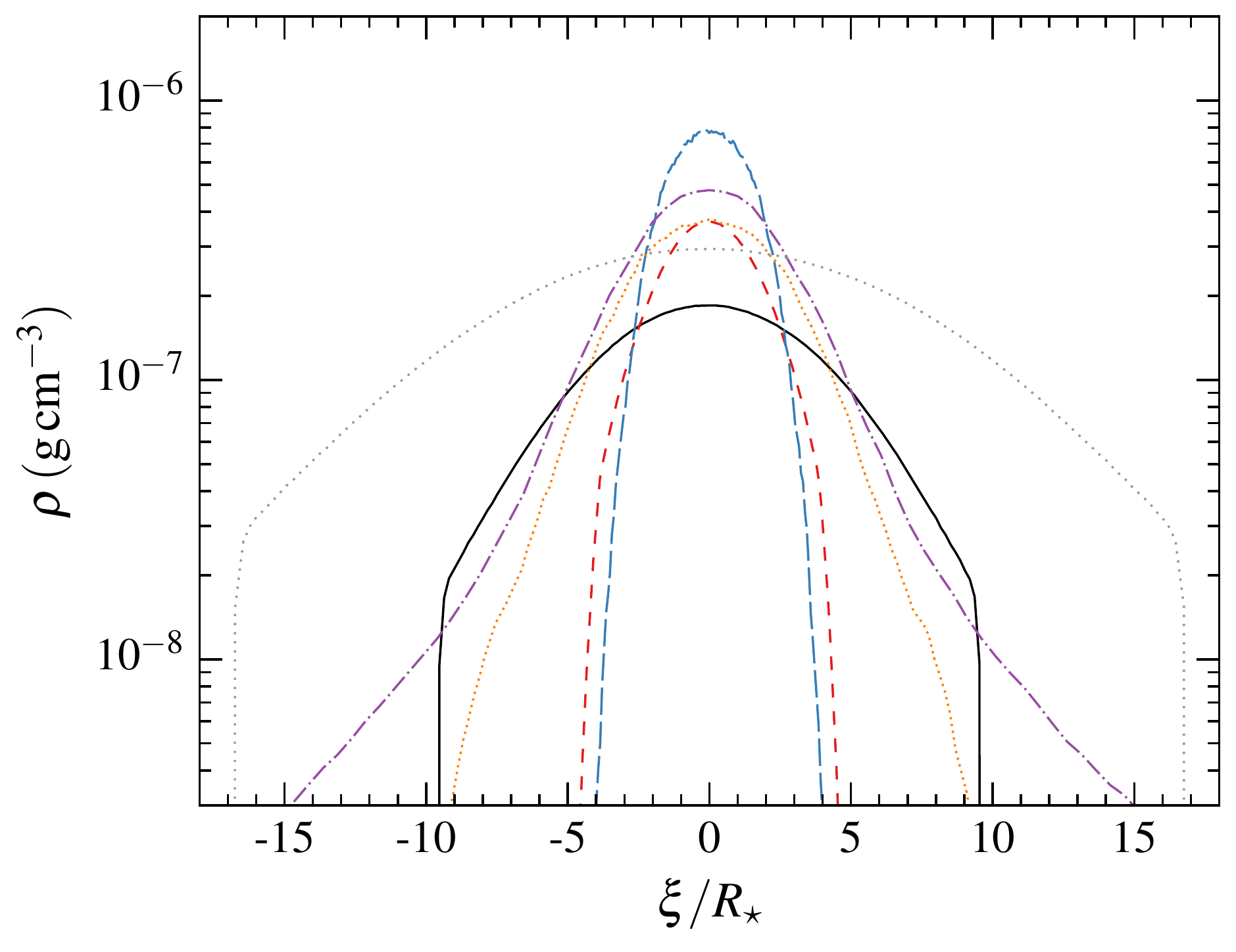}
\caption{Density profiles along the vertical (upper panel) and in-plane directions (lower panel) at different times $(t - \tp)/\tstar = $ -197 (black solid line), -15.6 (red dashed line), 15.6 (blue long-dashed line), 197 (orange dotted line) and 1622 (purple dash-dotted line) that are also shown in the snapshots of Fig. \ref{fig:density}. The grey dotted line corresponds to $(t - \tp)/\tstar = -1622$ that is before the start of the two-dimensional simulation. It is obtained by extrapolating from the initial density profile at $(t - \tp)/\tstar = -197$ using the scaling $H \propto R ^{1/2}$ identified during the infalling phase for the transverse widths evolution based on the upper panel of Fig. \ref{fig:hrhovsr_infall} and the analytical evolution of the mass of the stream element given by equation \eqref{eq:mass_change}. The chronological ordering of the different curves is indicated schematically in the top-left corner of the upper panel. The three earliest profiles are taken at times exactly opposite to those of the three latest ones, which means that they are obtained when the stream element is at the same distance from the black hole of $R/\rt =$ 9.3, 51.2 and 158 before and after pericenter passage, respectively.}
\label{fig:rhovzvszxi_large}
\end{figure}

Although the pressure gradients are close to purely vertical immediately after the nozzle shock, the resulting bounce rapidly leads to a more circular transverse density distribution (see Fig. \ref{fig:density} at $(t-\tp)/\tstar = 15.6$).\footnote{In the last snapshot of Fig. \ref{fig:density}, we note the presence of irregular density features in the outer layers of the stream element close to the orbital plane of its center of mass. We attribute them to an irregular distribution of the low number of particles present in this region, which results in artificial fluctuations of the gas density. Because these features are only present for a small fraction of the total mass, we nevertheless do not expect this effect to change the results of our simulation.} As a result, pressure forces start additionally acting to make the stream expand along the in-plane direction. This causes the gas to get deflected away from the center of mass trajectory, which induces a spread in the orientation of its different major axes with a range of precession angle $\psi_{\rm h} \approx \pm 10^{-5} \pi$. This influence of pressure can be directly seen by comparing the hydrodynamical (green dash-dotted line) and ballistic (blue dotted line) trajectories displayed in Fig. \ref{fig:trajectory}. While ballistic motion predicts an intersection with the center of mass trajectory at the location of the purple point, this collision is prevented by the deflection induced by pressure forces. This provides an explanation for the absence of in-plane collapse in the hydrodynamical evolution, as we described before. As explained more in Appendix \ref{ap:intersections}, an intersection is nevertheless still expected very close to apocenter. This can be seen from the flattening of the red dashed curve in Fig. \ref{fig:hvsr_outflow} at the largest radii, which is expected to be followed by an in-plane compression. This type of interactions already predicted by \citet{kochanek1994} (see his figure 5) may significantly affect the subsequent stream evolution, particularly by broadening its angular momentum distribution.

We have seen that the influence of pressure is not limited to the nozzle shock, but also affects the gas motion during most of its later evolution towards apocenter. This is because of the entropy jump that increases the gas internal energy past pericenter despite similar densities (see Fig. \ref{fig:uvst}). As the gas moves away from the black hole, the ratio of tidal to pressure forces scales as $F_{\rm t}/F_{\rm p} \approx G \mh H^2/(u R^3) \propto R^{-5/3}$ like during the infalling phase since the stream properties evolve similarly as long as $R\lesssim 10 \rt$ (see Fig. \ref{fig:hvsr_outflow}). However, the value of this ratio shortly after pericenter passage is of $F_{\rm t}/F_{\rm p}|_{R\gtrsim\rp} \approx 10$, that is more than an order of magnitude lower than before due to the entropy increase at the nozzle shock. Here, we have used $u\approx 10^{-3} v^2_{\star}$ and $H\approx 0.1 \rstar$ to get the numerical value, which corresponds to a time $(t - t_{\rm p})/\tstar \approx 1$. As the stream moves outward, pressure forces can therefore become comparable to the tidal force, thus affecting the gas motion. A consequence is that self-gravity is not expected to become dynamically important again since it will likely be dominated by pressure forces even after the density has reached $\rho \gtrsim \rho_{\rm sg}$ near apocenter.\footnote{Remarkably, this long-term influence of pressure forces is not present in deep stellar disruptions due to the strong stretching experienced by the debris after pericenter that degrades the injected internal energy through adiabatic losses \citep{guillochon2009}. The situation is different following the nozzle shock because the stream gets instead compressed along the longitudinal direction (see Section \ref{sec:treatment}) as it recedes from the black hole that prevents any such reduction of pressure forces.}

The density profiles of the stream element are shown in Fig. \ref{fig:rhovzvszxi_large} along the vertical (upper panel) and in-plane (lower panel) directions. They are obtained at different times $(t - \tp)/\tstar = $ -1622 (grey dotted line), -197 (black solid line), -15.6 (red dashed line), 15.6 (blue long-dashed line), 197 (orange dotted line) and 1622 (purple dash-dotted line) that are also displayed in the snapshots of Fig. \ref{fig:density}. The fact that these times have exactly opposite signs implies that the stream element is at the same  distance from the black hole of $R/\rt = $ 9.3, 51.2, 158 on opposite sides with respect to pericenter passage. At the lowest two radii, the gas is more concentrated\footnote{Our finding that the gas is more concentrated after the nozzle shock appears to be consistent with the semi-analytical calculations of \citet{lynch2021} (see their figure 14) focusing on the same mechanism in the context of eccentric accretion discs.} and features a more extended density profile along the vertical direction after pericenter passage than before. We suggest that this difference can be traced back to the bounce following the nozzle shock, during which the expansion of the gas closest to the equatorial plane is hampered by matter still moving inward (see Fig. \ref{fig:density_wave}). As a result, the outward motion of this central part of the stream is slowed down compared to matter near the boundary that expands instead in complete vaccuum. At the largest radius close to apocenter, the central concentration is reduced while the gas gets more compressed along the in-plane direction than during the infalling phase, which is consistent with the widths evolution displayed Fig. \ref{fig:hvsr_outflow} (solid black and red dashed line) at $R\gtrsim 100 \rt$.

Apart from the above differences in the exact mass distribution within the stream element, we find that the density profiles of Fig. \ref{fig:rhovzvszxi_large} are overall similar when considered at the same distance from the black hole. This is because the stream element does not undergo significant \textit{net} expansion due to the impact of pressure forces at the nozzle shock and later in the evolution. This conclusion is also confirmed by the widths computed in Fig. \ref{fig:hvsr_outflow} that only vary by a near-unity factor between the infalling phase and the recession towards apocenter. This qualitative evolution drastically differs from that obtained in several previous works, which find that the stream gets strongly inflated during pericenter passage.

\section{Discussion}
\label{sec:discussion}

\subsection{Convergence study}
\label{sec:convergence}

\begin{figure}
\centering
\includegraphics[width=\columnwidth]{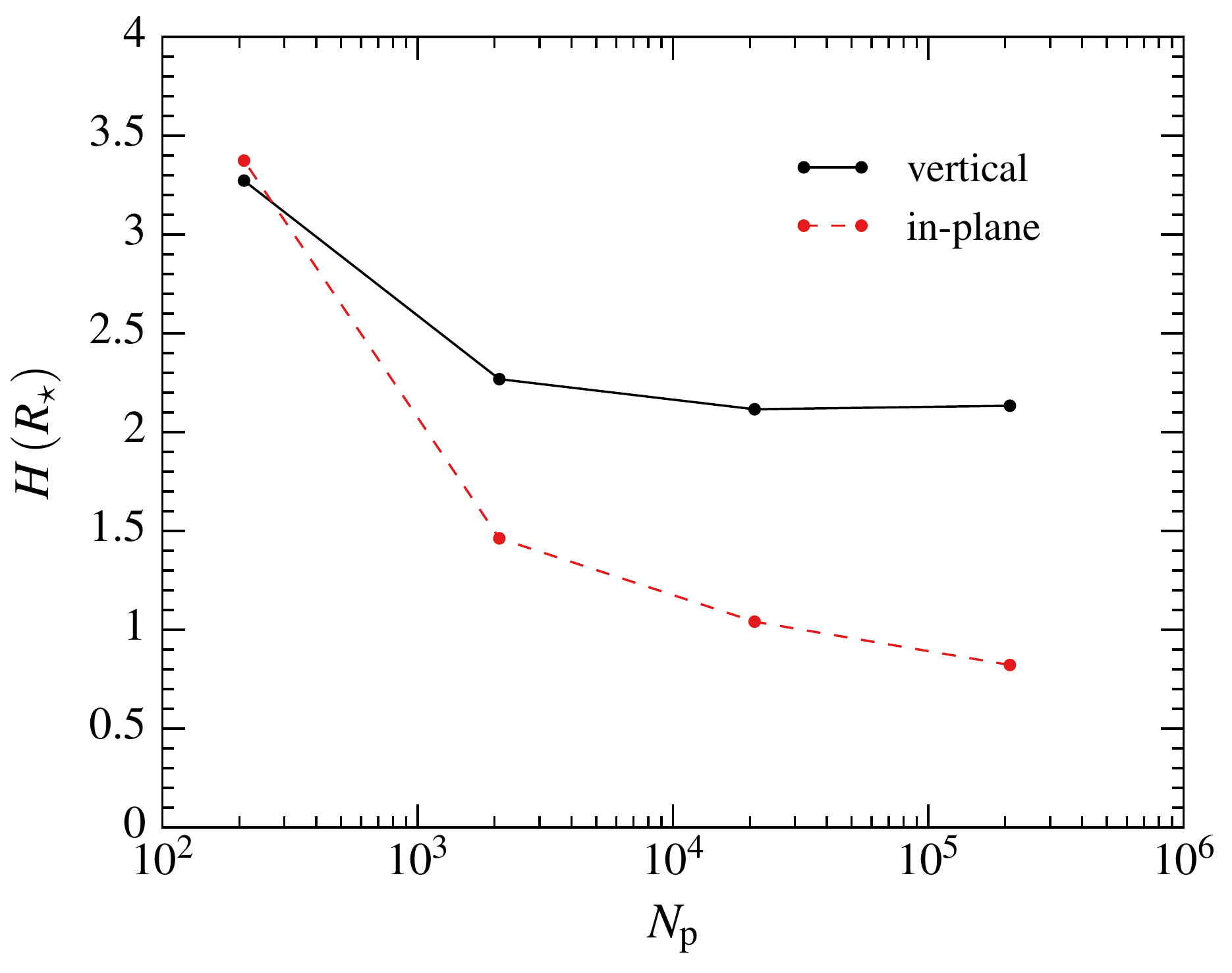}
\caption{Vertical (black solid line) and in-plane (red dashed line) widths as a function of number of particles, successively set to $N_{\rm p} \approx 2 \times 10^2$, $2 \times 10^3$, $2 \times 10^4$ and $2 \times 10^5$ as displayed with the points of the same colours. These widths are defined as the distances from the center of mass that enclose half of the mass in each direction. They are computed at a time  $(t - \tp)/\tstar = $ 197 when the stream element has receded back to a distance of $R = 51.2 \rt$ from the black hole after pericenter passage.}
\label{fig:hvsnp}
\end{figure}

We now carry out a convergence study to evaluate the dependence of the results on the resolution used. To this aim, we show in Fig. \ref{fig:hvsnp} the vertical (solid black line) and in-plane (red dashed line) widths as a function of number of particles used to describe the stream element. This number is successively set to $N_{\rm p} \approx 2 \times 10^2$, $2 \times 10^3$, $2 \times 10^4$ and $2 \times 10^5$, corresponding to the points of the same colours. The highest particle number is that used in the simulation presented throughout this paper. As before, the widths are defined by the distance from the center of mass that contains half of the mass in each direction. Here, they are evaluated after the nozzle shock at a time $(t - \tp)/\tstar = 197$ when the gas has receded back to a distance $R = 51.2 \rt$ from the black hole. The widths are essentially identical at the largest particles numbers, which demonstrates that the results presented in the paper are not expected to change if resolution is further increased. At lower particle numbers, the stream becomes both thicker and more circular, which we attribute to an artificially weaker vertical compression at the nozzle shock that makes the bounce more isotropic.

The particle number required for numerical convergence can be evaluated from $N^{\rm 2D}_{\rm p} \approx m/M_{\rm p}$, where the mass of the stream element is given by $m = \dot{M} l / v_{\rm c}$ (see Section \ref{sec:initial}) while the particle mass relates to the resolution length $h_{\rm res}$ in two dimensions as $M_{\rm p} \approx \Sigma \, h^2_{\rm res}$. Writing the surface density as $\Sigma = \rho l$ and making use of $\rho \approx \dot{M}/(H_{\rm z}H_{\perp} v_{\rm c})$, the particle number is then given by 
\be
N^{\rm 2D}_{\rm p} \approx  \frac{H_{\rm z}H_{\perp}}{h^2_{\rm res}} \gtrsim \epsilon^{-2} \frac{H_{\perp}}{H_{\rm z}} \approx 10^4 \left( \frac{\epsilon}{0.1} \right)^{-2} \left( \frac{H_{\rm z}}{10^{-3} \rstar} \right)^{-1} \left( \frac{H_{\perp}}{0.1 \rstar} \right),
\ee
where the inequality is obtained by imposing that the resolution length is less than a small fraction $\epsilon \approx 0.1$ of the minimal vertical width, i.e. $h_{\rm res} \lesssim \epsilon H_{\rm z}$. This condition is the most constraining at maximal compression where our simulation finds that the vertical and in-plane widths decrease to $H_{\rm z} \approx 10^{-3} \rstar$ and $H_{\perp} \approx 0.1 \rstar$ (see upper panel of Fig. \ref{fig:hrhovsr_infall}). It implies that the number of particles must be $N^{\rm 2D}_{\rm p} \gtrsim 10^4$ to sufficiently resolve the vertical size of the stream at pericenter within a two-dimensional simulation, which is consistent with the value where convergence starts in Fig. \ref{fig:hvsnp}.

Another approach to study the nozzle shock could rely on a single global three-dimensional simulation that first follows the stellar disruption, then the later evolution of the returning stream at pericenter. Like for the two-dimensional case, the number of particles required to resolve the stream at pericenter in three dimensions can be estimated. It is given by $N^{\rm 3D}_{\rm p} \approx \mstar / M_{\rm p}$, where the particle mass relates to the resolution length as $M_{\rm p} \approx \rho h^3_{\rm res}$. The condition $h_{\rm res} \lesssim \epsilon H_{\rm z}$ then yields
\be
N^{\rm 3D}_{\rm p}  \gtrsim \frac{\mstar H_{\perp} v_{\rm p}}{\epsilon^3 \dot{M} H^2_{\rm z}} = 10^{14} \left( \frac{\epsilon}{0.1} \right)^{-3} \left( \frac{H_{\rm z}}{10^{-3} \rstar} \right)^{-2} \left( \frac{H_{\perp}}{0.1 \rstar} \right),
\ee
where the density has been expressed as before, adopting its value at pericenter by setting $v_{\rm c } = v_{\rm p} \approx 0.2 c$. The extremely high number of particles $N^{\rm 3D}_{\rm p}  \approx 10^{14}$ necessary to resolve the compression cannot be reached with current computational capacities. Three-dimensional simulations of this process are so far limited to much lower values of $N^{\rm 3D}_{\rm p} \lesssim 10^7$, for which the vertical extent of the stream is effectively resolved with one particle or less. This lack of resolution likely causes a large overestimate of the stream thickness after pericenter passage, as expected based on an extrapolation of our convergence study to lower particle numbers (see Fig. \ref{fig:hvsnp}). Although we focused here on particle-based methods, the very large resolution necessary to accurately follow the nozzle shock is likely also problematic for grid-based codes. In this case, it implies that the computational domain must be divided into many cells that also increases computational demands.

Another source of artefacts in three-dimensional simulations is due to the possibility that a fraction of the orbital kinetic energy gets artificially dissipated in addition to the much lower transverse component. Due to the large ratio of $v^2_{\rm p}/v^2_{\rm z} \approx 10^4$ between these two energies, even a very small conversion of the orbital kinetic energy into heat can dramatically affect the transverse evolution. The important lack of resolution suffered by three-dimensional simulations near pericenter likely makes them prone to overestimating this energy transfer. This effect may accelerate the subsequent gas expansion, resulting in an artificially large width of the stream as it gets away from the black hole.  This is indeed the qualitative trend seen in three-dimensional simulations carried out by early works \citep{lee1996_tvd,ayal2000} and more recent investigations at higher resolution. While these authors find that the outgoing stream reaches a thickness comparable to its distance from the black hole, our two-dimensional study predicts similar transverse widths before and after pericenter passage (see Fig. \ref{fig:rhovzvszxi_large}). We suggest that the above effects affecting three-dimensional simulations are at the origin of these differences.

\subsection{Validity of the two-dimensional approach}
\label{sec:validity}

Our two-dimensional simulation neglects the mutual interaction between successive stream elements and we check here the validity of this assumption. A first possibility is that the returning stream gets heated before passing through the nozzle shock due to pressure waves travelling upstream from matter already undergoing this interaction. The ratio of the sound speed $c_{\rm s} = (9 u /10)^{1/2}$ at which this signal travels to the orbital velocity  $v_{\rm p} \approx (2 G  \mh /\rp)^{1/2}$ at pericenter is of about $c_{\rm s}/v_{\rm p} \approx 10^{-3} \ll 1$, making use of the internal energy $u \approx 10^{-2} v^2_{\star}$ found near the nozzle shock (see Fig \ref{fig:uvst}). This implies that the shocked matter is unable to communicate with the upstream gas. A given section of stream therefore only starts experiencing significant heating when it arrives at the nozzle shock. Our neglect of this longitudinal energy transfer is therefore fully justified.

Longitudinal pressure gradients can also develop downstream from the nozzle shock, leading to the acceleration of highly compressed matter towards the already expanding gas along the direction of motion. To evaluate the importance of this effect, we calculate the longitudinal pressure force $F_{\rm \, p \parallel} = -\nabla P \cdot \vect{e}_{\parallel}/\rho$ by estimating the gradient from the pressure variation of the stream element at successive times divided by the offset in position of its center of mass. This leads to a maximal value of $F_{\rm  p \,\parallel} \approx  0.01 v^2_{\star}/R_{\star} >0$ that is reached immediately downstream from the nozzle shock due a reduction in pressure. To determine its importance, this force must be compared to that in the vertical direction, which can be directly estimated from $F_{\rm p ,z} \approx u/H_{\rm z} = 100 \, v^2_{\star}/R_{\star}$. Here, we have used the gas properties found in the simulation when the gas is vertically compressed to a small width $H_{\rm z} \approx 10^{-3} \rstar$ (see upper panel of Fig. \ref{fig:hrhovsr_infall}). We therefore find that the longitudinal component of the pressure force is much lower than in the vertical direction with a ratio $F_{\rm p  \, \parallel} / F_{\rm \, p ,z} \approx 10^{-4} \ll 1$. Its influence on the dynamics can then be neglected as assumed in our two-dimensional approach.\footnote{The impact of longitudinal pressure gradients appears lower than in the situation of deep stellar disruptions, for which three-dimensional simulations find that they can significantly affect the dynamics \citep{guillochon2009}. This is likely because the inflow rate of the compressed star varies along the longitudinal direction due to its density profile, which enhances pressure leakage compared to the stream whose inflow rate through the nozzle shock is uniform.}\footnote{The configuration we studied is similar to that of ``oblique shocks'' \citep[e.g.][]{matzner2013}, particularly studied in the context of supernovae shock breakout.}

\subsection{Extrapolation to other parameters}
\label{sec:extrapolation}

We now evaluate the dependence of our results on the choice of parameters. The strength of the nozzle shock can be estimated from the vertical component of the stream velocity near pericenter where the trajectories would intersect in the ballistic limit. This speed is approximately given by $v^{\rm max}_{\rm z} \approx  v_{\rm p} \vartheta$ where $\vartheta$ denotes the inclination angle between the orbital plane of the collapsing gas and that of the stream center of mass. As explained in Section \ref{sec:ballistic}, gas motion becomes close to ballistic from the moment when the tidal force starts dominating the dynamics. For the stream element we considered, this occurs shortly after apocenter passage, which we argue remains valid in general due to the sharp decrease in density experienced by the gas as it starts moving inward (see lower panel of Fig. \ref{fig:hrhovsr_infall}). Therefore, hydrostatic equilibrium is maintained down to a distance larger than the semi-major axis, i.e. $R_{\rm eq} \gtrsim a$. The inclination angle is then given by the ratio of the average stream width to its distance from the intersection line at this location, that is $\vartheta \approx \tan \vartheta = H_{\rm eq} / d_{\rm eq}$. This distance can be geometrically expressed as $d_{\rm eq} = ((\rp R_{\rm a})/(2a/R_{\rm eq} -1))^{1/2}$, where $R_{\rm a} \approx 2 a$ denotes the apocenter distance. Making use of $R_{\rm eq} \gtrsim a$ yields $d_{\rm eq} \gtrsim b \approx (\rp a)^{1/2}$ that is comparable to the semi-minor axis $b = (\rp  R_{\rm a})^{1/2}$ of the stream element, as can also be seen from Fig. \ref{fig:trajectory} for our parameters. The stream width at this location is estimated from $H_{\rm eq} \approx H_{\rm t} (R_{\rm eq} / \rt)^{1/2} \approx H_{\rm t} (a / \rp)^{1/2} \beta^{-1/2}$ according to the scaling followed shortly after stellar disruption, introducing the normalization $H_{\rm t} \approx 0.1 \rstar$.

Combining these calculations leads to a maximal collapse speed of
\be
\frac{v^{\rm max}_{\rm z}}{v_{\star}} = \frac{H_{\rm eq} v_{\rm p}}{d_{\rm eq} v_{\star}} \approx 0.1  \beta \left(\frac{H_{\rm t}}{0.1 \rstar}\right) \left(\frac{d_{\rm eq}}{b}\right)^{-1},
\label{eq:vzmax}
\ee
where we have also used the relation $v_{\rm p} \approx v_{\star} \beta^{1/2} \rt  / \rstar$. This estimate is consistent with the vertical velocities found in our simulation and also similar to that obtained by \citet{guillochon2014-10jh} apart for the lower normalization. The independence of this speed on the black hole mass ultimately ties to the coincidence of having identical scalings $H \propto R^{1/2}$ for outward and inward stream motion (see upper panel of Fig. \ref{fig:hrhovsr_infall}).\footnote{The estimate of equation \eqref{eq:vzmax} remains approximately valid for parts of the stream whose entire dynamics is specified by the tidal force since they are also expected to obey the same scaling $H \propto R$ \citep{coughlin2016-structure} while moving inward and outward.} Note however that a dependence appears in the case where $R_{\rm eq} \approx  R_{\rm a}$, which results in $d_{\rm eq} \approx  R_{\rm a}$ such that equation \eqref{eq:vzmax} leads to a lower speed given by $v^{\rm max}_{\rm z}/v_{\star} \approx 0.01 \beta^{1/2} (H_{\rm t} / 0.1 \rstar) (\mh/10^6 \mstar)^{-1/6} $ due to the reduced inclination angle. We nevertheless consider this situation unlikely since it requires a significant degree of fine-tuning such that the tidal force becomes dominant exactly at apocenter. As a result, the energy dissipated by the nozzle shock mainly depends on the sound speed of the star being disrupted and the penetration factor $\beta$, but we do not expect this effect to drastically affect the qualitative picture obtained from our study.

The specific stream element followed in our simulation is mostly chosen for numerical convenience (see Section \ref{sec:initial}). We find that its initial vertical velocity profile is close to homologous, although this may change for other parts of the stream. Deviations from homology can come from oscillations around hydrostatic equilibrium close to the moment when the tidal force overwhelms self-gravity. It may also be due to a transverse density profile that significantly differs from the binormal distribution we use. Such a difference is visible for the most bound tip of the returning gas that features a dense core offset with respect to more tenuous surrounding matter. As discussed by \citet{stone2013} in the context of deep stellar disruptions, non-homologous compression implies that gas at initially different vertical distances reaches the equatorial plane  at distinct locations along the trajectory. This desynchronization may cause fluid elements to cross each other multiple times away from the stellar orbital plane, leading to internal collisions. This implies that the energy dissipation at the nozzle shock could be more gradual than found in our simulation where this interaction is confined to a small region. We argue that this can lead to a more isotropic bounce compared to the gas motion we predict that is largely limited to the vertical direction.

The stream element considered initially has a range of pericenter distances that causes its in-plane width to remain $H_{\perp}\gtrsim \rstar$ during the nozzle shock (see Fig. \ref{fig:hvsr_outflow}). Physically, this is due to a spread in angular momentum imparted by pressure forces during the previous evolution of the stream around hydrostatic equilibrium. However, this in-plane width would be reduced for the parts of the stream entirely dominated by the tidal force since their prior evolution occurs at constant angular momentum. This effect may result in more in-plane compression or even crossing of trajectories along this direction. The nozzle shock could therefore become more isotropic with a larger amount of gas being significantly deflected along the equatorial plane than what our simulation predicts. Settling these uncertainties would require to improve our understanding of the orbital properties of the different parts of the returning stream including its small angular momentum, which we defer to the future.

\subsection{Role of relativistic effects}
\label{sec:relativistic}

While our simulation is entirely Newtonian, we now discuss how general relativity could affect the results. The most important effect is caused by relativistic apsidal precession that acts at pericenter to rotate in the direction of motion the ellipse followed by the stream center of mass by an angle \citep{hobson2006}
\be
\Delta \phi \approx  \frac{3 \pi\rg}{\rp} = 0.06  \pi \, \beta \left( \frac{\mh}{10^6 \msun} \right)^{2/3},
\label{eq:apsidal}
\ee
making use of the near-parabolic nature of the trajectory and assuming a solar-like star. This precession implies that the stream crosses the line where its different orbital planes intersect (blue long-dashed segment in Fig. \ref{fig:trajectory}) earlier than in the Newtonian case. If the gas moved ballistically at the nozzle shock, the stream could therefore get vertically compressed a second time shortly after leaving pericenter. This effect was discovered in the context of deep stellar disruptions by \citet{luminet1985}, who predict several vertical collapses near pericenter when general relativity is accounted for. However, our work finds that the nozzle shock additionally induces a rotation of the intersection line that would likely delay this second vertical compression for weakly relativistic encounters with $\Delta \phi \ll \pi$.\footnote{If the vertical collapse taking place for deep stellar disruptions also induces the modification of orbital planes we find in the present work, this conclusion may also apply to this situation.} Nevertheless, we still expect successive compressions to take place for strong precession with $\Delta \phi \gtrsim 2 \pi$ since it induces at least one entire near-circular revolution at pericenter before the gas goes back to apocenter. Another consequence of apsidal precession is to cause the formation of a self-crossing shock between two parts of the stream, which we discuss in Section \ref{sec:crossing}.

While the center of mass precesses by the angle given by equation \eqref{eq:apsidal}, fluid elements composing the stream that pass at different distances from the black hole have slightly different precession angles. This induces a differential precession of these elements with respect to the center of mass by an angle that can be estimated as
\be
\Delta \phi_{\rm d} \approx  \Delta \phi \frac{\xi_{\rm p}}{\rp} = \pm 6 \times 10^{-4}  \pi \, \beta^2  \left( \frac{|\xi_{\rm p}|}{\rstar} \right) \left( \frac{\mh}{10^6 \msun} \right)^{1/3},
\label{eq:apsidal_diff}
\ee
where $\xi_{\rm p}$ represents the distance from the center of mass at pericenter, which we set in the numerical value to $\xi_{\rm p} \approx \pm \rstar$ as motivated by our simulation (see the red dashed line in Fig. \ref{fig:hvsr_outflow}). This differential precession
compensates the small spread in apsidal angles imposed by the initial conditions, as discussed in Appendix \ref{ap:intersections}. Because it makes the gas trajectories diverge, we expect this effect to delay in-plane intersections compared to the location predicted by Newtonian dynamics (purple point in Fig. \ref{fig:trajectory}). Due to the small precession angles involved, it is however unlikely to prevent this collapse to eventually occur close to apocenter.

If the black hole has a non-zero spin inclined by an angle $i\neq 0$ with respect to the stellar angular momentum, the center of mass trajectory experiences a change of orbital plane at pericenter due to the additional nodal relativistic precession. This causes the gas to have its angular momentum vector shifted during pericenter passage by an angle \citep{stone2019}
\be
\Delta \Omega \approx \sqrt{2} \pi \, a_{\rm h} \sin i \left(\frac{\rg}{\rp} \right)^{3/2} = 0.004 \pi \,a_{\rm h} \sin i \, \beta^{3/2} \left( \frac{\mh}{10^6\msun} \right),
\label{eq:nodal}
\ee
where $a_{\rm h}$ denotes the dimensionless spin parameter. All the gas within the stream experiences a change of trajectory similar to that of the center of mass. However, small variations between fluid elements passing at the same distance from the black hole but moving along different orbital planes are expected due to distinct values of the inclination angle $i$. This differential nodal precession likely leads to small modifications of the transverse speeds involved in the nozzle shock, which we do not expect to qualitatively affect our results.\footnote{In the context of deep stellar disruptions, a similar effect has been proposed \citep{leloudas2016} to widen the energy spread of the debris due to a partial alignment of the bounce velocity with the direction of orbital motion. A potentially significant difference is that the estimates of this work appear to rely on an inclination angle between the bounce and longitudinal velocity vectors of order the precession angle $\Delta \Omega$ given by equation \eqref{eq:nodal}. Because not only the center of mass but also all the matter within the star or stream element precesses by a similar amount, we expect these two velocities to be inclined by a smaller angle that likely reduces this impact of black hole spin.} A complete study of these effects on the nozzle shock requires to generalize our numerical method to account for general relativity, which we defer to the future.

\begin{figure}
\centering
\includegraphics[width=\columnwidth]{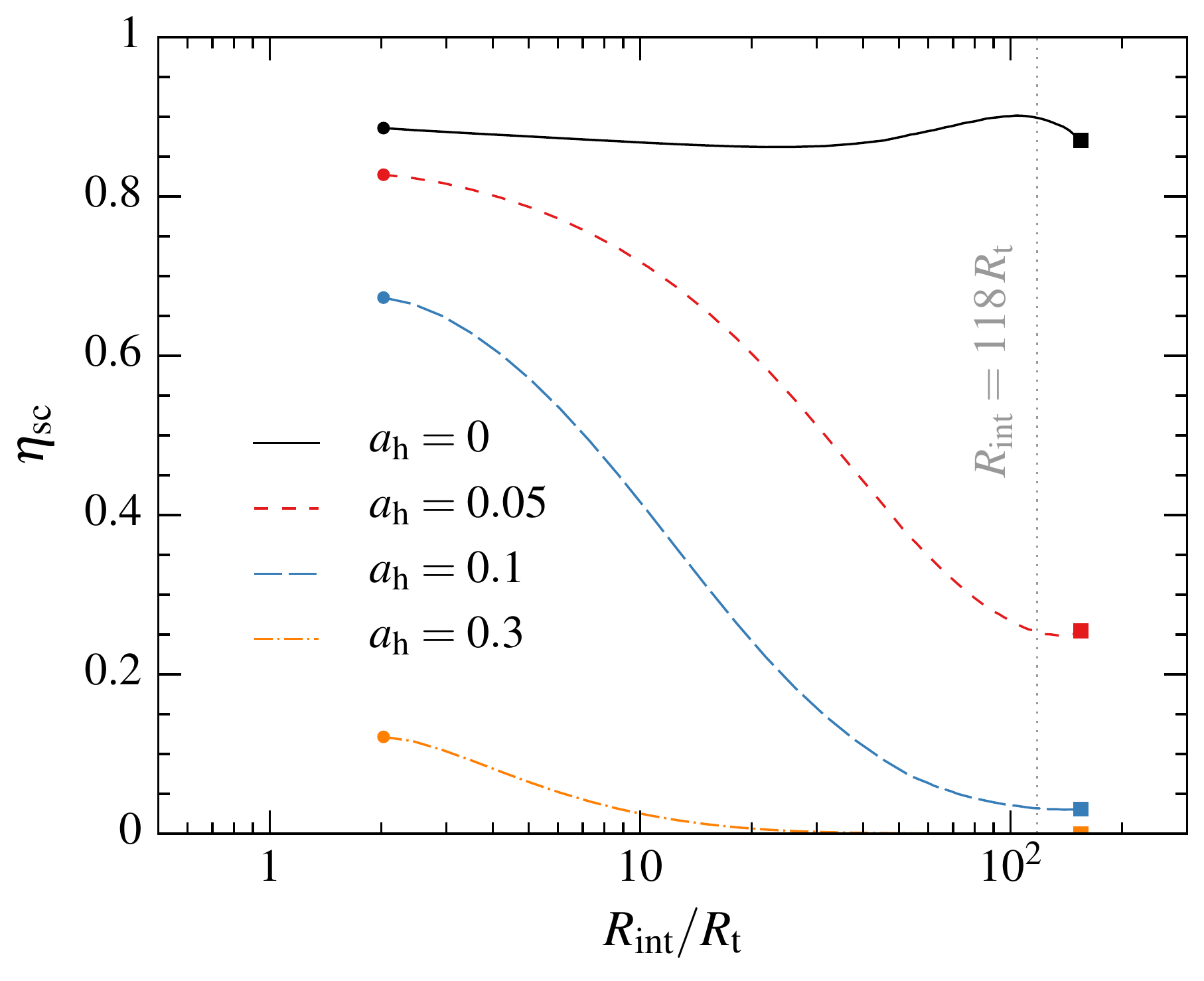}
\caption{Efficiency of the self-crossing shock obtained from equation \eqref{eq:efficiency} as a function of the intersection radius where this collision occurs for different black hole spin parameters $a_{\rm h}=0$ (black solid line), 0.05 (red dashed line), 0.1 (blue long-dashed line) and 0.3 (orange dash-dotted line). The vertical grey dotted line indicates the location of the intersection radius $\rint = 118 \rt$ resulting from apsidal precession for the orbital parameters of the stream element considered.}
\label{fig:etascvsrint}
\end{figure}

\subsection{Consequences on the self-crossing shock}
\label{sec:crossing}

Due to relativistic apsidal precession, the part of the stream that has passed pericenter is put on a collision course with the component still moving inward. If the black hole has no spin, this collision is unavoidable that results in a self-crossing shock at the intersection radius given by \citep{dai2015}
\be
R_{\rm int} = \frac{R_{\rm p} (1+e)}{1-e \cos(\Delta \phi/2)},
\label{eq:rint}
\ee
where the apsidal precession angle $\Delta \phi$ is given by equation \eqref{eq:apsidal}. For the stream element considered and our choice of parameters, this intersection radius is $R_{\rm int} = 118 \rt$ that is beyond the semi-major axis $a \approx 87 \rt$ of the trajectory due to weak precession. Through its impact on the gas at pericenter, the nozzle shock could lead to different properties for the two components colliding at the self-crossing shock, which may affect the strength of this interaction.

Even though our simulation does not include relativistic precession, we can evaluate this effect by comparing the properties of the stream element at the same distance from the black hole before and after pericenter.\footnote{Note that this computation assumes that the nozzle shock acts similarly on elements falling back to the black holes at different times, while we have seen in Section \ref{sec:extrapolation} that its impact may vary. Additionally, we neglect the different trajectories and fallback rates of the interacting streams, which could become significant when the intersection radius is close to apocenter.} This calculation can be done for an intersection radius ranging from pericenter to apocenter, extrapolating the properties of the infalling stream at $R\geq R_{\rm in}$ with the same method as used to produce the grey dotted line in Fig. \ref{fig:rhovzvszxi_large}. Quantitatively, this is done by defining an efficiency for the self-crossing shock given by
\be
\eta_{\rm sc} = \frac{1}{m_{\rm tot}} \int_{z_{\rm min}}^{z_{\rm max}}  \frac{\diff m_{\rm tot}}{\diff z} \tilde{\eta}(z)  \diff z,
\label{eq:efficiency}
\ee
where $m_{\rm tot} = m_{\rm in}+m_{\rm out}$ denotes the total mass of the colliding stream elements that move inward and outward when reaching the intersection point. The integral is computed by dividing the two distributions into slices containing matter at the same vertical distance that therefore collide with each other. The interacting mass is weighted by an efficiency parameter $\tilde{\eta}(z) = 4 \rho_{\rm in}(z) \rho_{\rm out}(z) / (\rho_{\rm in}(z) +\rho_{\rm out}(z))^2$ computed from the averaged densities in each slice for the two stream components. It represents the fraction of the incoming kinetic energy lost during the collision assuming inelasticity and velocities of equal magnitude but opposite directions, such that $\tilde{\eta}(z) = 1$ for $\rho_{\rm in}(z) = \rho_{\rm out}(z)$ while $\tilde{\eta}(z) \ll 1$ for widely different densities. The integrated efficiency given by equation \eqref{eq:efficiency} has the same physical meaning but averaged over the total mass of the intersecting stream elements.\footnote{Even if parts of the stream components can avoid a direct collision, they may however still be heated by the passage of a shockwave generated by a partial self-crossing shock involving the rest of the matter. This implies that their trajectories can be indirectly affected by the interaction, which is not taken into account in our calculation of the efficiency of equation \eqref{eq:efficiency}.}

This self-crossing shock efficiency is shown in Fig. \ref{fig:etascvsrint} with the black solid line as a function of the intersection radius  that ranges from near pericenter (circles) to apocenter (squares). It additionally assumes that the black hole is non-rotating with $a_{\rm h} =0$ such that the center of mass of the two streams evolve on the same orbital plane. The grey vertical dotted line indicates the value $\rint = 118 \rt$ corresponding to our choice of parameters. We find that $\eta_{\rm sc} \approx 0.9$ at all distances, with the small reduction compared to unity coming from the fact that the stream element is more centrally concentrated after pericenter passage than before. Its central density is larger by a factor $\rho_{\rm out}(0)/\rho_{\rm in}(0) \approx 2$ (upper panel of Fig. \ref{fig:rhovzvszxi_large}) that explains the value of the efficiency found.\footnote{Close to apocenter, the efficiency also displays a slight decrease due to faster vertical expansion of the outgoing stream that results in some of its mass passing beyond the sharper boundary of the infalling component.} We therefore conclude that the self-crossing shock remains very efficient despite the modifications induced by the nozzle shock, which results for the low level of net stream expansion it induces.

If the black hole rotates, the change of orbital plane caused by relativistic nodal precession results in a vertical offset 
\be
\Delta z =  \rint \sin \gamma \, \Delta \Omega,
\label{eq:offset}
\ee
between the stream components at the intersection point. Here, the angle $\Delta \Omega$ is obtained from equation \eqref{eq:nodal} while $\gamma$ is measured between the direction joining the intersection point to the black hole and the line along which the two orbital planes intersect.\footnote{This angle is a priori random but has been implicitly set to $\gamma = \pi/2$ is previous estimates that maximizes the vertical offset. This offset is only overestimated by a factor of order unity for most values of this angle except when $\gamma \ll \pi$ that leads to a strong reduction with $\Delta z \approx 0$ despite the presence of nodal precession.} This effect may prevent a fraction of the incoming gas from efficiently colliding, which can be evaluated by comparing the offset to the vertical widths of the stream components. We assume for now a common value for the latter given by
\be
H_{\rm int} = H_{\rm t} \left( \frac{R_{\rm int}}{\rt} \right)^{\kappa},
\label{eq:width}
\ee
with an exponent $\kappa = 1/2$ and a normalization set to $H_{\rm t} = 0.1 \rstar$ as motivated by our simulation (see black solid and dotted lines in Fig. \ref{fig:hvsr_outflow}). The ratio of the vertical offset to stream width is then given by
\be
\begin{split}
\frac{\Delta z}{H_{\rm int}} & =  \frac{\Delta \Omega \rt}{\rstar} \beta^{\kappa -1} \left( \frac{H_{\rm t}}{\rstar} \right)^{-1}  \left( \frac{\rint}{\rp} \right)^{1-\kappa}  \\
& \approx 100 \, a  \sin i \, \beta \left( \frac{\mh}{10^6 \msun} \right)^{4/3}  \left( \frac{H_{\rm t}}{0.1 \rstar} \right)^{-1} \left( \frac{\rint}{118 \rp} \right)^{1/2},
\end{split}
\label{eq:dzovh}
\ee
where we set $\gamma = \pi/2$ for simplicity and used the parameters of our simulation to get the numerical value.\footnote{Note that the this ratio may additionally be affected by black hole spin through a modification of the relativistic apsidal angle it induces compared to equation \eqref{eq:apsidal} that can either increase or decrease the intersection radius depending on the orientation of the black hole spin compared to the gas angular momentum vector.} Our estimate of this ratio is larger than found in past studies \citep{guillochon2015,jiang2016} that assumed a linear width evolution with $\kappa = 1$ and a larger normalization of the stream width of $H_{\rm t} \approx \rstar$. From our simulation, we find that the stream is both thinner near the black hole and expands slower,\footnote{As mentioned in Section \ref{sec:extrapolation}, parts of the stream with low densities are not confined by self-gravity that could result in a fast expansion with $\kappa \approx 1$, although potentially also decreasing $H_{\rm t}$. More work is need to pinpoint the impact of black hole spin for this gas that nevertheless represent a much lower fraction of the total mass.} which results in an increase of the ratio by about two orders of magnitude. This implies that nodal precession may strongly affect the self-crossing shock even for slowly-spinning black holes with $a_{\rm h} \gtrsim 0.01$ while past estimates require near extremal rotation for $\mh \approx 10^6 \msun$.

A more quantitative evaluation of this impact of black hole spin can be done by computing the self-crossing shock efficiency from equation \eqref{eq:efficiency}, introducing the vertical offset of equation \eqref{eq:offset} between the centers of the two colliding stream components. This efficiency is shown in Fig. \ref{fig:etascvsrint} for several spin parameters, assuming $\gamma = \pi/2$ and $\sin i = 1$ for simplicity. For $a_{\rm h} = 0.05$ (red dashed line), we find that the efficiency is strongly reduced to $\eta_{\rm sc} \lesssim 0.5$ for an intersection radius $\rint \gtrsim 50 \rt$, as predicted by equation \eqref{eq:dzovh}. The self-crossing shock is almost as efficient as for $a_{\rm h}=0$ (black solid line) with $\eta_{\rm sc} \gtrsim 0.8$ close to pericenter due to the reduced impact of the sub-linear width increase. Increasing the black hole spin to $a_{\rm h} = 0.1$ (blue long-dashed line) further decreases the efficiency but with a similar rise at low radii. For $a_{\rm h} = 0.3$ (orange dash-dotted line), the self-crossing shock is inefficient at all radii with $\eta_{\rm sc} \lesssim 0.1$ because most of the gas misses the collision due to a large vertical offset.

The above calculations suggest that the self-crossing shock is efficient with $\eta_{\rm sc} \gtrsim 0.9$ as long as the black hole has a negligible spin. This conclusion lends support to the assumption of identical stream components used in local simulations of this interaction \citep{lee1996,kim1999,jiang2016,lu2020}. In this situation, the shocked gas is launched into a large-scale quasi-spherical outflow that can have a significant unbound fraction. When the bound matter returns near the black hole, an accretion disc may promptly form with a direction of rotation potentially opposite to that of the original star \citep{bonnerot2020-realistic,bonnerot2021-light}.\footnote{Note that this evolution differs from that found in global simulations of disc formation \citep[e.g.][]{shiokawa2015,sadowski2016}, for which the nozzle shock appears to result in faster expansion of the outgoing stream. This may be due to either the different parameters used or insufficient resolution near pericenter that can both result in faster expansion, as explained in Sections \ref{sec:convergence} and \ref{sec:extrapolation}.} If the efficiency is reduced to $\eta_{\rm sc} \gtrsim 0.1$, the interaction likely produces a non-spherical outflow that contains less unbound matter. Nevertheless, we still expect a significant amount of gas to have their trajectories deflected that could initiate the formation of an accretion disc. In the future, we intend to better characterize the outcome of the self-crossing shock if the collision is significantly offset.

If the efficiency is $\eta_{\rm sc} \ll 1$, most of the stream components miss each other on the first pass. In this situation, the stream continues to evolve on a largely unaffected trajectory for several orbital periods until a delayed collision eventually occurs \citep{guillochon2015}. Similarly to the nozzle shock, we expect the stream to experience several episodes of strong compression that may lead to additional dissipation. Studying this stage of evolution is necessary to precisely determine the time of the first intersection and the properties of the stream components involved. This could be achieved through a generalization of the two-dimensional approach presented here, which we intend to do in the future.

\subsection{Impact of viscous dissipation}

\label{sec:viscosity}

The stream element experiences shearing as it revolves around the black hole, implying that viscous effects could act to dissipate part of its orbital kinetic energy. As a result, the gas may experience additional heating compared to that caused by the nozzle shock only. Although viscosity is not included in our simulation, we aim here at estimating its impact based on the shearing properties we find. Only considering shearing along the perpendicular direction,\footnote{There may be other sources of viscous dissipation due to shearing along the vertical direction, but we neglect its influence because of the lower velocity gradients involved.} the associated viscous heating rate is given by the integral $\dot{E}_{\rm vis} =  \int_{\mathcal{V}} \nu  (\diff v_{\parallel}/\diff \xi)^2 \rho \diff V$ \citetext{equation (11.31) of \citealt{clarke2007}} over the volume $\mathcal{V}$ swept by the stream element throughout its evolution. This integral can be simplified by writing the differential volume as $\diff V = v_{\rm c} \diff \xi  \diff z \diff t$ combined with the definition of the inflow rate $\dot{M} = \int_{-\infty}^{+\infty} \int_{-\infty}^{+\infty} \rho v_{\rm c} \diff \xi \diff z$, which leads to $\dot{E}_{\rm vis} \approx \nu_{\rm p} \dot{M}/\tau_{\rm sh}$, approximating the viscosity by its value $\nu \approx \nu_{\rm p}$ at pericenter where shearing is the strongest. Here, we have introduced the timescale $\tau_{\rm sh} = \int_{t =0}^{t_{\rm max} } (\diff v_{\parallel}/\diff \xi)^2 \diff t = 0.88 \tstar$ computed directly from the shearing profile, where $t=t_{\rm max}$ corresponds to the end of the simulation. This value of $\tau_{\rm sh} \approx t_{\star}$ comes from $\diff v_{\parallel}/\diff \xi \approx v_{\star}/\rstar = 1/ \tstar$ near pericenter (see Fig. \ref{fig:vparavsxi}) where the stream element stays for a timescale of order $t_{\star}$. The viscous heating rate can then be estimated as
\be
\label{eq:edotvis}
\frac{\dot{E}_{\rm vis}}{\mdot v^2_{\star}} = \frac{ \nu_{\rm p}}{\tau_{\rm sh} v^2_{\star}} \approx   10^{-5} \left(\frac{\nu_{\rm p}}{10^{-5} \rstar v_{\star} } \right),
\ee
showing that it competes with that at the nozzle shock where $u \approx 10^{-2} v^2_{\star}$ only if $\nu_{\rm p} \gtrsim 10^{-2} \rstar v_{\star}$.

To estimate the value of the viscosity, we make use of the prescription $\nu = \alpha c_{\rm s} H_{\rm z}$, involving the local sound speed and vertical extent of the stream. At pericenter, this estimate leads to $\nu_{\rm p}/(\rstar v_{\star}) \approx 10^{-5} (\alpha/0.1) (u / 10^{-2} v_{\star})^{1/2} (H_{\rm z}/10^{-3} \rstar)$, setting the internal energy and stream vertical width to their values at the nozzle shock, and the viscosity parameter to $\alpha \approx 0.1$. The obtained viscosity is $\nu_{\rm p}/  \rstar v_{\star} \approx 10^{-5} \ll 10^{-2}$, implying that the resulting heating rate of equation \eqref{eq:edotvis} is much lower than that due to the shock. The magneto-rotational instability \citetext{MRI, \citealt{balbus1991}} could provide a physical source of viscosity after saturation has been reached. However, this requires several orbital periods while in our situation the stream remains near pericenter for only about half an orbit \citep{guillochon2014-10jh,chan2018}. For this reason, saturation is unlikely to be attained that may strongly reduce the viscosity compared to the above value based on $\alpha \approx 0.1$. An improved evaluation of this effect would require to carry out a dedicated magneto-hydrodynamical simulation of the stream passage at pericenter.

\subsection{Influence of magnetic fields}

\label{sec:magnetic}

Following the disruption, the stellar magnetic field gets transported with the stream, which has been investigated through magneto-hydrodynamics simulations \citep{guillochon2017-magnetic,bonnerot2017-magnetic}. These studies find that the magnetic field lines get aligned with the longitudinal direction of stream elongation. This is  because magnetic flux conservation induces a strong decrease of the transverse components of the field due to stretching, while the longitudinal magnetic field strength gets only weakly reduced due to slow expansion. During the strong compression at the nozzle shock, the strength of this longitudinal field component is increased from its original stellar value $B_{\parallel \star} \approx 1 \gauss$ to 
\be
B_{\parallel} \approx B_{\parallel \star} \frac{\rstar^2}{H_{\rm z} H_{\perp} } = 10^3 {\rm G} \left( \frac{B_{\parallel \star}}{1 \gauss} \right) \left( \frac{H_{\rm z}}{10^{-3} \rstar} \right)^{-1} \left( \frac{H_{\perp}}{\rstar} \right)^{-1},
\label{eq:magnetic}
\ee
where we have used $H_{\rm z} \approx 10^{-3} \rstar$ and $H_{\perp} \approx \rstar$ as found in the simulation (see Fig. \ref{fig:hrhovsr_infall}). This increase in magnetic strength results in an enhanced magnetic pressure that may compete with that due to the gas only.

To estimate this effect, we compare the two pressures obtained from $P_{\rm mag} = B^2_{\parallel}/(8 \pi)$ using equation \eqref{eq:magnetic} and $P_{\rm gas} = 2 \rho u /3$. Evaluating the density at pericenter with $\rho = \mdot / (\pi H_{\rm z} H_{\perp} v_{\rm p} )$, the pressure ratio is given by
\be
\begin{split}
\frac{P_{\rm mag}}{P_{\rm gas}} & = \frac{3 B^2_{\parallel \star} \rstar^4 v_{\rm p}}{16 H_{\rm z} H_{\perp} u \mdot} \\
& \approx 7 \times 10^{-6} \left( \frac{B_{\parallel \star}}{1 \gauss} \right)^2 \left( \frac{H_{\perp}}{\rstar} \right)^{-1} \left( \frac{H_{\rm z}}{10^{-3} \rstar} \right)^{-1}  \left( \frac{u}{10^{-2} v^2_{\star}} \right)^{-1}.
\end{split}
\ee
This implies that a dynamical impact of magnetic pressure on the nozzle shock requires a stellar magnetic field of $B_{\parallel \star} \gtrsim 400 \gauss$ that can be reached inside solar-type stars as evidence by observations of sunspots. Note that even larger strengths $B_{\parallel \star} \gtrsim 10^5 \gauss$ are required for magnetic pressure to affect the previous stream evolution \citep{guillochon2017-magnetic,bonnerot2017-magnetic}. The impact on the nozzle shock becomes more important as the fallback rate drops, also due to the arrival of more magnetized gas originating from near the stellar core. This effect may also be stronger if the penetration factor is increased to $\beta>1$ due to a larger velocity at pericenter and a stronger vertical compression. Magnetic pressure could lead to a faster expansion of the stream past pericenter, which we intend to investigate in the future through a magneto-hydrodynamics simulation of the nozzle shock.

\subsection{Effect of radiative processes}

\label{sec:radiative}

We now estimate the impact of radiative processes on the gas, which is assumed to evolve adiabatically in the simulation. The time required for radiation to diffuse out from the stream along its vertical direction can be determined from $t_{\rm diff} \approx H_{\rm z} \tau / c$, where $\tau = \rho \kappa_{\rm s} H_{\rm z}$ denotes the optical depth and the density is given by $\rho = \dot{M} / (\pi H_{\rm z} H_{\perp} v_{\rm c})$. This leads to a diffusion timescale of 
\be
\frac{t_{\rm diff}}{t_{\star}} = \frac{\dot{M}\kappa_{\rm s}}{\pi v_{\rm c} c t_{\star}} \frac{H_{\rm z}}{H_{\perp}} \approx 50 \left( \frac{H_{\rm z}}{H_{\perp}} \right) \left( \frac{v_{\rm c}}{v_{\rm p}} \right)^{-1}.
\label{eq:diffusion}
\ee
To determine the impact of radiative losses\footnote{As shown in Section \ref{sec:treatment}, the internal energy or pressure in the stream is dominated by thermal motion rather than radiation. Nevertheless, radiation energy losses directly result in gas cooling because temperature equilibrium between these two components is reached almost instantaneously for the properties of the gas considered.} on the nozzle shock, the diffusion timescale must be compared to the time $t_{\rm exp} = H_{\rm z} / v_{\rm z}$ required for the stream element to expand significantly. Making use of $v_{\rm c} = v_{\rm p}$ in equation \eqref{eq:diffusion}, their ratio is $t_{\rm diff} / t_{\rm exp} \approx (\dot{M}\kappa_{\rm s} / \pi c H_{\perp}) (v_{\rm z} / v_{\rm p}) \approx 50 \gg 1$ where the vertical width cancels out. Here, we adopted the numerical values $v_{\rm z} \approx 0.1 v_{\star}$ and $H_{\perp} \approx 0.1 \rstar$ reached at the point of maximal compression. This implies that diffusion has a negligible impact on the gas dynamics near pericenter, which justifies our assumption of gas adiabaticity.\footnote{This inefficient diffusion implies that the emerging luminosity is lower than the heating rate $\dot{E}_{\rm sh} \approx \dot{M} u \approx 10^{39} \ergpers$ at the nozzle shock, so that it is unlikely to be distinguishable from the host galaxy emission.} When the stream recedes from the black hole at later times, the diffusion timescale decreases relative to the dynamical timescale $t_{\rm dyn} = R/v_{\rm c}$ with a ratio $t_{\rm diff} / t_{\rm dyn} \approx 50 \, (R/\rp)^{-1}$. Near apocenter, these timescales become similar such that radiative cooling may become important. This effect is enhanced if the mass fallback rate is lower than the value near peak we adopted, which is expected for the more tenuous tip of the stream.

As it passes through the nozzle shock, the hydrogen that had recombined during the previous evolution \citep[e.g.][]{kasen2010} gets reionized due to the large temperatures reached. This however does not affect the hydrodynamics of the interaction since the specific energy required for ionization is $E_{\rm H}/m_{\rm p} \approx 5 \times 10^{-4} v^2_{\star}$ where $E_{\rm H} = 13.6 \, \rm eV$, which is much lower than that $u \approx 10^{-2} v^2_{\star}$ injected due to dissipation. As the gas expands at larger distances, its temperature decreases back to $T \lesssim 10^4 \kelvin$ such that hydrogen recombines again, which is associated with a jump in entropy. This effect may result in additional expansion compared to what our adiabatic simulation predicts, if radiative cooling can be neglected.

\section{Summary}
\label{sec:summary}

We have carried out the first dedicated study of the return of the debris stream towards the black hole following a stellar disruption. This is realized by first performing a three-dimensional simulation of the disruption, from which we obtain the hydrodynamical properties of a given section of stream. These properties are then used to initialize a two-dimensional simulation that follows the subsequent transverse evolution of this stream element in the frame co-moving with its ballistic center of mass and co-rotating with the longitudinal direction. Using this numerical technique, we are able to study with sufficient numerical resolution the approach of this gas towards the black hole, its passage at pericenter where the nozzle shock takes place, and its recession back to large distances. The results of this investigation and the main points of the accompanying discussion can be summarized as follows.

\begin{enumerate}

\item During its infall towards the black hole, the transverse widths of the stream element decrease approximately as $H \propto R^{1/2}$, which is specified by the dominant tidal force. Due to a combination of this homologous compression and longitudinal stretching, its density increases according to $\rho \propto R^{-1/2}$ (see Fig. \ref{fig:hrhovsr_infall}). 

\item When it gets close to the black hole, the stream is strongly compressed along the vertical direction due to an intersection of its different orbital planes. Its vertical width sharply decreases down to $H_{\rm z} \approx 10^{-3} \rstar$ while its in-plane extent remains of $H_{\perp} \approx 0.1 \rstar$. This point of maximal compression is reached slightly before pericenter, as predicted from ballistic motion. 

\item The strong compression near pericenter results in the formation of the nozzle shock that dissipates the kinetic energy associated with vertical motion. As a result, a shockwave is launched from the equatorial plane that rapidly sweeps through the collapsing gas. The infalling motion of this matter is reverted, which makes it bounce back to larger vertical distances.

\item While the rest of its evolution is entirely adiabatic, the gas undergoes a sharp increase of its specific internal energy to a maximal value of $u \approx 0.1 v^2_{\star}$ during the nozzle shock (see Fig. \ref{fig:uvst}). This dissipation is associated with an entropy jump by about two orders of magnitude, causing an overall increase of pressure forces that therefore become important at later times.

\item Throughout its evolution, the stream element experiences shearing along its longitudinal direction due to the relative acceleration between its parts located at different distances from the black hole. This effect is the strongest at pericenter where the shearing profile becomes close to that expected from a Newtonian circular flow.

\item After pericenter passage, the gas expands with transverse widths increasing as $H \propto R^{1/2}$ close to the black hole. At larger distances, in-plane expansion stalls to eventually result in a collapse near apocenter while the vertical width increases faster as $H_{\rm z} \propto R$ (see Fig. \ref{fig:hvsr_outflow}). Contrary to the infalling phase, the gas evolution significantly deviates from ballistic motion due to the influence of pressure forces.

\item The nozzle shock does not result in significant net expansion of the gas during pericenter passage, but rather modifies its transverse density profile. The outgoing stream element is slightly more centrally concentrated with a larger extent than when it was at the same distance from the black hole before pericenter passage. 

\item Numerical convergence is reached for a particle number $N^{\rm 2D}_{\rm p} \approx 10^5$ used to describe the stream element in our two-dimensional approach (see Fig. \ref{fig:hvsnp}). We estimate that resolving the nozzle shock within a three-dimensional simulation of stellar disruption would require $N^{\rm 3D}_{\rm p} \approx 10^{14}$ particles, which does not appear computationally feasible.

\item Our simulation finds that the stream element has a lower vertical width that increases slower with radius than previously assumed. This implies that relativistic nodal precession may be able to prevent the self-crossing shock induced by apsidal precession even for slowly-spinning black holes with $a_{\rm h} \gtrsim 0.1$ (see Fig. \ref{fig:etascvsrint}).

\item If the stream is magnetized, the nozzle shock can be affected by the additional support from magnetic pressure against vertical collapse. We estimate that this effect becomes important for stellar magnetic field strength along the direction of stream elongation of about $B_{\parallel\star} \gtrsim 400 \gauss$, which may be reached for solar-type stars.

\end{enumerate}

\appendix

\section{Equations of motion}
\label{ap:equations}

We explain here how to derive equations \eqref{eq:xiacc} and \eqref{eq:vparadot} that are used to follow the transverse evolution of the stream element along its trajectory in the frame co-moving with its center of mass and co-rotating with the local longitudinal direction of stream elongation. We start by defining the vector
\be
\vect{r}_{\chi} = \chi \vect{e}_{\parallel} + \xi \vect{e}_{\perp},
\label{eq:rchi}
\ee
which gives the relative position of a part of the stream element with respect to the center of mass. Note that we restrict for clarity the motion to the equatorial plane by setting $z=0$, which is possible since the vertical evolution is completely decoupled. The two orthogonal unit vectors are defined as indicated in the sketch of Fig. \ref{fig:sketch} with the parallel one aligned with the center of mass velocity, that is $\vect{e}_{\parallel} = \vect{v}_{\rm c} / v_{\rm c}$. Note that there is an offset in the parallel direction, which must be conserved in the calculation that follows but will be then set to $\chi = 0$ to obtain the equations of motion for an element that remains on the surface orthogonal to the stream longitudinal direction. Taking the time derivative of equation \eqref{eq:rchi} in the inertial frame provides the velocity of the fluid element at this location, which can be written
\be
\vect{v}^{\chi}_{\rm f} \equiv \dot{\vect{r}}_{\chi} = \dot{\chi} \vect{e}_{\parallel} + \chi \dot{\vect{e}}_{\parallel} +  \dot{\xi} \vect{e}_{\perp} + \xi \dot{\vect{e}}_{\perp}.
\label{eq:vchif}
\ee
In the following, we make this expression more explicit in order to relate it to the transverse component of the velocity needed to evolve the gas on the orthogonal surface.

Due to their rotation, the unit vectors have a derivative in the inertial frame that can be obtained from $\dot{\vect{e}}_{\parallel} = \diff (\vect{v}_{\rm c} / v_{\rm c}) / \diff t =  (\vect{a}_{\rm c} \cdot \vect{e}_{\perp} / v_{\rm c}) \vect{e}_{\perp} = \Omega \vect{e}_{\perp}$, where 
\be
\Omega = -\frac{G \mh \sin \delta}{v_{\rm c} R^2},
\label{eq:omega}
\ee
denotes therefore the angular frequency of rotation. It is obtained from the acceleration $\vect{a}_{\rm c} = \dot{\vect{v}}_{\rm c} = -G \mh \vect{e}_{\rm r} / R^2$ experienced by the center of mass from the black hole and making use of $\sin \delta = \vect{e}_{\rm r}\cdot \vect{e}_{\perp}<0$ according to the definition of this angle shown in Fig. \ref{fig:sketch}. Similarly, one can show that $\dot{\vect{e}}_{\perp} = - \Omega \vect{e}_{\parallel}$. The derivative of the longitudinal coordinate above can also be written as
\be
\dot{\chi} = V + \lambda \chi,
\label{eq:chidot}
\ee
by separating between its component at the orthogonal surface where $\chi = 0$ whose value $V$ is computed below and that due to local stretching or compression. Mass conservation imposes that the longitudinal offset varies proportionally to the center of mass speed, which yields $\lambda = \dot{v}_{\rm c} /v_{\rm c} = \vect{a}_{\rm c} \cdot \vect{e}_{\parallel} / v_{\rm c}$. Including the gravitational attraction similarly to above, this leads to
\be
\lambda = -\frac{G \mh \cos \delta}{v_{\rm c} R^2},
\ee
which denotes the rate of longitudinal stretching, using $\cos \delta = \vect{e}_{\rm r}\cdot \vect{e}_{\parallel}$

The velocity of the fluid element given by equation \eqref{eq:vchif} can then be rewritten
\be
\vect{v}^{\chi}_{\rm f} = V \vect{e}_{\parallel} + \chi \boldgreek{\Gamma}  + \vect{v},
\label{eq:vchif_explicit}
\ee
substituting the above relations and additionally defining the quantity 
\be
\boldgreek{\Gamma} = \lambda \vect{e}_{\parallel} + \Omega \vect{e}_{\perp},
\ee
that combines the angular frequency and the rate of stretching. Here, we have also introduced the transverse component of the velocity
\be
\vect{v} = \dot{\xi} \vect{e}_{\perp} -  \xi \Omega \vect{e}_{\parallel},
\label{eq:vtrans}
\ee
which is confined to the orthogonal surface. It is this velocity which is used to follow the transverse evolution of the stream element as it orbits the black hole in the two-dimensional simulation presented in the paper. As explained in Section \ref{sec:treatment}, it describes the motion of fluid elements that change with time since the orthogonal surface is continuously crossed by gas due to shearing. The expression for the transverse velocity can be obtained by inverting equation \eqref{eq:vchif_explicit} and setting $\chi = 0$, which gives
\be
\vect{v} = \vect{v}_{\rm f} -V \vect{e}_{\parallel},
\label{eq:vtrans_final}
\ee
where $\vect{v}_{\rm f} = \lim_{\chi \rightarrow 0} \vect{v}^{\chi}_{\rm f}$ is the velocity of a fixed fluid element at the orthogonal surface. Projecting this relation on the parallel direction and using equation \eqref{eq:vtrans} leads to
\be
V = v_{\rm f \parallel} + \xi \Omega,
\label{eq:bigv}
\ee
where we have defined the projected velocity $v_{\rm f \parallel} = \vect{v}_{\rm f} \cdot \vect{e}_{\parallel}$, whose evolution is determined below.

Equation \eqref{eq:vchif_explicit} can be written $\vect{v}^{\chi}_{\rm f} = \vect{v}_{\rm f} + \chi \boldgreek{\Gamma}$, which can be derived with respect to time to obtain the acceleration of the fluid element in the inertial frame
\be
\vect{a}^{\chi}_{\rm f} \equiv \dot{\vect{v}}^{\chi}_{\rm f} = \vect{a}_{\rm f} + \dot{\chi} \boldgreek{\Gamma} + \chi \dot{\boldgreek{\Gamma}},
\ee
where we have defined the acceleration $\vect{a}_{\rm f} = \dot{\vect{v}}_{\rm f}$. Setting $\chi = 0$ and inverting the above relation then leads to 
\be
\vect{a}_{\rm f} = \vect{a}_{\rm i} - V \boldgreek{\Gamma},
\label{eq:accf_final}
\ee
making use of equation \eqref{eq:chidot} and realizing that $\lim_{\chi \rightarrow 0} \vect{a}^{\chi}_{\rm f} =  \vect{a}_{\rm i}$ is the inertial acceleration from external forces. The above equations \eqref{eq:vtrans_final} and \eqref{eq:accf_final} can be shown to be identical to equations (2.3) and (2.4) of \citet{kochanek1994}, respectively. However, we use different notations that we find more convenient for the specific purpose of the present paper. For our problem, the inertial acceleration is specified by pressure gradients and the tidal force from the black hole, both evaluated at the orthogonal surface.

We want to follow the transverse gas evolution in the frame co-rotating with the longitudinal direction of the stream element, whose angular frequency is given by
\be
\boldgreek{\Omega} = \Omega \vect{e}_{\rm z},
\ee
as specified by equation \eqref{eq:omega}. In this frame, the location on the orthogonal surface is determined by $\vect{r}' = \xi \vect{e}_{\perp}$ with a corresponding velocity $\vect{v}' = \dot{\xi} \vect{e}_{\perp}$ still restricting the motion to the equatorial plane where $z=0$. The transverse acceleration in this frame is given by $\vect{a}' = \ddot{\xi} \vect{e}_{\perp}$, which is determined from a change of frame by
\be
\vect{a}' = \vect{a} + \vect{a}_{\rm ni},
\label{eq:aprime}
\ee
where $\vect{a} = \dot{\vect{v}}$ is the derivative of the transverse velocity and the non-inertial acceleration is given by 
\be
\vect{a}_{\rm ni} = -2 \boldgreek{\Omega} \times \vect{v}' - \boldgreek{\Omega} \times (\boldgreek{\Omega} \times \vect{r}') - \boldgreek{\dot{\Omega}} \times \vect{r}',
\ee
that contains the centrifugal, Coriolis and Euler forces since the rotation occurs at a time-dependent angular frequency. The acceleration of equation \eqref{eq:aprime} can be computed from the time derivative of equation \eqref{eq:vtrans_final} using equations \eqref{eq:bigv} and \eqref{eq:accf_final}, which leads after some algebra to
\be
\ddot{\xi} = \vect{a}_{\rm i} \cdot \vect{e}_{\perp} - 2 V \Omega + \xi \Omega^2,
\label{eq:xiddot}
\ee
and $\vect{a}'\cdot \vect{e}_{\parallel}=0$ as expected since only the transverse motion is considered by construction. To determine $V$, we also need to know the parallel component of the fluid element velocity, which evolves as $\dot{v}_{{\rm f} \,\parallel} = \diff (\vect{v}_{\rm f} \cdot \vect{e}_{\parallel}) / \diff t = \vect{a}_{\rm f} \cdot \vect{e}_{\parallel} + \Omega \vect{v}_{\rm f} \cdot \vect{e}_{\perp}$ that yields
\be
\dot{v}_{{\rm f} \,\parallel} = \vect{a}_{\rm i} \cdot \vect{e}_{\parallel} + \dot{\xi} \Omega - V \lambda,
\label{eq:vfparadot}
\ee
according to equations \eqref{eq:vtrans_final} and \eqref{eq:accf_final}. Equations \eqref{eq:xiddot} and \eqref{eq:vfparadot} are used to write equations \eqref{eq:xiacc} and \eqref{eq:vparadot} in Section \ref{sec:treatment}, where the inertial acceleration has been explicitely written from the tidal and pressure forces. For simplicity, the parallel component of the velocity is denoted by $v_{\parallel}$ throughout the paper instead of $v_{\rm f \parallel}$, only used in this appendix.

\section{Width scaling}
\label{ap:scaling}

Neglecting pressure forces, the vertical distance of a given part of the stream element element obeys the equation
\be
\ddot{z} = -\frac{G \mh}{R^3} z,
\label{eq:zaccbal}
\ee
which is obtained by simplifying equation \eqref{eq:zacc}. Note that this equation is also valid for the in-plane  distance $\xi$ from the center of mass as long as the coefficient of equation \eqref{eq:xiacc_bal} can be approximated as $B_{\perp} \approx -1$. Approximating the trajectory of the center of mass as radial and parabolic, the center of mass position can be analytically determined by solving $\dot{R} = \sqrt{2 G \mh / R}$, which yields $R =  (9 G \mh t^2 /2)^{1/3}$ that possesses an explicit dependence on time. Injecting this expression in equation \eqref{eq:zaccbal}, one can show\footnote{This can be done by injecting the ansatz $z =z_{\rm in} (R/R_{\rm in})^{\omega}$ inside equation \eqref{eq:zaccbal}, which leads to a quadratic equation with two roots $\omega = 1$ and $\omega = 1/2$ that correspond to two independent solutions of the second-order differential equation.} that the vertical position accepts the general solution
\be
z = \frac{z_{\rm in}}{q+1} \left[ q \frac{R}{R_{\rm in}} + \left(\frac{R}{R_{\rm in}} \right)^{1/2}   \right],
\label{eq:zaccbalsol}
\ee
with $q = -(A_{\rm z,in} + 1/2)/ (A_{\rm z,in} + 1)$, under the two initial conditions $z = z_{\rm in}$ and $\dot{z} = v_{\rm z,in} = A_{\rm z,in} z_{\rm in} v_{\rm c,  in}/ R_{\rm in}$ at $R = R_{\rm in}$. This solution is equivalent to the last two of equations 31 derived by \cite{sari2010} but using specific initial conditions. Here, $q$ represents the weight of the two solutions, which is set by the initial homology factor restricted to the range $-1 \leq A_{\rm z,in} \leq 0$, that corresponds to $q \geq -1/2 $. Note that the above choice with $A_{\rm z,in}\leq0$ corresponds to the situation where the gas gets compressed over time at it approaches the black hole, but a similar solution exists for an initially expansing stream that moves outward. If $A_{\rm z,in} = -1$, then $q \rightarrow + \infty$ and the scaling is linear with $z\propto R$ at all distances. If $-1< A_{\rm z,in} \leq -1/2$, then $q \geq 0$ and is finite. The solution then initially scales close to linearly until a finite radius $R = \sqrt{q} R_{\rm in}$, below which the scaling becomes parabolic with $z\propto R^{1/2}$. If $-1/2 < A_{\rm z,in} \leq 0$, then $-1/2 \leq q < 0$ and the scaling rapidly becomes parabolic after the gas has moved to $R \lesssim R_{\rm in}$ with a transient phase of slower compression.

To apply this result to the infalling phase of the stream element, the most convenient choice is to set the initial conditions at the moment when the tidal force starts dominating self-gravity. Because the gas is still close to hydrostatic equilibrium, the transverse velocities are almost zero, that is $A_{\rm z,in} = A_{\rm z,eq} \approx 0$\footnote{Note that due to variations in the stream density imparted by stretching, the stream element may experience slight compression or expansion while being in hydrostatic equilibrium. This is different from the case of stellar disruption where the star does not have a vertical motion upon reaching its tidal radius since its density is fixed.} at a distance $R_{\rm in} = R_{\rm eq}$ from the black hole. This implies that $q \approx 0$ in equation \eqref{eq:zaccbalsol} such that both transverse widths are expected to reach the scaling $H \propto R^{1/2}$ slightly later when the gas has moved inward to $R\lesssim R_{\rm eq}$, as we indeed find in the hydrodynamical simulation we carried out (see upper panel of Fig. \ref{fig:hrhovsr_infall}). Note that the two scalings identified above can also be qualitatively understood for the vertical width in terms of motion relative to the line where the different orbital planes of the stream element intersect. If its center of mass moves almost parallel to this line like in the infalling phase, the scaling is expected to be parabolic as we just found. Motion orthogonal to this line results instead in a linear scaling with $H_{\rm z} \propto R$, which agrees with the evolution after pericenter passage when the gas reaches large radii (see Fig. \ref{fig:hvsr_outflow}) due to the rotation of the intersection line from the blue long-dashed line to the green dotted one in Fig. \ref{fig:trajectory}.

\section{Orbit intersections}
\label{ap:intersections}

The gas inside the stream element evolves on different orbital planes that are inclined with respect to that of its center of mass. These planes intersect along the same line if the transverse velocity profile is homologous, either corresponding to compression or expansion. We find this description to be a good approximation, but deviations from a homologous profile are possible (see Section \ref{sec:extrapolation}), which would cause the orbital planes to intersect at distinct locations along the trajectory in a desynchronized way \citep{stone2013}. The orbital plane intersection line is parallel to the center of mass velocity at the location where the gas experiences no transverse motion. This is because the vertical extent of the stream is then maximized, implying that its center of mass has reached the largest possible distance from the intersection line where its velocity is necessarily tangential. For the returning debris stream, this condition is satisfied at $R=R_{\rm eq}$ when the tidal force starts to become dominant because the gas is close to hydrostatic equilibrium. As long as $R_{\rm eq} \geq a$, the orientation of the center of mass velocity at this location implies that the stream collapses vertically slightly \textit{before} pericenter passage (see Fig. \ref{fig:trajectory}). We emphasize that this conclusion does not hold for the similar process of strong stellar compression where this collapse generally occurs after pericenter passage \citep{luminet1985}. This difference stems from the fact that the star follows a parabolic orbit instead of a elliptic one for the bound debris. The equilibrium condition is in this case attained at the tidal radius where the center of mass velocity is always oriented such that the parallel intersection line crosses the trajectory past pericenter.\footnote{Note however that if the star is already undergoing compression when reaching its tidal radius, it is possible that the collapse occurs before pericenter \citetext{see panel (d) in figure 7 of the paper by \citealt{stone2013} for an example}.}

The different trajectories of the stream element may also intersect with that of the center of mass along the equatorial plane, which results in an in-plane collapse of this gas. In his pioneering work, \citet{kochanek1994} claims that such intersections occur if the eccentricity satisfies $e> 7-4q$ (given in the text of his section 3.2) where the parameter $q$ specifies the slope of the angular frequency profile that can be written as $q = A_{\parallel} +1$ using our notation. We find that this condition can be obtained by imposing that gas with an offset $\xi = \xi_{\rm p} > 0$ at pericenter has an apocenter distance larger than that of the center of mass, which indeed causes the two trajectories to cross. This can be proved by showing that the relative variation in apocenter distance is given by $\Delta R_{\rm a}/R_{\rm a} = (\xi_{\rm p} / \rp) (2 (2 A_{\parallel} -1 )/(1-e)-1)$, so that $\Delta R_{\rm a}/R_{\rm a} > 0$ yields $e>3-4A_{\parallel} = 7-4q$, as stated above. This condition appears to be marginally satisfied in our situation because the homology factor reaches at pericenter a maximal value of $A_{\parallel} \gtrsim (3 -e)/4 \approx 0.5025$ (see Fig. \ref{fig:vparavsxi}) using the eccentricity $e \approx 0.99$ of the stream element. This confirms that ballistic trajectories are expected to intersect along the equatorial plane, which occurs at the purple point in Fig. \ref{fig:trajectory}. In addition to the difference in apocenter distances, we note that the precise location of this collision is affected by a small spread in apsidal angle $\psi_{\rm i} \approx \mp 10^{-5} \pi$ between the ballistic trajectories imposed by the initial conditions that tends to make them converge past pericenter. As explained in Section \ref{sec:recession}, the gas gets deflected by pressure forces that instead causes a divergence of these outgoing trajectories, additionally enhanced by differential apsidal precession (see Section \ref{sec:relativistic}). Nevertheless, these small modifications do not prevent the in-plane collapse but merely pushes its location even closer to apocenter.

\section*{Acknowledgments}

We thank Roseanne Cheng, Phil Hopkins, Chris Matzner, Ramesh Narayan, Martin Pessah, Sterl Phinney, and Tony Piro for useful discussions. We acknowledge the use of \textsc{splash} \citep{price2007} for producing most of the figures in this paper. This research benefited from interactions at the ZTF Theory Network Meeting, partly funded by the National Science Foundation under Grant No. NSF PHY-1748958. The research of CB was funded by the Gordon and Betty Moore Foundation through Grant GBMF5076. This project has received funding from the European Union’s Horizon 2020 research and innovation programme under the Marie Sklodowska-Curie grant agreement No 836751. WL was supported by the David and Ellen Lee Fellowship at Caltech and the Lyman Spitzer, Jr. Postdoctoral Fellowship at Princeton University. The authors thank the Yukawa Institute for Theoretical Physics at Kyoto University, where this work was initiated during the YITP-T-19-07 International Molecule-type Workshop "Tidal Disruption Events: General Relativistic Transients.” CB is particularly grateful to Richard D. Saxton for recognizing the knotty problem of the nozzle shock at this conference.

\section*{Data availability}

The data underlying this article will be shared on reasonable request to the corresponding author. A public version of
the \textsc{gizmo} code is available at \url{http://www.tapir.caltech.edu/~phopkins/Site/GIZMO.html}.

%%%%%%%%%%%%%%%%%%%%%%%%%%%%%%%%%%%%%%%%%%%%%%%%%%



\bibliographystyle{mnras} 
\bibliography{biblio}

\begin{thebibliography}{}
\makeatletter
\relax
\def\mn@urlcharsother{\let\do\@makeother \do\$\do\&\do\#\do\^\do\_\do\%\do\~}
\def\mn@doi{\begingroup\mn@urlcharsother \@ifnextchar [ {\mn@doi@}
  {\mn@doi@[]}}
\def\mn@doi@[#1]#2{\def\@tempa{#1}\ifx\@tempa\@empty \href
  {http://dx.doi.org/#2} {doi:#2}\else \href {http://dx.doi.org/#2} {#1}\fi
  \endgroup}
\def\mn@eprint#1#2{\mn@eprint@#1:#2::\@nil}
\def\mn@eprint@arXiv#1{\href {http://arxiv.org/abs/#1} {{\tt arXiv:#1}}}
\def\mn@eprint@dblp#1{\href {http://dblp.uni-trier.de/rec/bibtex/#1.xml}
  {dblp:#1}}
\def\mn@eprint@#1:#2:#3:#4\@nil{\def\@tempa {#1}\def\@tempb {#2}\def\@tempc
  {#3}\ifx \@tempc \@empty \let \@tempc \@tempb \let \@tempb \@tempa \fi \ifx
  \@tempb \@empty \def\@tempb {arXiv}\fi \@ifundefined
  {mn@eprint@\@tempb}{\@tempb:\@tempc}{\expandafter \expandafter \csname
  mn@eprint@\@tempb\endcsname \expandafter{\@tempc}}}

\bibitem[\protect\citeauthoryear{Andalman, Liska, Tchekhovskoy, Coughlin  \&
  Stone}{Andalman et~al.}{2020}]{andalman2020}
Andalman Z.~L.,  Liska M. T.~P.,  Tchekhovskoy A.,  Coughlin E.~R.,   Stone N.,
   2020, preprint, 20, 1 (\mn@eprint {arXiv} {2008.04922})

\bibitem[\protect\citeauthoryear{Ayal, Livio  \& Piran}{Ayal
  et~al.}{2000}]{ayal2000}
Ayal S.,  Livio M.,   Piran T.,  2000, \mn@doi [\apj] {10.1086/317835}, 545,
  772

\bibitem[\protect\citeauthoryear{Balbus \& Hawley}{Balbus \&
  Hawley}{1991}]{balbus1991}
Balbus S.~A.,  Hawley J.~F.,  1991, \mn@doi [\apj] {10.1086/170270}, 376, 214

\bibitem[\protect\citeauthoryear{Bonnerot \& Lu}{Bonnerot \&
  Lu}{2020}]{bonnerot2020-realistic}
Bonnerot C.,  Lu W.,  2020, \mn@doi [\mnras] {10.1093/mnras/staa1246}, 495,
  1374

\bibitem[\protect\citeauthoryear{Bonnerot \& Stone}{Bonnerot \&
  Stone}{2021}]{bonnerot2021-review}
Bonnerot C.,  Stone N.~C.,  2021, \mn@doi [Space Sci. Rev.]
  {10.1007/s11214-020-00789-1}, 217

\bibitem[\protect\citeauthoryear{Bonnerot, Rossi, Lodato  \& Price}{Bonnerot
  et~al.}{2016}]{bonnerot2016-circ}
Bonnerot C.,  Rossi E.~M.,  Lodato G.,   Price D.~J.,  2016, \mn@doi [\mnras]
  {10.1093/mnras/stv2411}, 455, 2253

\bibitem[\protect\citeauthoryear{Bonnerot, Rossi  \& Lodato}{Bonnerot
  et~al.}{2017a}]{bonnerot2017-stream}
Bonnerot C.,  Rossi E.~M.,   Lodato G.,  2017a, \mn@doi [\mnras]
  {10.1093/mnras/stw2547}, 15, 1

\bibitem[\protect\citeauthoryear{Bonnerot, Price, Lodato  \& Rossi}{Bonnerot
  et~al.}{2017b}]{bonnerot2017-magnetic}
Bonnerot C.,  Price D.~J.,  Lodato G.,   Rossi E.~M.,  2017b, \mn@doi [\mnras]
  {10.1093/mnras/stx1210}, 469, 4879

\bibitem[\protect\citeauthoryear{Bonnerot, Lu  \& Hopkins}{Bonnerot
  et~al.}{2021}]{bonnerot2021-light}
Bonnerot C.,  Lu W.,   Hopkins P.~F.,  2021, \mn@doi [\mnras]
  {10.1093/mnras/stab398}, 1

\bibitem[\protect\citeauthoryear{Brassart \& Luminet}{Brassart \&
  Luminet}{2008}]{brassart2008}
Brassart M.,  Luminet J.~P.,  2008, \mn@doi [\aap]
  {10.1051/0004-6361:20078264}, 481, 259

\bibitem[\protect\citeauthoryear{Brassart \& Luminet}{Brassart \&
  Luminet}{2010}]{brassart2010}
Brassart M.,  Luminet J.-P.,  2010, \mn@doi [\aap]
  {10.1051/0004-6361/200913442}, 511, A80

\bibitem[\protect\citeauthoryear{Carter \& Luminet}{Carter \&
  Luminet}{1982}]{carter1982}
Carter B.,  Luminet J.~P.,  1982, \mn@doi [Nature] {10.1038/296211a0}, 296, 211

\bibitem[\protect\citeauthoryear{Chan, Krolik  \& Piran}{Chan
  et~al.}{2018}]{chan2018}
Chan C.-H.,  Krolik J.~H.,   Piran T.,  2018, \mn@doi [\apj]
  {10.3847/1538-4357/aab15c}, 856, 12

\bibitem[\protect\citeauthoryear{Cheng \& Bogdanovi{\'{c}}}{Cheng \&
  Bogdanovi{\'{c}}}{2014}]{cheng2014}
Cheng R.~M.,  Bogdanovi{\'{c}} T.,  2014, \mn@doi [\prd]
  {10.1103/PhysRevD.90.064020}, 90, 064020

\bibitem[\protect\citeauthoryear{Chornock et~al.,}{Chornock
  et~al.}{2014}]{chornock2014}
Chornock R.,  et~al., 2014, \mn@doi [\apj] {10.1088/0004-637X/780/1/44}, 780,
  44

\bibitem[\protect\citeauthoryear{Clarke \& Carswell}{Clarke \&
  Carswell}{2007}]{clarke2007}
Clarke C.,  Carswell B.,  2007, {Principles of Astrophysical Fluid Dynamics}.
Cambridge University Press, Cambridge

\bibitem[\protect\citeauthoryear{Coughlin \& Nixon}{Coughlin \&
  Nixon}{2015}]{coughlin2015-variability}
Coughlin E.~R.,  Nixon C.,  2015, \mn@doi [\apjl]
  {10.1088/2041-8205/808/1/L11}, 808, L11

\bibitem[\protect\citeauthoryear{Coughlin, Nixon, Begelman  \&
  Armitage}{Coughlin et~al.}{2016a}]{coughlin2016-structure}
Coughlin E.~R.,  Nixon C.,  Begelman M.~C.,   Armitage P.~J.,  2016a, \mn@doi
  [\mnras] {10.1093/mnras/stw770}, 17, 1

\bibitem[\protect\citeauthoryear{Coughlin, Nixon, Begelman, Armitage  \&
  Price}{Coughlin et~al.}{2016b}]{coughlin2016-pancakes}
Coughlin E.~R.,  Nixon C.,  Begelman M.~C.,  Armitage P.~J.,   Price D.~J.,
  2016b, \mn@doi [\mnras] {10.1093/mnras/stv2511}, 455, 3612

\bibitem[\protect\citeauthoryear{Dai, Escala  \& Coppi}{Dai
  et~al.}{2013}]{dai2013}
Dai L.,  Escala A.,   Coppi P.,  2013, \mn@doi [\apj]
  {10.1088/2041-8205/775/1/L9}, 775, L9

\bibitem[\protect\citeauthoryear{Dai, McKinney  \& Miller}{Dai
  et~al.}{2015}]{dai2015}
Dai L.,  McKinney J.~C.,   Miller M.~C.,  2015, \mn@doi [\apj]
  {10.1088/2041-8205/812/2/L39}, 812, L39

\bibitem[\protect\citeauthoryear{Evans \& Kochanek}{Evans \&
  Kochanek}{1989}]{evans1989}
Evans C.~R.,  Kochanek C.~S.,  1989, \mn@doi [\apj] {10.1086/185567}, 346, L13

\bibitem[\protect\citeauthoryear{Gezari et~al.,}{Gezari
  et~al.}{2012}]{gezari2012}
Gezari S.,  et~al., 2012, \mn@doi [Nature] {10.1038/nature10990}, 485, 217

\bibitem[\protect\citeauthoryear{Gezari, Cenko  \& Arcavi}{Gezari
  et~al.}{2017}]{gezari2017-15oi}
Gezari S.,  Cenko S.~B.,   Arcavi I.,  2017, \mn@doi [\apjl]
  {10.3847/2041-8213/aaa0c2}, 851, L47

\bibitem[\protect\citeauthoryear{Guillochon \& McCourt}{Guillochon \&
  McCourt}{2017}]{guillochon2017-magnetic}
Guillochon J.,  McCourt M.,  2017, \mn@doi [\apj]
  {10.3847/2041-8213/834/2/L19}, 834, L19

\bibitem[\protect\citeauthoryear{Guillochon \& Ramirez-Ruiz}{Guillochon \&
  Ramirez-Ruiz}{2013}]{guillochon2013}
Guillochon J.,  Ramirez-Ruiz E.,  2013, \mn@doi [\apj]
  {10.1088/0004-637X/767/1/25}, 767, 25

\bibitem[\protect\citeauthoryear{Guillochon \& Ramirez-Ruiz}{Guillochon \&
  Ramirez-Ruiz}{2015}]{guillochon2015}
Guillochon J.,  Ramirez-Ruiz E.,  2015, \mn@doi [\apj]
  {10.1088/0004-637X/809/2/166}, 809, 166

\bibitem[\protect\citeauthoryear{Guillochon, Ramirez-Ruiz, Rosswog  \&
  Kasen}{Guillochon et~al.}{2009}]{guillochon2009}
Guillochon J.,  Ramirez-Ruiz E.,  Rosswog S.,   Kasen D.,  2009, \mn@doi [\apj]
  {10.1088/0004-637X/705/1/844}, 705, 844

\bibitem[\protect\citeauthoryear{Guillochon, Manukian  \&
  Ramirez-Ruiz}{Guillochon et~al.}{2014}]{guillochon2014-10jh}
Guillochon J.,  Manukian H.,   Ramirez-Ruiz E.,  2014, \mn@doi [\apj]
  {10.1088/0004-637X/783/1/23}, 783, 23

\bibitem[\protect\citeauthoryear{Hayasaki, Stone  \& Loeb}{Hayasaki
  et~al.}{2013}]{hayasaki2013}
Hayasaki K.,  Stone N.,   Loeb A.,  2013, \mn@doi [\mnras]
  {10.1093/mnras/stt871}, 434, 909

\bibitem[\protect\citeauthoryear{Hayasaki, Stone  \& Loeb}{Hayasaki
  et~al.}{2016}]{hayasaki2016-spin}
Hayasaki K.,  Stone N.,   Loeb A.,  2016, \mn@doi [\mnras]
  {10.1093/mnras/stw1387}, 461, 3760

\bibitem[\protect\citeauthoryear{Hobson, Efstathiou  \& Lasenby}{Hobson
  et~al.}{2006}]{hobson2006}
Hobson M.,  Efstathiou G.,   Lasenby A.,  2006, {General Relativity: An
  Introduction for Physicists}.
Cambridge University Press, Cambridge

\bibitem[\protect\citeauthoryear{Holoien et~al.,}{Holoien
  et~al.}{2019}]{holoien2019}
Holoien T. W.~S.,  et~al., 2019, preprint (\mn@eprint {arXiv} {1904.09293})

\bibitem[\protect\citeauthoryear{Hopkins}{Hopkins}{2015}]{hopkins2015}
Hopkins P.~F.,  2015, \mn@doi [\mnras] {10.1093/mnras/stv195}, 450, 53

\bibitem[\protect\citeauthoryear{Hung et~al.,}{Hung et~al.}{2020}]{hung2020}
Hung T.,  et~al., 2020, preprint (\mn@eprint {arXiv} {2003.09427})

\bibitem[\protect\citeauthoryear{Jiang, Guillochon  \& Loeb}{Jiang
  et~al.}{2016}]{jiang2016}
Jiang Y.-F.,  Guillochon J.,   Loeb A.,  2016, \mn@doi [\apj]
  {10.3847/0004-637X/830/2/125}, 830, 125

\bibitem[\protect\citeauthoryear{Kasen \& Ramirez-Ruiz}{Kasen \&
  Ramirez-Ruiz}{2010}]{kasen2010}
Kasen D.,  Ramirez-Ruiz E.,  2010, \mn@doi [\apj]
  {10.1088/0004-637X/714/1/155}, 714, 155

\bibitem[\protect\citeauthoryear{Kim, Park  \& Lee}{Kim et~al.}{1999}]{kim1999}
Kim S.~S.,  Park M.,   Lee H.~M.,  1999, \mn@doi [\apj] {10.1086/307394}, 519,
  647

\bibitem[\protect\citeauthoryear{Kobayashi, Laguna, Phinney  \&
  Meszaros}{Kobayashi et~al.}{2004}]{kobayashi2004}
Kobayashi S.,  Laguna P.,  Phinney E.~S.,   Meszaros P.,  2004, \mn@doi [\apj]
  {10.1086/424684}, 615, 855

\bibitem[\protect\citeauthoryear{Kochanek}{Kochanek}{1994}]{kochanek1994}
Kochanek C.~S.,  1994, \mn@doi [\apj] {10.1086/173745}, 422, 508

\bibitem[\protect\citeauthoryear{Komossa et~al.,}{Komossa
  et~al.}{2008}]{komossa2008-j095}
Komossa S.,  et~al., 2008, \mn@doi [\apj] {10.1086/588281}, 678, L13

\bibitem[\protect\citeauthoryear{Lee \& Kim}{Lee \& Kim}{1996}]{lee1996_tvd}
Lee H.~M.,  Kim S.~S.,  1996, J. Korean Astron. Soc.

\bibitem[\protect\citeauthoryear{Lee, Kang  \& Ryu}{Lee et~al.}{1996}]{lee1996}
Lee H.~M.,  Kang H.,   Ryu D.,  1996, \mn@doi [\apj] {10.1086/177305}, 464, 131

\bibitem[\protect\citeauthoryear{Leloudas et~al.,}{Leloudas
  et~al.}{2016}]{leloudas2016}
Leloudas G.,  et~al., 2016, \mn@doi [Nat. Astron.] {10.1038/s41550-016-0002},
  1, 0002

\bibitem[\protect\citeauthoryear{Leloudas et~al.,}{Leloudas
  et~al.}{2019}]{leloudas2019}
Leloudas G.,  et~al., 2019, \mn@doi [\apj] {10.3847/1538-4357/ab5792}, 887, 218

\bibitem[\protect\citeauthoryear{Liptai, Price, Mandel  \& Lodato}{Liptai
  et~al.}{2019}]{liptai2019-spin}
Liptai D.,  Price D.~J.,  Mandel I.,   Lodato G.,  2019, preprint, 12, 1
  (\mn@eprint {arXiv} {1910.10154})

\bibitem[\protect\citeauthoryear{Lodato, King  \& Pringle}{Lodato
  et~al.}{2009}]{lodato2009}
Lodato G.,  King a.~R.,   Pringle J.~E.,  2009, \mn@doi [\mnras]
  {10.1111/j.1365-2966.2008.14049.x}, 392, 332

\bibitem[\protect\citeauthoryear{Lu \& Bonnerot}{Lu \& Bonnerot}{2020}]{lu2020}
Lu W.,  Bonnerot C.,  2020, \mn@doi [\mnras] {10.1093/mnras/stz3405}, 492, 686

\bibitem[\protect\citeauthoryear{Luminet \& Marck}{Luminet \&
  Marck}{1985}]{luminet1985}
Luminet J.-P.,  Marck J.-a.,  1985, \mn@doi [\mnras] {10.1093/mnras/212.1.57},
  212, 57

\bibitem[\protect\citeauthoryear{Lynch \& Ogilvie}{Lynch \&
  Ogilvie}{2021a}]{lynch2021}
Lynch E.~M.,  Ogilvie G.~I.,  2021a, \mn@doi [\mnras] {10.1093/mnras/staa3459},
  500, 4110

\bibitem[\protect\citeauthoryear{Lynch \& Ogilvie}{Lynch \&
  Ogilvie}{2021b}]{lynch2021_magnetic}
Lynch E.~M.,  Ogilvie G.~I.,  2021b, \mn@doi [\mnras] {10.1093/mnras/staa4026},
  501, 5500

\bibitem[\protect\citeauthoryear{Matzner, Levin  \& Ro}{Matzner
  et~al.}{2013}]{matzner2013}
Matzner C.~D.,  Levin Y.,   Ro S.,  2013, \mn@doi [\apj]
  {10.1088/0004-637X/779/1/60}, 779, 60

\bibitem[\protect\citeauthoryear{Panuelos, Wadsley  \& Kevlahan}{Panuelos
  et~al.}{2020}]{panuelos2020}
Panuelos J.,  Wadsley J.,   Kevlahan N.,  2020, \mn@doi [J. Comput. Phys.]
  {10.1016/j.jcp.2020.109454}, 414, 109454

\bibitem[\protect\citeauthoryear{Price}{Price}{2007}]{price2007}
Price D.~J.,  2007, \mn@doi [PASA] {10.1071/AS07022}, 24, 159

\bibitem[\protect\citeauthoryear{Ramirez-Ruiz \& Rosswog}{Ramirez-Ruiz \&
  Rosswog}{2009}]{ramirez-ruiz2009}
Ramirez-Ruiz E.,  Rosswog S.,  2009, \mn@doi [\apj]
  {10.1088/0004-637X/697/2/L77}, 697, L77

\bibitem[\protect\citeauthoryear{Rees}{Rees}{1988}]{rees1988}
Rees M.~J.,  1988, \mn@doi [Nature] {10.1038/333523a0}, 333, 523

\bibitem[\protect\citeauthoryear{Rosswog, Ramirez-Ruiz  \& Hix}{Rosswog
  et~al.}{2009}]{rosswog2009}
Rosswog S.,  Ramirez-Ruiz E.,   Hix W.~R.,  2009, \mn@doi [\apj]
  {10.1088/0004-637X/695/1/404}, 695, 404

\bibitem[\protect\citeauthoryear{Sadowski, Tejeda, Gafton, Rosswog  \&
  Abarca}{Sadowski et~al.}{2016}]{sadowski2016}
Sadowski A.,  Tejeda E.,  Gafton E.,  Rosswog S.,   Abarca D.,  2016, \mn@doi
  [\mnras] {10.1093/mnras/stw589}, 458, 4250

\bibitem[\protect\citeauthoryear{Sari, Kobayashi  \& Rossi}{Sari
  et~al.}{2010}]{sari2010}
Sari R.,  Kobayashi S.,   Rossi E.~M.,  2010, \mn@doi [\apj]
  {10.1088/0004-637X/708/1/605}, 708, 605

\bibitem[\protect\citeauthoryear{Saxton, Read, Komossa, Lira, Alexander  \&
  Wieringa}{Saxton et~al.}{2017}]{saxton2017}
Saxton R.~D.,  Read A.~M.,  Komossa S.,  Lira P.,  Alexander K.~D.,   Wieringa
  M.~H.,  2017, \mn@doi [\aap] {10.1051/0004-6361/201629015}, 598, A29

\bibitem[\protect\citeauthoryear{Shiokawa, Krolik, Cheng, Piran  \&
  Noble}{Shiokawa et~al.}{2015}]{shiokawa2015}
Shiokawa H.,  Krolik J.~H.,  Cheng R.~M.,  Piran T.,   Noble S.~C.,  2015,
  \mn@doi [\apj] {10.1088/0004-637X/804/2/85}, 804, 85

\bibitem[\protect\citeauthoryear{Short et~al.,}{Short et~al.}{2020}]{short2020}
Short P.,  et~al., 2020, preprint, 19, 1 (\mn@eprint {arXiv} {2003.05470})

\bibitem[\protect\citeauthoryear{Steinberg, Coughlin, Stone  \&
  Metzger}{Steinberg et~al.}{2019}]{steinberg2019}
Steinberg E.,  Coughlin E.~R.,  Stone N.~C.,   Metzger B.~D.,  2019, \mn@doi
  [Mon. Not. R. Astron. Soc. Lett.] {10.1093/mnrasl/slz048}, 485, L146

\bibitem[\protect\citeauthoryear{Stone, Sari  \& Loeb}{Stone
  et~al.}{2013}]{stone2013}
Stone N.,  Sari R.,   Loeb A.,  2013, \mn@doi [\mnras] {10.1093/mnras/stt1270},
  435, 1809

\bibitem[\protect\citeauthoryear{Stone, Kesden, Cheng  \& van Velzen}{Stone
  et~al.}{2019}]{stone2019}
Stone N.~C.,  Kesden M.,  Cheng R.~M.,   van Velzen S.,  2019, \mn@doi [Gen.
  Relativ. Gravit.] {10.1007/s10714-019-2510-9}, 51, 30

\bibitem[\protect\citeauthoryear{Zanazzi \& Ogilvie}{Zanazzi \&
  Ogilvie}{2020}]{zanazzi2020}
Zanazzi J.~J.,  Ogilvie G.~I.,  2020, \mn@doi [\mnras]
  {10.1093/mnras/staa3127}, 499, 5562

\bibitem[\protect\citeauthoryear{Zel'dovich \& Raizer}{Zel'dovich \&
  Raizer}{1967}]{zeldovich1967}
Zel'dovich Y.~B.,  Raizer Y.~P.,  1967, {Physics of shock waves and
  high-temperature hydrodynamic phenomena}.
Academic Press, New York

\bibitem[\protect\citeauthoryear{van Velzen et~al.,}{van Velzen
  et~al.}{2019}]{van_velzen2019}
van Velzen S.,  et~al., 2019, \mn@doi [\apj] {10.3847/1538-4357/aafe0c}, 872,
  198

\makeatother
\end{thebibliography}

%%%%%%%%%%%%%%%%%%%%%%%%%%%%%%%%%%%%%%%%%%%%%%%%%%

%%%%%%%%%%%%%%%%% APPENDICES %%%%%%%%%%%%%%%%%%%%%

%%%%%%%%%%%%%%%%%%%%%%%%%%%%%%%%%%%%%%%%%%%%%%%%%%

% Don't change these lines
\bsp	% typesetting comment
\label{lastpage}

%\appendix 
%\section{}

\end{document}